\documentclass[a4paper,12pt]{article}

\usepackage[pdftex]{graphicx}
\usepackage{subfigure}

\newcommand{\be}{\begin{equation}}
\newcommand{\ee}{\end{equation}}

\newcommand{\bea}{\begin{eqnarray}}
\newcommand{\eea}{\end{eqnarray}}

\newcommand{\p}{\partial}

\newcommand{\nn}{\nonumber \\}
\newcommand{\f}{\frac}
\newcommand{\w}{\wedge}

\begin{document}
\thispagestyle{empty}
\begin{flushright}
{\bf arXiv: 1108.5577}
\end{flushright}
\begin{center} \noindent \Large \bf
RG flow of transport quantities
\end{center}

\bigskip\bigskip\bigskip
\vskip 0.5cm
\begin{center}
{ \normalsize \bf Bum-Hoon Lee${}^{1,2}$,  Shesansu Sekhar Pal${}^2$ and Sang-Jin Sin${}^3$}


\vskip 0.5 cm
${}^1$Department of Physics, Sogang University, Seoul 121-742, South Korea\\
${}^2$Center for Quantum Spacetime, 
Sogang University, 121-742, Seoul, South Korea\\
${}^3$Department of Physics, Hanyang University, Seoul 133-791, South Korea
\vskip 0.5 cm
\sf bhl${\frame{\shortstack{AT}}}$sogang.ac.kr,  shesansu${\frame{\shortstack{AT}}}$gmail.com and sangjin.sin${\frame{\shortstack{AT}}}$gmail.com
\end{center}
\centerline{\bf \small Abstract}
The RG flow equation of various transport quantities are studied in arbitrary spacetime dimensions, in the fixed as well as fluctuating background geometry both for the  Maxwellian and DBI type of actions. The regularity condition on the flow equation of the conductivity at the horizon for the DBI  action reproduces naturally the leading order result of {\it Hartnoll et al.}, [{\it JHEP}, {\bf 04}, 120 (2010)].
Motivated by the result of {\it van der Marel et al.}, [{\it science}, {\bf 425}, 271 (2003], we studied, analytically, the conductivity versus frequency plane by dividing  it into three distinct parts: $\omega <T,~\omega >T$ and $\omega >> T$. In order to compare, we choose $3+1$ dimensional bulk spacetime for the  computation of the conductivity. 
In the $\omega <T$ range, the conductivity does not show up the Drude like form  in any spacetime dimensions. In the $\omega > T$ range and staying away from the  horizon, for the DBI action with unit dynamical exponent, non-zero magnetic field and charge density,  the conductivity goes as $\omega^{-2/3}$, whereas the phase of the conductivity, goes as, $ArcTan(Im\sigma^{xx}/Re\sigma^{xx})=\pi/6$ and  $ArcTan(Im\sigma^{xy}/Re\sigma^{xy})=-\pi/3$. There exists a universal quantity  at the horizon that is the phase angle of conductivity, which  either vanishes or an integral multiple of $\pi$. Furthermore, we calculate  the temperature dependence to the thermoelectric and the thermal conductivity at the horizon. The charge diffusion constant for the  DBI action  is studied.

\newpage
\tableofcontents
\section{Introduction}

Several aspects of the   non-Fermi Liquid (NFL)
has been computed successfully, using AdS/CFT correspondence \cite{jm}. 
In particular, the temperature dependence to the  longitudinal and Hall dc conductivities.
However, at the moment, it is  not yet possible to gather all  the properties  of NFL, theoretically, in one concrete example. Let us recall the properties of high $T_c$ superconductor that has been found experimentally in \cite{dvdm}. The entire graph of the conductivity vs frequency plane is separated into three distinct parts, depending on the range of the parameters. In this study, apart from frequency the other dimension full parameters  are temperature, $T$, and a scale $\Omega$, which we denote it as chemical potential, $\mu$. The  regions are classified depending on the value of the frequency, (1) low frequency, $\omega < T$,
(2) mid-frequency range, $T < \omega < \Omega\equiv\mu$, (3) high frequency range, $\omega > \Omega$.

In the low frequency range, the conductivity curve has been fitted to 
\be
\sigma \sim \f{1}{T[1+(\omega/T)^2]}=\f{1}{T}[1-(\omega/T)^2+\cdots],
\ee
which is reminiscent of the Drude type behavior. In fact, the scattering rate comes out to be $\tau^{-1}_{DvdM}\sim T$, see eq(\ref{time_dvdm}), which suggests  there exists only one scale, the temperature. Moreover, the phase angle of the conductivity, $arg(\sigma)$, depends non-trivially on the frequency. Based on this result, there follows a very surprising thing that is even though in NFL, the coupling is generically strong, but in the low frequency range, it somehow conspires to  give the Drude's theory of conductivity of a gas of free electrons.

In the mid frequency range, the conductivity goes as
\be
\sigma \simeq (-i\omega)^{\gamma-2},\quad \gamma\approx 1.35
\ee
up to an overall constants, but  
only for optimally and overdoped samples. Samples that are underdoped, which means with low $T_c$, does not strictly obey this equation, and the phase of the conductivity depends on the frequency. The scattering rate goes as, $\tau^{-1}_{DvdM}\sim \omega^{0.65}$. It follows from the experimental result that  in the high frequency limit, $\omega > T$, the samples that are optimally doped and overdoped the phase of the conductivity does not depend on the frequency. 
 
In the very high frequency range, there does not exists any universal fitting of the conductivity curve. However, it is suggested to be  of a non-universal form like
\be\label{mid_frequency}
\sigma \simeq (-i\omega)^{\gamma-2}(\Omega-i\omega)^{1-\gamma}.
\ee
The scattering rate does depends on the scale $\Omega$ and the fitting of the  the phase of the conductivity curve does not depend on the frequency.

There exists yet another reason to study the conductivity versus frequency plane by dividing it into different parts. 
The conductivity is defined, roughly, as the inverse frequency times the gradient of the logarithm of the gauge field, $A$, i.e., $\sigma\sim i \omega^{-1} \p_r Log~A$. For precise definition see, eq(\ref{def_conductivity}) and eq(\ref{def_sigma_plus_minus_dbi_b_metric_rho}), when the actions are of the Maxwellian and DBI type, respectively. Generically, it is very difficult to solve  the equation of motion to gauge field, analytically, in all regions of the parameter space, simultaneously.  So, we better divide the parameter space,   study them individually and then patch up the results.

Motivated by this experimental result and the inability to solve the equation of motion to guage field analytically, we do a theoretical study of the conductivity versus frequency plane by  dividing it into three distinct parts\footnote{The question about the form of the conductivity when $\omega\sim T$, remains unanswered because of the complexity of the RG flow  equation. }, as mentioned above, with a minor difference to the definition of the conductivity in the very high frequency region. The approach that we adopt to study the behavior of the conductivity  is by writing down a RG flow equation, which is a first order differential equation of Riccati type. Then the  boundary condition is imposed by looking at the regularity of the conductivity at the horizon. This in turn gives us, naturally, the in-falling boundary condition at the horizon for the gauge field.  It is interesting to note that the result of the dc conductivity calculation performed in \cite{hpst} is automatically reproduced from the regularity condition at the horizon. We found this is a big advantage over the other approaches that are available at present \cite{kob} and \cite{ssp}.

In the low frequency region, there exists a naturally  small parameter $\omega/T$, which is smaller than unity. So, it is expected that the conductivity can be expanded in a series form. The consistency of this series expansion with the boundary condition at the horizon suggests it to have a Taylor series expansion,  for the real part of the conductivity. Substituting, the real part of the conductivity into the RG flow equation gives a series form, eq(\ref{im_sigma_z_1_maxwell_d_dim})  for the imaginary part of the conductivity in the Einstein-Maxwell system, whose most dominating term at the small frequency goes as $Im\sigma\sim \omega^{-1}$.  Moreover, the series expansion for the real part of the conductivity  cannot be summed up to take a Drude like form. 

The definition of  the very high frequency region that we adopt  is a bit different from \cite{dvdm}. We shall use, $\omega\rightarrow\infty$, as our definition. The fluctuating gauge field obeys the Schr\"{o}dinger type equation with a very complicated looking potential energy. However, if we stay in  $3+1$ dimensional Einstein-Maxwell system with unit dynamical exponent then there is not any significant contribution  to the potential energy.
In which case, it is very clear that the conductivity remains constant. 

However, in the mid frequency region, both the flow equation for the conductivity and the equation of motion to the gauge field becomes very complicated. Hence it is very difficult to solve the equations analytically. However, by staying in a particular range of the radial coordinate, $r_h\ll r \leq r_{0}$, \cite{hpst}, where $r_h$ is the horizon and $r_{0}$ is another holographic energy scale, in such a way the quantity, $h\simeq 1$, see eq(\ref{RN_black_hole}).   It means to stay far away  from the horizon.

The results of our finding through the study of RG flow of conductivity, which is in some sense a top-down approach because we do not consider any specific form the background fields except the diagonal form. The findings can be summarized and compared with the experimental result, as follows. In doing the comparision,  we shall set the momentum to zero, i.e., $k=0$ and work in $d+1$ dimensional bulk spacetime.

\begin{center}
\begin{tabular}{ | l | l | l | l|}\hline
source& Low frequency & Mid-frequency & High frequency\\ \hline
Expt.  & $Re\sigma\sim \f{1}{T}-\f{\omega^2}{T^3}+\cdots$ &
$Re\sigma\sim (-i\omega)^{\gamma-2}$ & $Re\sigma\sim {\rm constant}$ ?\\ 
\cite{dvdm}&  & $\gamma\approx 1.35$ & $Arg(\sigma) \sim  \pi/2$\\ 
&$Im\sigma$ = Not Known & $Arg(\sigma) \sim 0.325 \pi$& $Im\sigma$ = Not Known\\ \hline\hline
 Einstein- & $Re\sigma\sim T^{d-3}[{\hat a}_0+{\hat a}_1 \f{\omega^2}{T^2}+\cdots$] & For $d=3$ & For $d=3$, \\ 
Maxwell& eq(\ref{re_sigma_z_1_maxwell_d_dim}), and  & &$Re\sigma= \Sigma_A$ \\ 
type of action& $Im\sigma \sim T^{d-3}h(1-u)^{3-d}\times $&$\sigma\sim (i\omega)^{-1}$ &$Arg(\sigma) =  n\pi$, \\ 
with $z=1$& $[\f{\p_uLog{\hat a}_0}{\omega/T}+\p_u({\hat a}_1/{\hat a}_0)(\omega/T)$& &for $n=0,1,\cdots$. \\ 
&$+\cdots]$, eq(\ref{im_sigma_z_1_maxwell_d_dim}) & &$Im\sigma = 0$ \nn \hline\hline
DBI& $Re\sigma\sim {\tilde a}_0+{\tilde a}_1 {\hat\omega}^2+\cdots$ &For $z=1,~B\neq 0$ & $Re\sigma\sim \Sigma_A$\\ 
type of action&  & $\sigma\sim \omega^{-2/3}$& $Arg(\sigma) =  n\pi$\\ 
& $Im\sigma \sim \f{{\tilde b}_0}{{\hat\omega}}+{\tilde b}_1{\hat\omega}+\cdots $& $tan^{-1}(\f{Im\sigma^{xx}}{Re\sigma^{xx}})=\f{\pi}{6}$ &$Im\sigma = 0$ \\ 
& eq(\ref{series_low_frequency_long_hall_dbi_b})& $tan^{-1}(\f{Im\sigma^{xy}}{Re\sigma^{xy}})=-\f{\pi}{3}$& \\ \hline\hline
\end{tabular}
\end{center}
We denote the real and imaginary parts of the conductivity as $Re\sigma$ and  $Im\sigma$, respectively and  have dropped the arguments of the functions ${\hat a}_i$'s, which generically depends on $u$, $q\sim \mu/T$ and  magnetic field, if present. To determine the form of ${\hat a}_i$'s and  ${\tilde a}_i$'s explicitly, we have to  solve the necessary differential equations. Moreover, the ${\hat a}_i$'s and  ${\tilde a}_i$'s are dimensionless.

There are five interesting relations that comes up from the study of RG flow, two are at zero frequency and three are at infinite frequency. (1) The real part of the conductivity at zero frequency becomes independent of the frequency, i.e., $Re\sigma(\omega=0)=a_0$, (2) The imaginary  part of the conductivity at zero frequency goes as, $Im\sigma(\omega\rightarrow 0)=b_0 \omega^{-1}$,  (3) The real part of the conductivity, the  imaginary part of the  conductivity and the phase of the conductivity becomes independent of the frequency at infinite frequency provided we ignore the contribution of the potential energy term in the Schr\"{o}dinger equation for the gauge field.

In the condensed matter literature \cite{ds}, it was argued  that at zero frequency and at infinite frequency, the real part of the conductivity does not depend on the frequency, which is consistent with our result.

There exists a universal quantity at the horizon, namely, the phase angle of the conductivity, defined as the ratio of the imaginary part of the conductivity to the real part of the conductivity, vanishes. This results follows by demanding  the regularity condition on the flow equation to  the conductivity.

We provide a generic formula for the charge diffusion constant for the DBI type of action in $d+1$ dimensional bulk Lifshitz spacetime with dynamical exponent, $z=1$ as
\be
D=\f{L^{2-z}}{(d-1-z)} \bigg(\f{TL}{\alpha}\bigg)^{\f{z-2}{z}}\sqrt{1+\rho^2\bigg(\f{\alpha}{TL}\bigg)^{\f{2d-2}{z}}}{}_2F_1\Bigg[\f{3}{2},\f{1-d+z}{2-2d},\f{3-3d+z}{2-2d},-\rho^2\bigg(\f{TL}{\alpha}\bigg)^{\f{2-2d}{z}}\Bigg],
\ee
where $\rho$ is the charge density, $T$ is the Hawking temperature and $\alpha$ is some numerical constant.

The paper is organized as follows. In section 2, we shall derive the flow equation of the conductivity in the fixed but  generic  background geometry (not necessarily restricted to have the rotational symmetry) with Maxwell action\footnote{This type of RG flow equation is derived   in \cite{il}.  The flow equation for conductivity in $4+1$ dimensional RN AdS black hole  is also found in \cite{bd}.} and study some of its properties. In section 3, we study the flow equation of the  conductivity in the Einstein-Hilbert-Maxwell action by including the back reaction of the background geometry. In section 4, we study the flow equation of the conductivity by considering  the    fixed geometry in the DBI action. Then include the back reaction in section 5 but without the magnetic field.  In section 6, we study the flow equation  of the thermoelectric and thermal conductivity for a system described by Einstein-Maxwell action,
then we conclude in section 7.  The details of the numerical analysis is relegated to Appendices. In Appendix A, we do the analysis for the Einstein-Maxwell system, in Appendix B for the DBI action. In Appendix C, we discuss the relation between the in-falling boundary condition and the regularity condition.  In Appendix D, we calculate the scattering time using a simple form of the Green function.

\section{Flow equation: Fixed geometry}

In this case the gauge field is allowed to fluctuate but not the spacetime geometry and then using the Ohm's law, we shall calculate the conductivity. 

For simplicity we shall start with a trivial gauge field configuration, upon which we shall do the fluctuation. In which case the action is simply the Maxwellian type
\be
S=-\f{1}{4}\int d^{d+1}x \f{\sqrt{-g}}{g^2_{YM}}F_{MN}F^{MN}.
\ee 
We are considering the situation where the YM's coupling, $g_{YM}$, need not be a constant and can depend on the radial coordinate, $r$. The background geometry is assumed to be diagonal and depends only on the radial
coordinate $r$ 
\be\label{metric}
ds^2_{d+1}=g_{MN}dx^Mdx^N=-g_{tt}(r)dt^2+g_{xx}(r)dx^2+\sum^{d-2}_{a=1}g_{ab}(r)
dy^ady^b+g_{rr}(r)dr^2.
\ee
The equation of motion to gauge field and the expression to  currents   are\footnote{The current is written for a constant-r slice. In what follows, we shall try to find the flow equation to the transport quantity as go we from one slice to another. }
\be\label{current}
\p_M\bigg[\f{\sqrt{-g}}{g^2_{YM}}F^{MN} \bigg]=0,~~~
J^{\mu}=-\f{\sqrt{-g}}{g^2_{YM}}F^{r\mu}.
\ee

The convention for the indices are the capital Latin letters run over the $d+1$ dimensional bulk spacetime whereas the Greek letters for the $d$ dimensional field theory, the small Latin letters run over the spatial directions  and the holographic direction is denoted as $r$. It is easy to convince oneself that only in the zero momentum limit the equations of motion to the spatial components of the gauge field, $A_i$,  are same.  Considering the convention of momentum as $k^{\mu}=(\omega,k,0,0,\cdots,0)$, and with the gauge choice $A_r=0$, the equations of motion to gauge field, in Fourier space, are
\bea\label{eom_A_fixed_geometry}
&& \p_r\bigg(\f{\sqrt{-g}}{g^2_{YM}}\f{A'_t}{g_{rr}g_{tt}}\bigg)-\f{\sqrt{-g}}{g^2_{YM}} \f{(k^2A_t+\omega k A_x)}{g_{xx}g_{tt}}=0,\nn &&
 \p_r\bigg(\f{\sqrt{-g}}{g^2_{YM}}\f{A'_x}{g_{rr}g_{xx}}\bigg)+\f{\sqrt{-g}}{g^2_{YM}} \f{(\omega^2A_x+\omega k A_t)}{g_{xx}g_{tt}}=0,\nn
&&\p_r\bigg(\f{\sqrt{-g}}{g^2_{YM}}\f{g^{ab}A'_b}{g_{rr}}\bigg)+\f{\sqrt{-g}}{g^2_{YM}} g^{ab}A_b \f{(\omega^2 g_{xx}- k^2 g_{tt})}{g_{xx}g_{tt}}=0,~~~
\omega g_{xx}A'_t+k g_{tt}A'_x=0.
\eea

Let us consider the variable $E_{||}\equiv\omega A_x+k A_t$, which obeys the following equation  
\be\label{eom_e_pll}
\p_r\bigg[\f{\sqrt{-g}}{g^2_{YM}} \f{E'_{||}}{g_{rr}(\omega^2 g_{xx}-k^2g_{tt})} \bigg]+\f{\sqrt{-g}E_{||}}{g^2_{YM}g_{tt}g_{xx}}=0
\ee

The electrical conductivity along x direction  can be computed, using Ohm's law,  as 
\be\label{conductivity_pll}
\sigma^{xx}=\f{J^x}{E_{x}}=i \f{\sqrt{-g}}{g^2_{YM}} \f{\omega }{g_{rr}(\omega^2 g_{xx}-k^2 g_{tt})}\f{E'_{||}}{E_{||}},
\ee
where we have evaluated $J^x$ using eq(\ref{current}) and the electric field, $E_{x}=\p_xA_t-\p_t A_x=iE_{||}$, in Fourier space.
The  flow equation for 
$\sigma^{xx}$ can be derived using  eq(\ref{eom_e_pll}) 
\be\label{flow_fixed_geometry_pll}
\p_r \sigma^{xx}=i\omega\sqrt{\f{g_{rr}}{g_{tt}}}\Bigg[\f{(\sigma^{xx})^2}{\Sigma_A}\bigg(1-\f{k^2}{\omega^2} \f{g_{tt}}{g_{xx}} \bigg)-\Sigma_A \Bigg],
\ee
where $\Sigma_A=\sqrt{\f{-g}{g_{rr}g_{tt}}}\f{1}{g^2_{YM}g_{xx}}$. Similarly, the flow equation for the transverse part of the conductivity, $\sigma^{yy}\equiv \f{J^y}{E_y}$ (denoting $y_1\equiv y$) is
\be\label{flow_fixed_geometry_pendi}
\p_r \sigma^{yy}=i\omega\sqrt{\f{g_{rr}}{g_{tt}}}\Bigg[\f{(\sigma^{yy})^2}{{\widetilde\Sigma}_A}-{\widetilde\Sigma}_A\bigg(1-\f{k^2}{\omega^2} \f{g_{tt}}{g_{xx}} \bigg) \Bigg],
\ee
where $~E_y=\p_yA_t-\p_tA_y=i\omega A_y$, as there is no momentum along the $y$ direction and ${\widetilde\Sigma}_A=\sqrt{\f{-g}{g_{rr}g_{tt}}}\f{1}{g^2_{YM}g_{yy}}$.
The flow equations (\ref{flow_fixed_geometry_pll}) and (\ref{flow_fixed_geometry_pendi}) are  the generalized flow equation as compared to those derived in \cite{il}, which deals with only  the rotational invariant background geometry. These flow equations  are of the Riccati type  differential equations. These are first order differential equations and requires one boundary condition, in order to find the solution. The RG flow equations should be regular from the horizon to the boundary and the 
imposition of the the regularity criteria at the horizon fixes the boundary condition to be \cite{il}
\be\label{bc_fixed_geometry}
\sigma^{xx}(horizon)=(\Sigma_A)_{horizon},~~~
\sigma^{yy}(horizon)=({\widetilde\Sigma}_A)_{horizon}.
\ee

It is nice to check that the flow equations eq(\ref{flow_fixed_geometry_pll}) and  eq(\ref{flow_fixed_geometry_pendi}) satisfy the time reversal symmetry of \cite{dvdm}, $\sigma(\omega)=\sigma^{\star}(-\omega)$, in the sense that both $\sigma(\omega)$ and $\sigma^{\star}(-\omega)$ satisfy the same differential equation. Here $\star$ means the complex conjugation.

{\bf A consequence}:\\

Interestingly, there follows a very important consequence 
from these flow equations. It is known that in $2+1$ dimensional field theory the electrical conductivity does not depend on the frequency at zero momentum and this result holds true for any value of frequency \cite{hkss}. More importantly, it is  shown to hold in the probe approximation i.e., for  a non fluctuating background geometry.

At zero momentum and at non vanishing frequency the flow equation of $\sigma^{xx}\equiv \sigma$ reduces to the following form 
\be\label{reduced_flow}
\p_r \sigma=i\omega\sqrt{\f{g_{rr}}{g_{tt}}}\Bigg[\f{\sigma^2}{\Sigma_A}-\Sigma_A\Bigg],~~~
\Sigma_A=\f{1}{g^2_{YM}}\sqrt{\f{\prod^{d-2}_1g_{aa}}{g_{xx}}}
\ee
For the background metric of the diagonal form as written in   eq(\ref{metric}), which asymptotes to $AdS_{d+1}$  with the structure 
\be
ds^2(AdS_{d+1})=-r^2 f(r)dt^2+\f{dr^2}{r^2g(r)}+r^2\sum^{d-1}_1 dx^idx^i
\ee
gives,
$\Sigma_A=\f{r^{d-3}}{g^2_{YM}}$, where we have set the $AdS$ radius to unity for convenience. Substituting all these ingredients into eq(\ref{reduced_flow}) results
\be
\p_r\sigma=i\f{\omega}{r^2 \sqrt{f(r)g(r)}}\Bigg[\f{\sigma^2 g^2_{YM}}{r^{d-3}} -\f{r^{d-3}}{g^2_{YM}}\Bigg].
\ee 

From this equation it is easy to convince that for any $3+1$ dimensional bulk spacetime that asymptotes to $AdS$, there exists  an exact   constant conductivity  solution provided  the YM's coupling, $g_{YM}$, is constant. 
\be\label{constant_cond}
\sigma=\pm\f{1}{g^2_{YM}}.
\ee 

So, the frequency independent conductivity in $2+1$ dimensional field theory is an exact result only in the probe approximation as stated earlier provided the coupling is constant. Essentially, it is the real part of the conductivity that goes as the inverse square of the YM's coupling constant and the imaginary part of the conductivity vanishes identically.  For $g^2_{YM}=4\pi$, it reproduces the result of the conductivity found using membrane paradigm \cite{pw}.

\subsection{Sum rules}

The spectral density is defined to be the imaginary part of the retarded Green function: $\rho(\omega,k)=-Im G_R(\omega,k)/\pi$. The Green's function is related to the conductivity as $G_R=-i \omega\sigma$, which means in the special case where there exists an exact solution of the  conductivity i.e., eq(\ref{constant_cond}), results
\be
\rho(\omega,k=0)=-\f{Im G_R}{\pi}=\f{\omega}{\pi g^2_{YM}},
\ee
where we have chosen the positive sign of the conductivity in eq(\ref{constant_cond}), so as to make the spectral function positive.
Hence, the sum rule, which is the sum over all frequencies  of spectral function 
\be
\int^{\infty}_{0} d\omega \bigg(\rho(\omega,k=0)- \f{\omega}{\pi g^2_{YM}}\bigg) =0.
\ee

A form of it is derived in another way in \cite{ghk}.

\subsection{The flow equation for the phase of the conductivity }

In this subsection, we shall analyze the flow equations eq(\ref{flow_fixed_geometry_pll}) and eq(\ref{flow_fixed_geometry_pendi}), We shall  find the flow equation for the phase angle of the conductivity in each case and see whether there exists any universal structure or not.

Equating the real and imaginary parts of the conductivity, $\sigma^{xx}\equiv Re\sigma^{xx}+i~ Im\sigma^{xx}$, results in
\bea
\p_r\bigg(Re\sigma^{xx}\bigg)&=&-\f{2\omega}{\Sigma_A} \sqrt{\f{g_{rr}}{g_{tt}}}\bigg(Re\sigma^{xx}\bigg)~\bigg(Im\sigma^{xx}\bigg),\nn
\p_r\bigg(Im\sigma^{xx}\bigg)&=&\omega\sqrt{\f{g_{rr}}{g_{tt}}}\Bigg[\bigg(\f{(Re\sigma^{xx})^2-(Im\sigma^{xx})^2}{\Sigma_A}\bigg)\bigg(1- \f{k^2}{\omega^2}\f{g_{tt}}{g_{xx}}\bigg)- \Sigma_A\Bigg].
\eea
Now using the first of the flow equation into the second gives
\bea
\p_r\Bigg[\Sigma_A \sqrt{\f{g_{tt}}{g_{rr}}} \f{\p_r Re\sigma^{xx}}{Re\sigma^{xx}} \Bigg]&=&-2\omega^2 \sqrt{\f{g_{rr}}{g_{tt}}}
\Bigg[\f{(Re\sigma^{xx})^2}{\Sigma_A}\bigg(1- \f{k^2}{\omega^2}\f{g_{tt}}{g_{xx}}\bigg)-\nn &&\f{\Sigma_A}{4\omega^2} \f{g_{tt}}{g_{rr}}\bigg(\f{\p_r Re\sigma^{xx}}{Re\sigma^{xx}}\bigg)^2\bigg(1- \f{k^2}{\omega^2}\f{g_{tt}}{g_{xx}}\bigg)-\Sigma_A\Bigg].
\eea
Similarly, equating the real and imaginary part of the conductivity, $\sigma^{yy}$, gives
\bea
\p_r\bigg(Re\sigma^{yy}\bigg)&=&-\f{2\omega}{{\widetilde\Sigma}_A} \sqrt{\f{g_{rr}}{g_{tt}}}\bigg(Re\sigma^{yy}\bigg)~\bigg(Im\sigma^{yy}\bigg),\nn
\p_r\bigg(Im\sigma^{yy}\bigg)&=&\omega\sqrt{\f{g_{rr}}{g_{tt}}}\Bigg[\f{(Re\sigma^{yy})^2-(Im\sigma^{yy})^2}{{\widetilde\Sigma}_A}- {\widetilde\Sigma}_A\bigg(1- \f{k^2}{\omega^2}\f{g_{tt}}{g_{xx}}\bigg)\Bigg].
\eea
The decoupled differential flow equation for $Re\sigma^{yy}$ is
\be
\p_r\Bigg[{\widetilde\Sigma}_A \sqrt{\f{g_{tt}}{g_{rr}}} \f{\p_r Re\sigma^{yy}}{Re\sigma^{yy}} \Bigg]=-2\omega^2 \sqrt{\f{g_{rr}}{g_{tt}}}
\Bigg[\f{(Re\sigma^{xx})^2}{{\widetilde\Sigma}_A}-\f{{\widetilde\Sigma}_A}{4\omega^2} \f{g_{tt}}{g_{rr}}\bigg(\f{\p_r Re\sigma^{yy}}{Re\sigma^{yy}}\bigg)^2-{\widetilde\Sigma}_A\bigg(1- \f{k^2}{\omega^2}\f{g_{tt}}{g_{xx}}\bigg)\Bigg].
\ee
Now defining the phase angle of the conductivity as, $tan\theta^{xx}=Im\sigma^{xx}/Re\sigma^{xx}$, gives us the flow for phase angle as
\bea\label{flow_phase_probe_pll}
\p_r tan\theta^{xx}&=&\f{\omega}{Re\sigma^{xx}}\sqrt{\f{g_{rr}}{g_{tt}}}\Bigg[\f{(Re\sigma^{xx})^2+(Im\sigma^{xx})^2}{\Sigma_A}-\f{(Re\sigma^{xx})^2-(Im\sigma^{xx})^2}{\Sigma_A}\bigg(\f{k^2}{\omega^2}\f{g_{tt}}{g_{xx}}\bigg)- \Sigma_A\Bigg],\nn
&=& \f{\omega}{Re\sigma^{xx}}\sqrt{\f{g_{rr}}{g_{tt}}}\Bigg[\f{(Re\sigma^{xx})^2 sec^2\theta^{xx}}{\Sigma_A}-\f{(Re\sigma^{xx})^2}{\Sigma_A}\f{2cos2\theta^{xx}}{1+cos2\theta^{xx}}\bigg(\f{k^2}{\omega^2}\f{g_{tt}}{g_{xx}}\bigg)- \Sigma_A\Bigg].
\eea
Similarly for the perpendicular phase angle, $tan\theta^{yy}=Im\sigma^{yy}/Re\sigma^{yy}$ 
\bea\label{flow_phase_probe_pendi}
\p_r tan\theta^{yy}&=&\f{\omega}{Re\sigma^{yy}}\sqrt{\f{g_{rr}}{g_{tt}}}\Bigg[\f{(Re\sigma^{yy})^2+(Im\sigma^{yy})^2}{{\widetilde\Sigma}_A}- {\widetilde\Sigma}_A\bigg(1- \f{k^2}{\omega^2}\f{g_{tt}}{g_{xx}}\bigg)\Bigg],\nn
&=&\f{\omega}{Re\sigma^{yy}}\sqrt{\f{g_{rr}}{g_{tt}}}\Bigg[\f{(Re\sigma^{yy})^2 sec^2\theta^{yy}}{{\widetilde\Sigma}_A}- {\widetilde\Sigma}_A\bigg(1- \f{k^2}{\omega^2}\f{g_{tt}}{g_{xx}}\bigg)\Bigg]
\eea

In the zero momentum limit,  the flow equation for the conductivities, $\sigma^{xx}$ and $\sigma^{yy}$, becomes same. So also  the phase angle of the conductivity, $\theta^{xx}$ and $\theta^{yy}$.

\subsection{Behavior at finite momentum and high frequency }

It is always of interest to find any universal structure, if any. In the probe brane approximation, the flow equation of the conductivity does show up a universal feature at finite momentum and at very high frequency, $k\rightarrow$ finite, $\omega\rightarrow\infty$. To proceed, let us demand that at some critical frequency, $\omega\rightarrow\omega_c$, the quantities $\f{d Re\sigma}{d\omega}$ and  $\f{d Im\sigma}{d\omega}$ evaluated at this critical frequency vanishes.
It means 
\be\label{constraint_I_im_sigma_probe}
\f{\Sigma_A}{2\omega^2_c}\sqrt{\f{g_{tt}}{g_{rr}}}\bigg(\f{\p_r Re \sigma^{xx}}{Re \sigma^{xx}}\bigg)_{\omega_c}=0,\quad \f{{\widetilde\Sigma}_A}{2\omega^2_c}\sqrt{\f{g_{tt}}{g_{rr}}}\bigg(\f{\p_r Re \sigma^{yy}}{Re \sigma^{yy}}\bigg)_{\omega_c}=0
\ee
and demanding $\bigg(\f{d Re\sigma}{d\omega}\bigg)_{\omega_c}=0$,  gives
\be\label{constraint_II_re_sigma_probe}
\bigg(\f{\p_r Re\sigma^{xx}}{Re\sigma^{xx}}\bigg)^2_{\omega_c}=\f{4\omega^4_cg_{rr}g_{xx}}{\Sigma^2_A k^2g^2_{tt}}\bigg((Re\sigma^{xx})_{\omega_c}^2-
\Sigma^2_A\bigg),~~~
\f{4\omega_c}{\Sigma_A }\sqrt{\f{g_{tt}}{g_{rr}}}\bigg((Re\sigma^{yy})_{\omega_c}^2-
{\widetilde\Sigma}^2_A\bigg)=0.
\ee

Now using the first equation of eq(\ref{constraint_II_re_sigma_probe}) into the first equation of eq(\ref{constraint_I_im_sigma_probe})
\be
\f{g_{xx}}{k^2g_{tt}}\bigg((Re\sigma^{xx})_{\omega_c}^2-
\Sigma^2_A\bigg)=0,
\ee
which for finite momentum, $k\neq\infty$, gives
\be
Re\sigma^{xx}(r,\omega_c)=\pm \Sigma_A.
\ee

Substituting it in the first equation of eq(\ref{constraint_I_im_sigma_probe}), gives the critical frequency to be very large, i.e., $\omega_c\rightarrow\infty$. Similarly, at non-zero critical frequency,  the second equation of eq(\ref{constraint_II_re_sigma_probe}) gives
\be
Re\sigma^{yy}(r,\omega_c)=\pm {\widetilde\Sigma}_A,
\ee
which upon using the second equation of eq(\ref{constraint_I_im_sigma_probe}), gives the large critical frequency. In summary, the real and imaginary parts of the conductivity at very high frequency goes as
\bea
&&Lim_{\omega\rightarrow\omega_c=\infty}Re\sigma^{xx}(r,\omega_c)=\pm \Sigma_A,\quad Lim_{\omega\rightarrow\omega_c=\infty}Im\sigma^{xx}(r,\omega_c)=0,\nn
&& Lim_{\omega\rightarrow\omega_c=\infty}Re\sigma^{yy}(r,\omega_c)=\pm {\widetilde\Sigma}_A,\quad Lim_{\omega\rightarrow\omega_c=\infty}Im\sigma^{yy}(r,\omega_c)=0.
\eea 

Differentiating eq(\ref{flow_phase_probe_pll}) and eq(\ref{flow_phase_probe_pendi}) with respect to frequency and evaluating it at the critical frequency, $\omega_c\rightarrow\infty$, by demanding that it vanishes, gives the phase angle as
\be\label{phase_probe_pendi_pll}
Lim_{\omega\rightarrow\omega_c=\infty}\theta^{xx}=n\pi,\quad 
Lim_{\omega\rightarrow\omega_c=\infty}\theta^{yy}=n\pi,
\ee
where $n$ is an integer including zero. It means the phase angle becomes constant at very high frequency and takes a universal value, which is an integer multiple of $\pi$, i.e., eq(\ref{phase_probe_pendi_pll}).

\subsection{An exact result}

In \cite{mst}, the authors have found an exact solution to the gauge field equation of motion, eq(\ref{eom_A_fixed_geometry}), in an asymptotically $AdS_5$ black hole  for zero momentum. The $4+1$ dimensional spacetime geometry is of the form eq(\ref{RN_black_hole}) with zero charge. The solution with the in-falling boundary condition and for constant YM's coupling reads
\be
E({\bar x})={\bar x}^{-i{\bf w}/2}(2-{\bar x})^{-{\bf w}/2}(1-{\bar x})^{\f{(1+i){\bf w}}{2}} {}_2F_{1}\bigg(1-\f{(1+i){\bf w}}{2},-\f{(1+i){\bf w}}{2},1-i{\bf w},\f{{\bar x}}{2({\bar x}-1)}\bigg),
\ee
where ${\bar x}=1-r^2_h/r^2,~{\bf w}=\omega/(2\pi T)$ and $T$ is the temperature of the black hole. Substituting this solution into eq(\ref{conductivity_pll}), gives the exact form of the conductivity at any location on the holographic energy scale 
\bea
\sigma({\bar x}, {\bf w}, k=0)&=&\f{iT\pi L}{g^2_{YM}}\Bigg[\f{[1+i(1-{\bar x})](1+i)}{2({\bar x}-1)}+\nn&&\f{{\bar x}(2-{\bar x})({\bf w}+i-1)}{4(i+{\bf w})({\bar x}-1)^2}\f{{}_2F_{1}\bigg(2-\f{(1+i){\bf w}}{2},1-\f{(1+i){\bf w}}{2},2-i{\bf w},\f{{\bar x}}{2({\bar x}-1)}\bigg)}{{}_2F_{1}\bigg(1-\f{(1+i){\bf w}}{2},-\f{(1+i){\bf w}}{2},1-i{\bf w},\f{{\bar x}}{2({\bar x}-1)}\bigg)}\Bigg],
\eea
where $\sigma^{xx}=\sigma^{yy}\equiv\sigma$ in the zero momentum limit. It  clearly follows  the trend of dimensional analysis done in \cite{ks}. Moreover, the exact form of the function that multiplies the temperature, in this example is known. Unfortunately, the direct evaluation of the conductivity at the boundary ${\bar x}=1$, diverges, because  we have not regularized the current. On evaluating the conductivity at the horizon, instead, gives 
\be
\sigma({\bar x}=0, {\bf w}, k=0)=\f{T\pi L}{g^2_{YM}},
\ee
which matches precisely with the regularity condition of conductivity at the horizon, i.e., eq(\ref{bc_fixed_geometry}).
The regularized expression to the Green's function at the boundary  is given in \cite{mst}
\be
G_R(\omega)=\f{N^2_cT^2}{8} \bigg[i{\bf w}+{\bf w}^2\bigg(\psi((1-i){\bf w}/2)+\psi(-(1+i){\bf w}/2)\bigg)\bigg],
\ee
where $\psi$ is the digamma function, $\psi(z)\equiv\Gamma'(z)/\Gamma(z)$. From which we can read out both the real and imaginary part of the conductivity, using $Re\sigma=-ImG_R/\omega$ and $Im\sigma=ReG_R/\omega$, in the high frequency limit, $\omega\rightarrow\infty$
\be
Re\sigma=\f{N^2_c\omega}{32\pi},\quad Im\sigma=\f{N^2_c\omega}{32\pi^2}~Log~\bigg(\f{\omega^2}{8\pi^2T^2}\bigg)
\ee

The phase angle of the conductivity,  $tan~\theta\equiv Im\sigma/Re\sigma$, in the high frequency limit, $\omega\rightarrow\infty$
\be
tan~\theta\rightarrow\infty,\quad \Longrightarrow \quad \theta=\pi/2.
\ee  
This looks to be a puzzle as found in the last section, 
because of the mismatch of both the result to the conductivity as well as the phase angle of the conductivity, in the $\omega\rightarrow\infty$ limit. This is mainly due to restricting ourselves to a region, $r_h\ll r\leq  r_0$, where we can ignore the contribution of the potential energy, as discussed in detail in section 3.2.

The exact result to the real part of the conductivity is
\be
Re\sigma=\f{N^2_c \omega}{16} \f{sinh~\pi{\bf w}}{cosh~\pi{\bf w}-cos~\pi{\bf w}}.
\ee

It is easy to notice that this result to the conductivity is not of the  the Drude like form.  
In \cite{ghk}, the authors found the Green function of the gauge field in $AdS_3$ Schrawschild  black hole. The phase of the conductivity in the   $\omega\rightarrow\infty$ limit  gives $\theta=\pi/2$ and the exact form of the real part of the conductivity, for zero momentum, is
\be
Re\sigma=\f{N^2_c \pi^2}{2\omega} coth(\pi {\bf w}/2).
\ee
Once, again it does not show the  Drude like form. From the study of these two examples, it looks like the phase angle of the conductivity in the fixed geometry case shows, $\theta=\pi/2$, for even dimensional field theory spacetime.

Since, we have  exact expressions of the conductivity means we can evaluate it at any value of frequency. At $\omega=T$, the conductivity  goes linear in temperature and inverse in temperature  for the $3+1$ dimensional and $1+1$ dimensional dual field theory, respectively. Which means, it simply follows the trend, $Re\sigma\sim T^{(d-3)/z}$.

\subsection{Connection with \cite{flr}}

In this subsection, we shall see the RG flow equations derived in 
\cite{flr} are same as those derived in \cite{il}, i.e., eq(\ref{flow_fixed_geometry_pll}) and eq(\ref{flow_fixed_geometry_pendi}) for rotational invariant background geometry. Let us recall the flow equations  from  \cite{flr}, especially when there exists a Neumann type boundary condition at the boundary.  Then  eq(4.21) and eq(4.24) of \cite{flr} can be rewritten in the form of eq(\ref{flow_fixed_geometry_pll}) and eq(\ref{flow_fixed_geometry_pendi}), provided 
\be
f_T=-i\omega\f{\sqrt{g_{rr}}g_{xx}}{\sqrt{-g}}\sigma^{yy},~~~h_L=\f{i}{\omega}\f{\sqrt{g_{rr}} g_{tt}g_{xx}}{\sqrt{-g}}\sigma^{xx}.
\ee

This behavior of $f_T$ and $h_L$ is generated by matching both the flow equations.  On comparing with  the eq(4.25)  and eq(4.26) of  \cite{flr}, we find\footnote{Similarly, we know the flow equation for the shear viscosity, i.e., eq(60) in \cite{il}. It matches precisely with the flow equation for the massless scalar field eq(2.18) in \cite{flr}, provided we set $f=-i\omega \f{\sqrt{g_{rr}}}{\sqrt{-g}}\chi$.}
\be 
\sigma^{yy}=\f{\prod^i_T}{i\omega A^i_T},\quad \sigma^{xx}=\f{i\prod^L }{E^L},
\ee
where $E^L=\omega A^L+kA_0$ and the $\prod$'s are momenta as defined  in \cite{flr}.

\section{Flow equation: Fluctuating geometry}

It is interesting to know whether eq(\ref{constant_cond}) still holds true when  both the background geometry and gauge fields fluctuate.  As we shall see under certain conditions eq(\ref{constant_cond}) still holds true but in an approximate sense. But before that we need to know the exact structure of the  flow equation for conductivity. In this regard,  let us assume that the dynamics  is described by a charged RN asymptotically AdS black hole in $d+1$ dimension. So, the action is
\be\label{ehm}
S=\f{1}{2\kappa^2}\int\sqrt{-g}[R-2\Lambda-\f{1}{4{\tilde g}^2_{YM}}F_{MN}F^{MN}],
\ee
where $\Lambda$ is the bulk cosmological constant. 
Once again we allow for position dependent i.e., $r$ dependent  YM's coupling. The YM's coupling, $g_{YM}$, is related to gravitational coupling as $g^2_{YM}=2\kappa^2 {\tilde g}^2_{YM}$. The equation of motion that follows are
\bea\label{eom}
&&R_{MN}-\f{2\Lambda}{d-1}g_{MN}+g_{MN} \f{F_{KL}F^{KL}}{4(d-1){\tilde g}^2_{YM}}-\f{F_{MK}{F_N}^{K}}{2{\tilde g}^2_{YM}}=0,\nn
&&\p_{M}\Bigg(\f{\sqrt{-g}}{{\tilde g}^2_{YM}}F^{MN} \Bigg)=0
\eea

There exists an exact solution to this equation of motion that is the charged RN black hole for which the metric is diagonal and only the zeroth component of the gauge potential is non zero. Generically, the solution can be written as
\be\label{ansatz_sol}
ds^2_{d+1}=g_{MN}dx^Mdx^N=-g_{tt}(r)dt^2+g_{xx}(r)dx^2+\sum^{d-2}_{a=1}g_{ab}(r)
dy^ady^b+g_{rr}(r)dr^2,~~~A=A_t(r) dt,
\ee 
which in the AdS/CFT language is interpreted as the solution dual to  ``R-charged" states. In order to calculate the conductivity, we need the expression to spatial currents and  the  equations of motion to the fluctuating fields.  Let us for simplicity, fluctuate the $t-x$ component of the metric, $g_{tx}(t,x,r)$, and
the $x$ component of the gauge field, $A_x(t,x,r)$. It turns out that both the fields, $A_x$ and $g_{tx}$ couples in a very non trivial way.
The relevant fluctuating equations for the metric component, $g_{tx}$, that follows from eq(\ref{eom}) with the convention to momentum $k^{\mu}=(w,k,0,\cdots,0)$ is
\be
R_{xr}=\f{i\omega}{2g_{tt}g_{xx}}\bigg[g'_{xx}g_{xt}-g_{xx}g'_{xt}\bigg]=-\f{F_{xt}F_{rt}g^{tt}}{2{\tilde g}^2_{YM}}=\f{i\omega A_x A'_t}{2{\tilde g}^2_{YM}g_{tt}},
\ee
and the quadratically fluctuated action for the  gauge field is
\be
S^{(2)}=-\f{1}{2\kappa^2}\int\f{\sqrt{-g}}{4{\tilde g}^2_{YM}}\bigg[\f{2A'^2_x}{g_{rr}g_{xx}}-\f{2\omega^2 A^2_x}{g_{tt}g_{xx}}+\f{2A'^2_t  A^2_x}{{\tilde g}^2_{YM}g_{rr}g_{tt}g_{xx}} \bigg]. 
\ee
From which there follows the  current at a constant-r slice  as
\be
J^x=-\f{\sqrt{-g}A'_x}{g^2_{YM}g_{rr}g_{xx}}.
\ee
where the calculation is done at zero momentum, i.e., $k=0$ and  Fourier transform is taken as $e^{-i\omega t}$. Upon simplifying, we obtain the fluctuating equations in $d+1$ dimension as
\bea\label{coupled_eom}
&&g'_{xt}-\f{g'_{xx}}{g_{xx}}g_{xt}+\f{A'_t}{{\tilde g}^2_{YM}}A_x=0, \nn 
&& \p_r\Bigg(\f{\sqrt{-g}A'_x}{g^2_{YM}g_{rr}g_{xx}} \Bigg)+
\f{\sqrt{-g}}{g^2_{YM}g_{tt}g_{xx}}\Bigg(\omega^2-\f{A^{'2}_t}{{\tilde g}^2_{YM}g_{rr}} \Bigg)A_x=0.
\eea

Let us calculate the flow equation for conductivity using Ohm's law: $\sigma^{xx}=\f{J^x}{E_x}$. The current, $J^x=-\f{\sqrt{-g}A'_x}{g^2_{YM}g_{rr}g_{xx}}$, and electric field, $E_x=i\omega A_x$,  are evaluated at zero momentum, $k=0$, resulting in the conductivity as 
\be\label{def_conductivity}
\sigma^{xx}=\f{i}{\omega}\f{\sqrt{-g}}{g^2_{YM}g_{rr}g_{xx}}\f{A'_x}{A_x}\equiv \f{i}{\omega}G(r)\f{A'_x}{A_x}.
\ee

From which there follows the flow equation, using eq(\ref{coupled_eom})
\be\label{flow_fluctuating_geometry_pll}
\p_r \sigma^{xx}=i\omega\sqrt{\f{g_{rr}}{g_{tt}}}\Bigg[\f{(\sigma^{xx})^2}{\Sigma_A}-\Sigma_A \bigg(1-\f{A^{'2}_t}{{\tilde g}^2_{YM}g_{rr}\omega^2}\bigg)\Bigg],
\ee
where $\Sigma_A:=\sqrt{\f{-g}{g_{rr}g_{tt}}}\f{1}{g^2_{YM}g_{xx}}$ and $G(r):=\f{\sqrt{-g}}{g^2_{YM}g_{rr}g_{xx}}$. It is easy to cross check the consistency of the flow equation eq(\ref{flow_fluctuating_geometry_pll}), i.e.,   the flow equation for neutral black holes for which gauge potential vanishes, $A_t=0$, is same as that found in the fixed background geometry case eq(\ref{flow_fixed_geometry_pll}).
Upon assuming that the conductivity is independent of the frequency, in the very high frequency limit, $\omega\rightarrow\infty$, which we shall justify latter, suggests that 
the last term in eq(\ref{flow_fluctuating_geometry_pll}) can be dropped in comparision to  the other two terms. Hence in either of the cases, $A_t=0$ or $A_t\neq 0$,  the solution of the conductivity, in the high frequency limit, of  $2+1$ dimensional field theory is still same as eq(\ref{constant_cond}). In summary, even away from the probe approximation but in the very high frequency limit one still  has the frequency independent conductivity in $2+1$ dimensional field theory.

The boundary condition that we shall impose is the regularity of the conductivity at the horizon, $r_h$, namely
\be\label{bc_flow_fluctuating_geometry_pll}
\sigma^{xx}(r_h,\omega)=\pm \Sigma_A(r_h),\quad \Longrightarrow Re\sigma^{xx}(r_h,\omega)=\pm \Sigma_A(r_h),\quad Im\sigma^{xx}(r_h,\omega)=0.
\ee 


\subsection{The Precise behavior of $Re\sigma^{xx}$  and $Im\sigma^{xx} $}

In \cite{gd}  another technique has been adopted to read out the behavior of the real part of the conductivity. In this subsection we shall employ that technique along with the flow equation eq(\ref{flow_fluctuating_geometry_pll}) to read out the frequency dependence to  $Im\sigma^{xx} $. The idea  essentially is to write down a conserved flux using the equation of motion to the fluctuated gauge field. We can write schematically  the fluctuated gauge field equation of motion eq(\ref{coupled_eom}) as 
\be\label{diff_eom_gauge_field}
A''_x+f_1(r) A'_x+f_2(r,~\omega)A_x=0,
\ee
where prime denotes derivative with respect to $r$, the functions $f_1(r)$ and $f_2(r,~\omega)$ are real and can very easily be read out. For completeness, $f_1=\p_r Log\bigg(\f{\sqrt{-g}}{g^2_{YM}g_{rr}g_{xx}} \bigg)$ and $f_2=\f{g_{rr}}{g_{tt}}\bigg(\omega^2-\f{A'^2_t}{{\tilde g}^2_{YM}g_{rr}} \bigg)$. Given this form of the equation of motion, we can define the conserved flux, for which $\p_r {\cal F}=0$, 
\be
{\cal F}=\f{i}{2} f(r) [A^{\star}_x A'_x-A_x A'^{\star}_x],
\ee
where the function $f(r)$ is related to $f_1(r)$ as $f(r)=e^{(\int^r f_1(r') dr')}$. Given this form of the conserved flux we can rewrite it in terms of the conductivity as ${\cal F}=\f{\omega f(r)}{G(r)}|A_x|^2 Re\sigma^{xx}$, where the function $G(r):=\f{\sqrt{-g}}{g^2_{YM}g_{rr}g_{xx}} $. It means  
\be\label{sol_re_sigma_flux}
Re\sigma^{xx}=\bigg(\f{G(r)}{f(r)}\bigg)\bigg(\f{1}{\omega |A_x|^2}\bigg){\cal F}(\omega),
\ee
Let us assume  that the flux, ${\cal F}$, is independent of the frequency as in \cite{hr} for the extremal black hole and in which case, 
all the important information about the  frequency dependence is stored in the second parenthesis. As an example, the solution to gauge field, in asymptotically $AdS_4$ spacetime, is found to have the frequency dependence as: $A_x~\sim~ \omega^{-\Delta_{\phi}+1/2} $, in the low frequency limit \cite{gd}. Using the formula 
eq(\ref{sol_re_sigma_flux}), we find that the real part of conductivity goes as $Re\sigma^{xx}\sim \omega^{2(\Delta_{\phi}-1)}$, which precisely matches with the  result of \cite{gd}. In order to go with the notation  of \cite{hr}, let the frequency dependence to the gauge field goes at low frequency as, $A_x\sim \omega^{-\nu}$, also reproduces the exact power law behavior of the  conductivity $Re\sigma^{xx}\sim \omega^{2\nu-1}$.  

Let us pause for a moment and tries to see whether the flux, ${\cal F}$, is really  independent of the frequency or not, generically. Recall, from the flow eq(\ref{flow_fluctuating_geometry_pll}), we needed to impose the regularity condition at the horizon, which is
\be\label{horizon_conductivity_maxwell_gauge_field}
\sigma^{xx}(r_h)=\pm \Sigma_A(r_h)=\f{i}{\omega}G(r_h)\bigg(\f{A'_x}{A_x}\bigg)_{r_h},
\ee 
where, in the second equality, we have used eq(\ref{def_conductivity}). Assuming the metric is not degenerate, which means close to the horizon, $g_{tt}=c_t(r-r_h)$ and $g_{rr}=c_r/(r-r_h)$, for some constant $c_t$ and $c_r$. Using this there follows that the gauge field close to the horizon can be written  as\footnote{See Appendix C for further details.} 
\be\label{sol_gauge_field_horizon}
A_x(r,\omega)=A_x(r_h,\omega)~(r-r_h)^{\mp i\omega \sqrt{c_r/c_t}}.
\ee

Imposition of the in-falling boundary condition at the horizon suggests us to choose the negative sign. Note that $f(r)=G(r)$, and using the fact that the flux is conserved  means the flux at the horizon is same as anywhere else
\be
{\cal F}(r_h,\omega)=\omega \bigg(|A_x|^2 Re\sigma^{xx}\bigg)_{r_h}.
\ee

Using eq(\ref{sol_gauge_field_horizon}) and the regularity condition for the conductivity at the horizon  gives
\be
{\cal F}(r_h,\omega)=\omega |A_x(r_h,\omega)|^2 \Sigma_A(r_h).
\ee
So, the imposition of the regularity condition on the  conductivity at the horizon, choosing the negative sign, makes  the in-falling boundary condition on the gauge field at the horizon.  This in turn gives via Ohms' law the  flux, which  is, roughly, linear in frequency  times the absolute-square of the value  of the gauge field at the horizon.

Using the fact that the flux, ${\cal F}$, is independent of the location of evaluation, means the conductivity at any location, $r_{\star}$, as
\be
Re\sigma^{xx}(r_{\star},\omega)=\f{Re\sigma^{xx}(r_h,\omega)|A_x(r_h,\omega)|^2}{|A_x(r_{\star},\omega)|^2}=\f{\Sigma_A(r_h)|A_x(r_h,\omega)|^2}{|A_x(r_{\star},\omega)|^2}.
\ee

Now equating the real and imaginary parts of the conductivity in the flow equation, eq(\ref{flow_fluctuating_geometry_pll}) we obtain
\bea\label{flow_diff_separate}
\p_r\bigg(Re\sigma^{xx}\bigg)&=&-\f{2\omega}{\Sigma_A} \sqrt{\f{g_{rr}}{g_{tt}}}\bigg(Re\sigma^{xx}\bigg)~\bigg(Im\sigma^{xx}\bigg),\nn
\p_r\bigg(Im\sigma^{xx}\bigg)&=&\omega\sqrt{\f{g_{rr}}{g_{tt}}}\Bigg[\f{(Re\sigma^{xx})^2-(Im\sigma^{xx})^2}{\Sigma_A}- \Sigma_A\bigg(1- \f{A^{'2}_t}{{\tilde g}^2_{YM}g_{rr}\omega^2}\bigg)\Bigg].
\eea

Form these two differential equation as it stands, is very  difficult to read out the frequency dependence to $Im\sigma^{xx}$. However, if we use eq(\ref{sol_re_sigma_flux}) in the first equation of eq(\ref{flow_diff_separate}), then the solution to $Im\sigma^{xx}$ is
\be\label{sol_im_sigma_sense}
Im\sigma^{xx}=\f{\Sigma_A}{2\omega }\sqrt{\f{g_{tt}}{g_{rr}}}\f{1}{|A_x|^2}\p_r\bigg(|A_x|^2 \bigg).
\ee

Eq(\ref{sol_re_sigma_flux}) and eq(\ref{sol_im_sigma_sense}) are, in some sense, the exact solutions to the real and imaginary part of the conductivity in any $d+1$ dimensional spacetime provided the flux
\be\label{flux_im_sigma}
{\cal F}(\omega)=\omega|A^2_x|\sqrt{\f{G }{\omega}\p_rIm\sigma^{xx}+(Im\sigma^{xx})^2+G^2 \f{g_{rr}}{g_{tt}}\bigg(1-\f{A'^2_t}{{\tilde g}^2_{YM}g_{rr}\omega^2}\bigg)}.  
\ee

The consistency suggests that the flux ${\cal F}(\omega)$, written down in eq(\ref{flux_im_sigma}) should be conserved. It means imposing $\p_r {\cal F}(\omega)=0$, gives us the 
second equation of eq(\ref{flow_diff_separate}) upon using eq(\ref{sol_re_sigma_flux}).

To find the precise low frequency behavior to both the real and imaginary part of conductivity requires us to know the solution to eq(\ref{diff_eom_gauge_field}). At low frequency, upon assuming  that the gauge field takes a power law form, i.e., Laurents expansion, like $A_x\sim \omega^{-\nu} a_x(r)+\cdots$ in frequency, for some function $a_x(r)$. Now, with this structure of the  solution to the gauge field and substituting it in the  right hand side of eq(\ref{sol_re_sigma_flux}) and eq(\ref{sol_im_sigma_sense}) gives us the solution as
\be\label{low_frequency_behavior_cond}
Re\sigma^{xx}\sim \omega^{2\nu-1}{\cal F}(\omega),~~~Im\sigma^{xx}\sim \f{1}{\omega}.
\ee

Consistency of this behavior of conductivity in the second equation of eq(\ref{flow_diff_separate}) suggests the exponent $\nu$ should be a positive number, i.e., $\nu \geq 0$, which is in complete agreement with \cite{hr}, where the authors found $\nu=\sqrt{V_0+1/4}$, for some constant $V_0$, close to the horizon. It means there can be a pole to $Re\sigma^{xx}(\sim \f{1}{\omega})$ for $\nu=0$ and zeros in frequency for $\nu >0$, provided the flux is constant. But, latter we shall see that there does not exists any pole structure to $Re\sigma^{xx}$ at very small frequency.

However, if we do not want to use the conservation of flux criteria because, {\em a priori}, it is not clear the precise dependence  of flux  on the frequency, which  suggests it is better to solve the flow equations eq(\ref{flow_diff_separate}) directly, 
Using the first equation of eq(\ref{flow_diff_separate}) we find 
\be\label{sol_im_sigma}
Im\sigma^{xx}=-\f{\Sigma_A}{2\omega} \sqrt{\f{g_{tt}}{g_{rr}}} \bigg(\f{\p_r  Re\sigma^{xx}}{Re\sigma^{xx}}\bigg)
\ee
and substituting it into the second equation of eq(\ref{flow_diff_separate})  gives
\be\label{sol_re_sigma}
\p_r\Bigg[\Sigma_A \sqrt{\f{g_{tt}}{g_{rr}}} \f{\p_r Re\sigma^{xx}}{Re\sigma^{xx}} \Bigg]=-2\omega^2 \sqrt{\f{g_{rr}}{g_{tt}}}
\Bigg[\f{(Re\sigma^{xx})^2}{\Sigma_A}-\f{\Sigma_A}{4\omega^2} \f{g_{tt}}{g_{rr}}\bigg(\f{\p_r Re\sigma^{xx}}{Re\sigma^{xx}}\bigg)^2-\Sigma_A\bigg(1- \f{A^{'2}_t}{{\tilde g}^2_{YM}g_{rr}\omega^2}\bigg)\Bigg].
\ee

Analytically, it is very difficult to solve eq(\ref{sol_re_sigma}) and determine the exact structure of
the conductivity. The  phase angle of the conductivity is defined as, $\tan~\theta=Im\sigma/Re\sigma$,  which flows as
\bea
\p_r\tan~\theta&=&\f{\omega}{Re\sigma^{xx}}\sqrt{\f{g_{rr}}{g_{tt}}}\Bigg[\f{(Re\sigma^{xx})^2+(Im\sigma^{xx})^2}{\Sigma_A}- \Sigma_A\bigg(1- \f{A^{'2}_t}{{\tilde g}^2_{YM}g_{rr}\omega^2}\bigg)\Bigg],\nn
&=& \f{\omega}{Re\sigma^{xx}}\sqrt{\f{g_{rr}}{g_{tt}}}\Bigg[\f{(Re\sigma^{xx})^2~sec^2\theta}{\Sigma_A}- \Sigma_A\bigg(1- \f{A^{'2}_t}{{\tilde g}^2_{YM}g_{rr}\omega^2}\bigg)\Bigg].
\eea

Now assuming an algebraic structure to, $Re\sigma^{xx}\sim \omega^{2\nu-1}$, gives 
\be\label{flow_phase}
\p_r\tan~\theta=\sqrt{\f{g_{rr}}{g_{tt}}}\Bigg[\f{\omega^{2\nu}~sec^2\theta}{\Sigma_A}- \f{\Sigma_A}{\omega^{2\nu-2}}+ \f{\Sigma_AA^{'2}_t}{{\tilde g}^2_{YM}g_{rr}\omega^{2\nu}}\Bigg].
\ee

We know from the experimental result, \cite{dvdm}, in the range $\omega >T$, i.e., the phase of the conductivity should not depend on the frequency and using the choice of $\gamma$, from eq(\ref{mid_frequency}), suggests $0 < \nu < 1/2.$ For this range of $\nu$, means the last term in eq(\ref{flow_phase}) drops out trivially in the high frequency range,  the first  and the second term makes the phase of the conductivity to depend on the frequency, which is inconsistent  with the experimental result and hence is ruled out.

Recall that the gauge field $A_t$ is solved by solving the Maxwell equation of motion eq(\ref{eom})
\be
A'_t=\rho\bigg( \f{ g_{rr}g_{tt}{\tilde g}^2_{YM}}{\sqrt{-g}}\bigg),
\ee
where $\rho$ is the constant of integration and we are interpreting it as the charge density. From eq(\ref{sol_im_sigma}), it is clear that for a power law type of frequency dependency to $Re\sigma^{xx}$ results in a pole to $Im \sigma^{xx}\sim \f{1}{\omega}$  at the origin of frequency. This is mostly true in the low frequency limit. 
Note that we can have  a frequency independent  structure to $Im \sigma^{xx}$, provided $Re\sigma^{xx}(r)\sim e^{-\omega b(r)}$, for some function $b(r)$.

In \cite{hh}, the authors found an exact analytic result of the conductivity in the low frequency and small magnetic field limit in $3+1$ dimensional bulk. In the zero magnetic field limit to their result, the conductivity goes as
\be
\sigma_{xx}=\sigma_{Q}+i\f{1}{\omega}\f{\rho^2}{(\epsilon+{\cal P})},
\ee 
where $\sigma_{Q}$ is a quantity that does not depend on the frequency, $\epsilon$ and ${\cal P}$ are the energy density and pressure, respectively. Comparing with eq(\ref{low_frequency_behavior_cond}), it follows that for a charged system, the imaginary part of the conductivity has an inverse dependence on frequency. It is highly plausible that this kind of behavior is generic and may holds true
 irrespective of the dimensionality of the spacetime.

\subsection{Universal behavior at high frequency: $\omega\rightarrow\infty$}

Here we shall confirm the result that as frequency becomes very large, $\omega\rightarrow\infty$, the conductivity becomes independent of frequency by calculating it in the usual way, i.e., via Schr\"{o}dinger equation with two assumptions. First one is that the potential energy is smaller than $\omega^2$ at very high frequency, which is precise in $d=3$, and the second one is a condition  derived latter.  Recall that the equation of motion to gauge field written in eq(\ref{diff_eom_gauge_field}) can be expressed in the Schr\"{o}dinger type of equation, $\f{d^2V}{d\xi^2}+[\omega^2-Q]V=0$, where $V$ and $Q$ are functions of $\xi$, by doing some change of variables as follows: define $d\xi/dr=\sqrt{g_{rr}/g_{tt}}$ and $V=A_x(g_{rr}/g_{tt})^{1/4}\sqrt{f}$. The precise form of $Q$ for the spacetime of the form $ds^2=-g_{tt}dt^2+g_{xx}dx^2_i+g_{rr}dr^2$ is
\be
Q(r(\xi))=\f{A'^2_t}{{\tilde g}^2_{YM}g_{rr}}-\f{g_{tt}f'^2}{4f^2 g_{rr}}-5\f{g_{tt}g'^2_{rr}}{16g^3_{rr}}+\f{g'_{rr}g'_{tt}}{8g^2_{rr}}+\f{3g'^2_{tt}}{16g_{rr}g_{tt}}+\f{g_{tt}f''}{2fg_{rr}}+\f{g_{tt}g''_{rr}}{4g^2_{rr}}-\f{g''_{tt}}{4g_{rr}}.
\ee
Generically, the potential energy for the explicit metric of the form eq(\ref{RN_black_hole}), diverges as $r\rightarrow\infty$, but for $d=3$, it vanishes. Explicitly
\be
Q=\f{(d^2+3-4d)r^2h^2}{4L^4}+\f{(d-3)r^3hh'}{2L^4}+\f{\mu^2(d-2)^2hr_h^{2(d-2)}}{L^2{\tilde g}^2_{YM}r^{2(d-2)}}.
\ee
In what follows we shall  restrict ourselves to $3+1$ dimensional bulk spacetime, In which  case the potential energy, $Q$, becomes smaller then $\omega^2$. So, in the high frequency limit, $V\propto e^{\pm i\omega \xi}$. Using this form of $V$, we can calculate 
\be
\f{d}{dr}Log A_x=\pm i\omega \sqrt{\f{g_{rr}}{g_{tt}}}+\sqrt{f}\bigg(\f{g_{rr}}{g_{tt}}\bigg)^{1/4}\f{d}{dr}\bigg[\f{1}{\sqrt{f}}\bigg(\f{g_{tt}}{g_{rr}}\bigg)^{1/4}\bigg], \quad f(r)=\f{\sqrt{-g}}{g^2_{YM}g_{rr}g_{xx}}.
\ee

Now, using eq(\ref{def_conductivity}), the conductivity results
\be
\sigma^{xx}(r_0)=i\Bigg(G\Bigg[\pm i \sqrt{\f{g_{rr}}{g_{tt}}}+\f{1}{\omega}\f{d}{dr}Log\bigg[\f{1}{\sqrt{f}}\bigg(\f{g_{tt}}{g_{rr}}\bigg)^{1/4}\bigg]\Bigg]\Bigg)_{r_0}\rightarrow\mp[ \Sigma_A(r_0).
\ee

In taking  the limit of $\omega\rightarrow\infty$, we have assumed that the quantity, $\Bigg|\Bigg(\f{d}{dr}Log\bigg[\f{1}{\sqrt{f}}\bigg(\f{g_{tt}}{g_{rr}}\bigg)^{1/4}\bigg]\Bigg)_{r_0}\Bigg|$, evaluated at $r_0$ is smaller than $\omega$, i.e., $\Bigg|\Bigg(\f{d}{dr}Log\bigg[\f{1}{\sqrt{f}}\bigg(\f{g_{tt}}{g_{rr}}\bigg)^{1/4}\bigg]\Bigg)_{r_0}\Bigg|<\omega$. If we calculate this quantity for the RN black hole,  eq(\ref{RN_black_hole}), it goes as $(3-d)/r_0 < \omega$, and which does makes sense in the spacetime dimensions that we are interested in .

\subsubsection{Checking it another way }

In this subsection, we shall show the existence of the universal behavior of the conductivity at very high frequency, $\omega\rightarrow\infty$, in another way. In particular,  the real and imaginary part of the conductivity becomes constant and zero, respectively. Let us demand that for certain critical  frequency, $\omega_c$, the conductivity does not depend on the frequency. It means $\f{d Re \sigma^{xx}}{d\omega}$ and $\f{d Im \sigma^{xx}}{d\omega}$ vanishes, as frequency approaches $\omega_c$. Differentiating eq(\ref{sol_re_sigma}) with respect to the  frequency, and imposing the condition that it vanishes as we  take the limit, $\omega\rightarrow\omega_c$, we find 
\be
-\f{4\omega_c}{\Sigma_A}\bigg(Re\sigma^{xx}(r,\omega_c)\bigg)^2 \sqrt{\f{g_{rr}}{g_{tt}}}+4\omega_c\Sigma_A \sqrt{\f{g_{rr}}{g_{tt}}} =0.
\ee

It means for non-zero $\omega_c$, the real part of the conductivity goes as
\be\label{re_sigma_high_frequency}
Lim_{\omega\rightarrow\omega_c\neq 0}~~~ Re\sigma^{xx} \longrightarrow \Sigma_A,
\ee

Differentiating eq(\ref{sol_im_sigma}) with respect to the  frequency, and imposing the condition that it should vanish as we  take the limit, $\omega\rightarrow\omega_c$, we find 
\be\label{im_sigma_high_frequency}
\bigg(\f{d Im \sigma^{xx}}{d\omega}\bigg)_{\omega_c}=0=\f{\Sigma_A}{2\omega^2_c}\sqrt{\f{g_{tt}}{g_{rr}}}(\p_r Log ~Re\sigma^{xx})_{\omega_c}-\f{\Sigma_A}{2\omega_c}\sqrt{\f{g_{tt}}{g_{rr}}}\p_r(\p_{\omega} Log~ Re\sigma^{xx})_{\omega_c}.
\ee
The last term vanishes. Now using eq(\ref{re_sigma_high_frequency}) in eq(\ref{im_sigma_high_frequency}), results 
\be
\f{1}{2\omega^2_c}\sqrt{\f{g_{tt}}{g_{rr}}}~\p_r \Sigma_A=0\quad \Longrightarrow\quad \omega_c\rightarrow\infty.
\ee
So, consistency suggests to take  the  critical frequency to be very high, i.e., $\omega_c\rightarrow\infty$. It means evaluating eq(\ref{sol_im_sigma}) at very high critical frequency gives

\be
Lim_{\omega\rightarrow\omega_c=\infty}~~~ Im\sigma^{xx} \longrightarrow 0.
\ee

Similarly, evaluating the phase angle of the conductivity, defined as,  $tan~\theta=Im\sigma^{xx}/Re\sigma^{xx}$, at high critical frequency gives  
\be
Lim_{\omega\rightarrow\omega_c=\infty}~~~ Re\sigma^{xx}~ sec~\theta=\Sigma_A,\quad\Longrightarrow \theta=n\pi, 
\ee
where $n$ is an integer including zero. So, at very high frequency the phase angle is either zero or $180^o$, which is  a universal result and  independent of the nature of the background geometry. However, it differs significantly from the experimental result, which is around $60^o$ \cite{dvdm}, which is $\theta_{expt}\simeq 0.325 ~\pi$.

\subsection{Low frequency: $\omega\rightarrow 0$ }

In this subsection we shall investigate the low frequency behavior of conductivity using the exact
symmetry properties obeyed by eq(\ref{flow_diff_separate}).  There is a symmetry that is the time reversal symmetry, $\sigma^{\star}(\omega)=\sigma(-\omega)$, \cite{dvdm}.  It means the real part of the conductivity is an even function of frequency whereas the imaginary part of the conductivity  is an odd function of frequency, i.e., $Re \sigma(-\omega)=Re \sigma(\omega)$ and $Im \sigma(-\omega)=-Im\sigma(\omega)$. In the low frequency limit, let us assume the following series expansion 
\be\label{series_re_sigma_low_frequency}
Re\sigma^{xx}(r, \omega)=a_0(r)+a_1(r)\omega^2+a_2(r)\omega^4+\cdots. \quad {\rm with }~a_0\neq 0
\ee
The functions $a_i$'s depends  on temperature, and density but its explicit dependence  are suppressed. Upon substituting it into eq(\ref{sol_im_sigma}) and expanding it in the low frequencies, results in
\be\label{series_im_sigma_low_frequency}
Im \sigma^{xx}(r, \omega)=\f{b_0(r)}{\omega}+b_1(r)\omega+b_2(r)\omega^3+\cdots,
\ee 
which means the most singular term to $Im \sigma^{xx}$ is $1/\omega$ at low frequency\footnote{Even though, we get the  $1/\omega$ behavior to the  $Im \sigma^{xx}$  at low frequency, surprisingly, $Re \sigma^{xx}$ does not show up the $\delta(\omega=0)$ as predicted by the  Kramers-Kronig relation. }. In assuming an expansion like that in  eq(\ref{series_re_sigma_low_frequency}), we have excluded the possibility to have a pole in $Re\sigma^{xx}\sim 1/\omega$, as it does not obey the condition $Re \sigma^{\star}(-\omega)=Re \sigma(\omega)$. Moreover, we have excluded the possibility to have any logarithmic dependence on frequency\footnote{ It is argued in \cite{ghk} the absence of branch cuts in the large N limit, strong coupling and at non-zero temperature.}.  The coefficients $b_i$'s that appear in the imaginary part of the conductivity are determined in terms of  $a_i$'s
\bea
b_0&=&-\f{\Sigma_A}{2}\sqrt{\f{g_{tt}}{g_{rr}}}~\p_r Log~ a_0,\quad b_1=-\f{\Sigma_A}{2}\sqrt{\f{g_{tt}}{g_{rr}}}~\p_r(a_1/a_0),\nn
b_2&=&\f{\Sigma_A}{4}\sqrt{\f{g_{tt}}{g_{rr}}}~\p_r\bigg(\f{a^2_1-2a_0a_2}{a^2_0} \bigg),\quad \cdots
\eea

Substituting the low frequency expansion of the real part of the conductivity i.e., eq(\ref{series_re_sigma_low_frequency}) into eq(\ref{sol_re_sigma}), and equating each power of frequency on both the left hand side and right hand side, gives 
\bea\label{coefficients_a_i_maxwell}
\p_r\bigg(\Sigma_A \sqrt{\f{g_{tt}}{g_{rr}}}\p_r Log~ a_0\bigg)&=&\f{\Sigma_A}{2}\sqrt{\f{g_{tt}}{g_{rr}}}~(\p_r Log~a_0)^2-\f{2\Sigma_A A'^2_t}{{\tilde g}^2_{YM}\sqrt{g_{tt}g_{rr}}},\nn
\p_r\bigg(\Sigma_A \sqrt{\f{g_{tt}}{g_{rr}}}\p_r(a_1/a_0)\bigg)&=&-\f{2}{\Sigma_A}\sqrt{\f{g_{rr}}{g_{tt}}}a^2_0+\Sigma_A\sqrt{\f{g_{tt}}{g_{rr}}}\p_r (\f{a_1}{a_0})\p_r Log~a_0+2\Sigma_A \sqrt{\f{g_{rr}}{g_{tt}}},\nn
\p_r\Bigg[\Sigma_A \sqrt{\f{g_{tt}}{g_{rr}}}\p_r\bigg(\f{a^2_1-2a_0a_2}{2a^2_0} \bigg)\Bigg]&=& \f{4}{\Sigma_A}\sqrt{\f{g_{rr}}{g_{tt}}} a_0 a_1-\f{\Sigma_A}{2}\sqrt{\f{g_{tt}}{g_{rr}}}\bigg[\bigg(\p_r(a_1/a_0)\bigg)^2-\nn&&2\p_r Log ~a_0~ \p_r\bigg(\f{a^2_1-2a_0a_2}{2a^2_0}\bigg)\bigg],\cdots.
\eea

These equations need to be  solved consistently with the boundary condition consistent with eq(\ref{bc_flow_fluctuating_geometry_pll}). It means at the horizon, we shall set, $a_0(r_h)=\Sigma_A(r_h), ~a_i(r_h)=0$ and $b_{i-1}(r_h)=0$, for $i=1,2,3,\cdots.$, in order  to obtain the 
series solution of the conductivity at low frequency. 

The series expansion to the real part of the conductivity,  eq(\ref{series_re_sigma_low_frequency}), even though is reminiscent of the Fermi liquid like behavior provided the function $a_1$ is non-zero. But, here we shall interpret it as the low frequency expansion of the conductivity in the region, $\omega < T$, where $T$ is the temperature of the black hole, which is same as  the temperature of the field theory.

Note that in the probe approximation, i.e., in the fixed background geometry case, the flow equation is described by eq(\ref{flow_fixed_geometry_pll}) and eq(\ref{flow_fixed_geometry_pendi}). In which case $A'_t$ vanishes, but still the imaginary part of the conductivity has a pole in frequency in the low frequency limit.  This
is not in contradiction with the  \cite{ss} and \cite{pss}, where  the Green's function has been found by doing a Taylor series expansion around zero frequency. Because, firstly, there the expansion is done in the $\omega > T$ region and secondly, the Green's function is related to the conductivity as $Re G^R=\omega Im\sigma$ and $Im G^R=-\omega Re\sigma$.

\begin{figure}[htb]
\centering
\subfigure[$Re{\tilde\sigma}$]{\includegraphics[width=3.0 in]{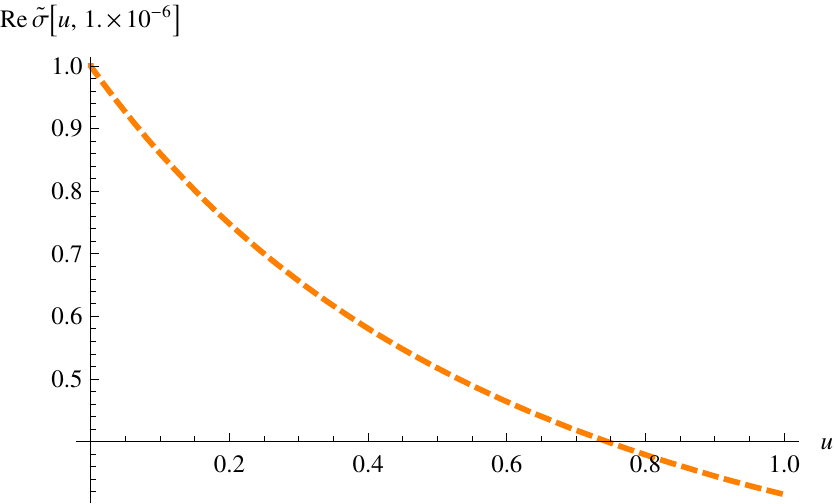}}
\subfigure[$Im{\tilde\sigma}$]{\includegraphics[width=3.0 in]{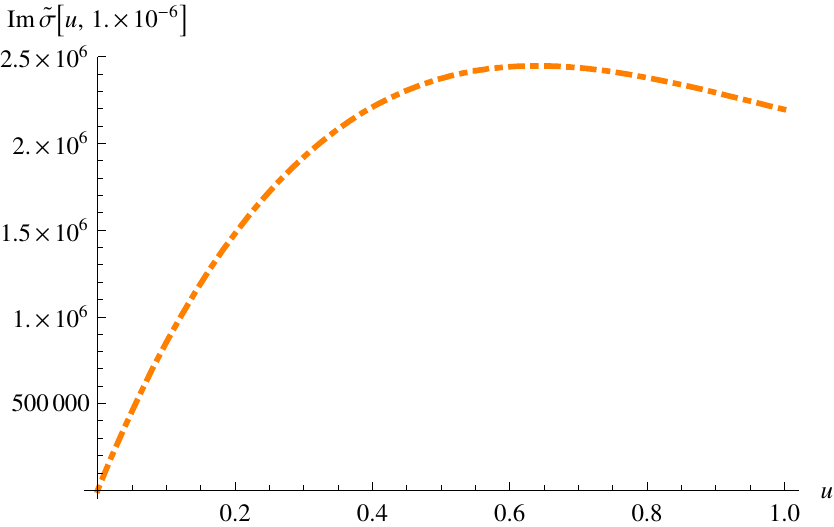}}
\caption{The flow for the $Re {\tilde\sigma}$ and $Im {\tilde\sigma}$, along the vertical direction, is plotted versus $u$  for $q=0.7,~ ~{\hat\omega}=10^{-6}, ~{\hat g}_{YM}=1$ and in $3+1$ dimensional RN black hole.}
\label{fig_1}
\end{figure}

\begin{figure}[htb]
\centering
\subfigure[$Re{\tilde\sigma}$]{\includegraphics[width=3.0 in]{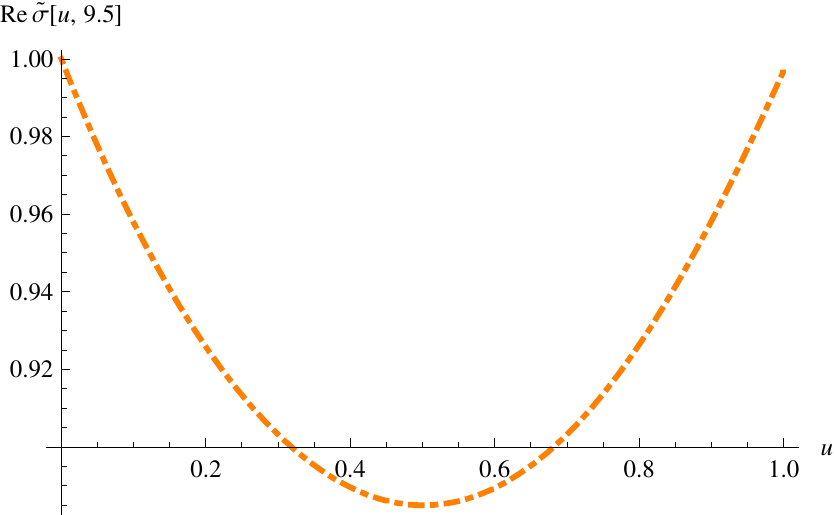}}
\subfigure[$Im{\tilde\sigma}$]{\includegraphics[width=3.0 in]{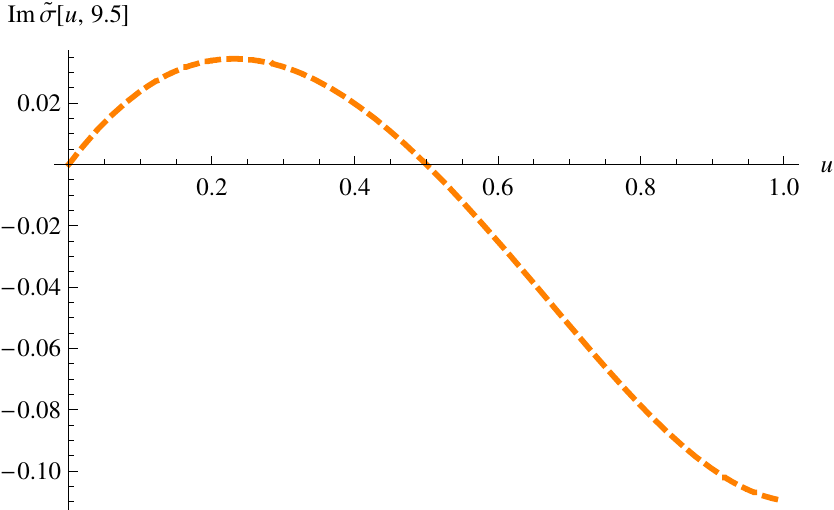}}
\caption{The flow for the $Re {\tilde\sigma}$ and $Im {\tilde\sigma}$, along the vertical direction, is plotted versus $u$  for $q=0.7,~ ~{\hat\omega}=9.5,~{\hat g}_{YM}=1$ and in $3+1$ dimensional RN black hole.}
\label{fig_2}
\end{figure}

Given the low frequency expansion in eq(\ref{series_re_sigma_low_frequency}) and eq(\ref{series_im_sigma_low_frequency}), it suggests that there exists an extremum frequency, $\omega_{ext}$, where the imaginary part of the conductivity extremizes. The extremum frequency occurs at, $\omega_{ext}\simeq\sqrt{b_0/b_1}$. However, for $b_0=0$, the extremum occurs at $\omega_{ext}\simeq\sqrt{-b_1/3b_2}$, which means either $b_1$ or $b_2$ is negative. Similarly, for the real part of the conductivity, the extremum occurs either at the origin of the frequency or at $\omega_{ext}\simeq\sqrt{-a_1/2a_2}$. This observation matches, qualitatively, with the result  that follows from  solving eq(\ref{flow_diff_separate}) numerically for the RN black hole with low charge over temperature, $q < 1$. In fig(\ref{fig_1}) and fig(\ref{fig_2}), we have plotted the RG flows of the conductivity. From numerics, there follows that 
for low $q$, the $Re{\tilde\sigma}$ does not has a zero whereas the 
$Im{\tilde\sigma}$ has a zero starting from  ${\hat\omega}=4.7998$.

\subsubsection{No signal of Drude form}

It is known from  the studies of the  free theory of electrons, the conductivity goes as $\sigma=\f{\sigma_0}{1-i\omega\tau}$, where $\sigma_0$ is a constant and $\tau$ is the relaxation time. As suggested in \cite{dvdm}, the CuO system at very low frequency region, $\omega < T$, shows the Drude from. So, it is naturally very interesting to see whether this is consistent with  the RG flow or not .  

If there is a  Drude form of the conductivity, then eq(\ref{series_re_sigma_low_frequency}) should be possible to resume 
\be
Re\sigma^{xx}=a_0+\sum_{n=1}a_n\omega^{2n}=a_0[1+\f{a_1}{a_0}\omega^2+\f{a_2}{a_0}\omega^4+\cdots]\approx\f{a_0}{[1-\f{a_1}{a_0}\omega^2]},
\ee
and the third equality is possible when $a_n/a_0=(a_1/a_0)^{n}$ for $n=2,3,4\cdots$. Let us check whether the condition on the coefficient $a_2=a^2_1/a_0$ is consistent with eq(\ref{coefficients_a_i_maxwell}) or not. 
Upon imposing this condition on the last equation of eq(\ref{coefficients_a_i_maxwell}), we ended up with
\be\label{another_a1_maxwell}
\f{\Sigma_A}{2}\sqrt{\f{g_{tt}}{g_{rr}}}\bigg(\p_r(a_1/a_0)\bigg)^2+2\Sigma_A\sqrt{\f{g_{rr}}{g_{tt}}}\f{a_1}{a_0}+\f{2}{\Sigma_A}\sqrt{\f{g_{rr}}{g_{tt}}} a_0 a_1=0.
\ee

This condition on $a_1$ is different from the second equation of eq(\ref{coefficients_a_i_maxwell}). So, generically, we do not necessarily get the Drude like behavior of the conductivity.

\subsubsection{Study of an example: In RN black hole}

In this subsection, we shall study an explicit example by considering the background geometry  of RN black hole and find the consequences of the flow. The details of which are explained in the Appendix A. Since, the conductivity, ${\tilde\sigma}$ is dimensionless, we can do a series expansion like that of eq(\ref{series_re_sigma_low_frequency})  as
\be\label{series_re_sigma_low_frequency_dim_less}
Re{\tilde\sigma}^{xx}(u, \omega)={\tilde a}_0(u)+{\tilde a}_1(u){\hat\omega}^2+{\tilde a}_2(u){\hat\omega}^4+\cdots. \quad {\rm with }~{\tilde a}_0\neq 0,
\ee
where the ${\tilde a}_i$'s are dimensionless quantities and as before the explicit dependence on the charge and temperature is suppressed. Substituting this expansion of the conductivity into
eq(\ref{z_1_flow_sigma_rn}), and equating each power of the frequency determines the first two coefficients, for constant YM's coupling, as
\bea
\p_{u}\bigg[h(1-u)^{3-d} \f{\p_{u}{\tilde a}_0}{{\tilde a}_0} \bigg]&=&\f{h}{2}(1-u)^{3-d}(\f{\p_{u}{\tilde a}_0}{{\tilde a}_0})^2-4q^2(d-1)(d-2)(1-u)^{d-1},\nn 
\p_{u}\bigg[h(1-u)^{3-d} \p_{u}({\tilde a}_1/{\tilde a}_0) \bigg]&=&-2\f{m^2_1}{h}(1-u)^{d-3}\bigg[ {\hat g}^4_{YM}{\tilde a}^2_0-\f{h^2(1-u)^{2(3-d)}}{2m^2_1}(\p_{u}{\tilde a}_0/{\tilde a}_0) \p_{u}({\tilde a}_1/{\tilde a}_0)\nn
&-&(1-u)^{2(3-d)}\bigg].
\eea

\begin{figure}[htb]
\centering
\subfigure[${\tilde b}_0(u)$]{\includegraphics[width=3.0 in]
{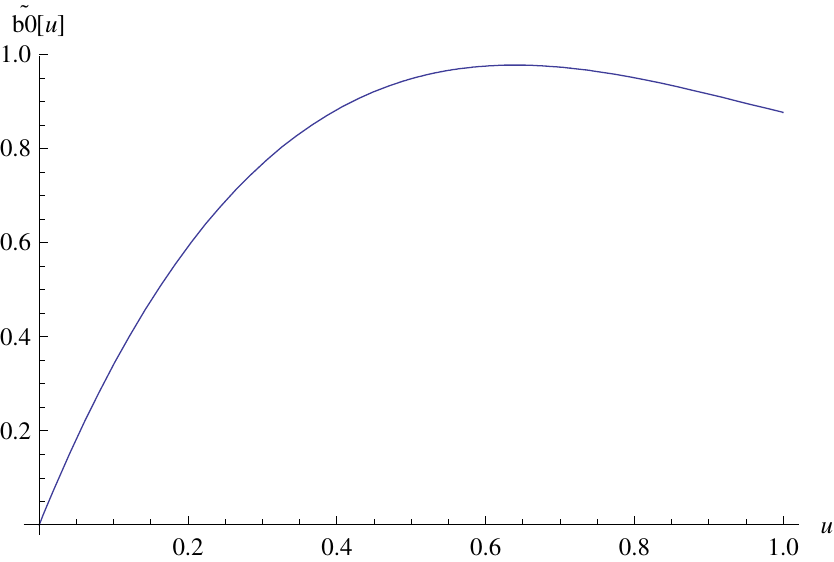}}
\subfigure[${\tilde a}_0(u)/{\tilde a}_0(r_h)$]{\includegraphics[width=3.0 in]
{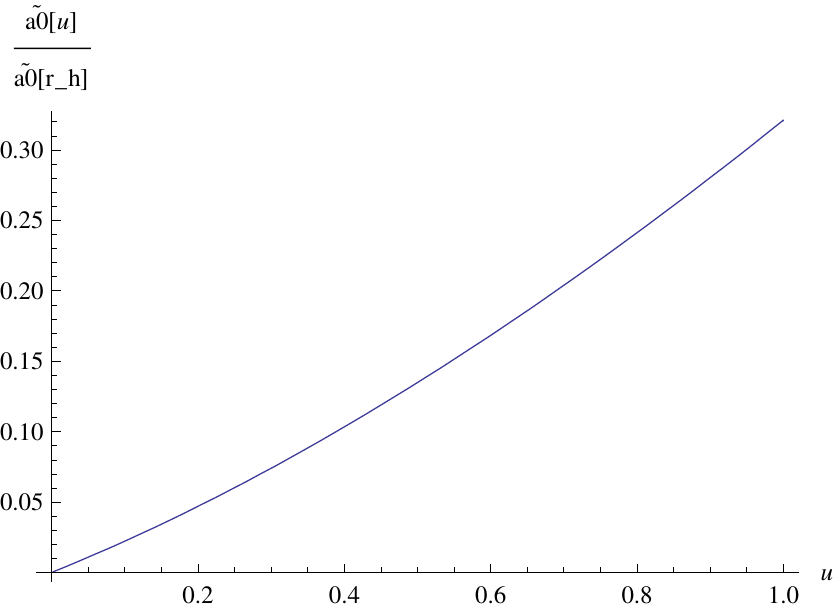}}
\caption{The behavior of the $ {\tilde b}_0(u)$ and 
$ {\tilde a}_0(u)/{\tilde a}_0(r_h)$, along the vertical direction, is plotted versus $u$  for $q=0.7$ and in $3+1$ dimensional RN black hole.}
\label{fig_3}
\end{figure}

These equations are very difficult to solve, analytically. So, using Mathematica, we numerically find the following behavior to  ${\tilde a}_0(u)$ and ${\tilde b}_0\equiv -h(1-u)^{3-d}\p_u Log {\tilde a}_0(u)$, which is  given in fig(\ref{fig_3}). From these there follows that both ${\tilde a}_0$ and ${\tilde b}_0$ are non-zero.

Let us define ${\tilde a}_i(u)={\tilde a}_0(r_h){\hat a}_i(u)$, for $i=0,1,2,\cdots$, where ${\tilde a}_0(r_h)=r^{d-3}_h/g^2_{YM}$. In which case the boundary condition at the horizon becomes, ${\hat a}_0(u_h)=1$, and ${\hat a}_{i+1}(u_h)=0$.  Now, the differential equation for ${\hat a}_i(u)$ are
\bea\label{eq_for_hat_a0_RN}
\p_{u}\bigg[h(1-u)^{3-d} \f{\p_{u}{\hat a}_0}{{\hat a}_0} \bigg]&=&\f{h}{2}(1-u)^{3-d}(\f{\p_{u}{\hat a}_0}{{\hat a}_0})^2-4q^2(d-1)(d-2)(1-\rho)^{d-1},\nn 
\p_{u}\bigg[h(1-u)^{3-d} \p_{u}({\hat a}_1/{\hat a}_0) \bigg]&=&-2\f{m^2_1}{h}(1-u)^{d-3}\bigg[ {\hat a}^2_0-\f{h^2(1-u)^{2(3-d)}}{2m^2_1}(\p_{u}{\hat a}_0/{\hat a}_0) \p_{u}({\hat a}_1/{\hat a}_0)\nn
&-&(1-u)^{2(3-d)}\bigg].
\eea

The equation for both ${\hat a}_0$, and ${\hat a}_1$ suggests that these function does not depend explicitly  on the temperature only, but can appear through the dimensionless charge, $q=Q/r^{d-1}_h\sim \mu/T$. 

In summary, the conductivity has an expansion like
\be\label{re_sigma_z_1_maxwell_d_dim}
Re\sigma^{xx}\simeq T^{d-3}\f{(4\pi L)^{d-3}}{d^{d-3}g^2_{YM}}[{\hat a}_0(u,q)+{\hat a}_1(u,q) (\omega/T)^2+\cdots],
\ee
where we have used $r_h/L\sim 4\pi L T/d$. The terms in the square bracket are dimensionless. Using this result to the real part of the conductivity in eq(\ref{sol_im_sigma}), gives
\be\label{im_sigma_z_1_maxwell_d_dim}
Im\sigma^{xx}\simeq-T^{d-3}\bigg(\f{4\pi L}{d}\bigg)^{d-3}\f{(1-u)^{3-d}}{2m_1}\f{h}{g^2_{YM}}\bigg[ \f{\p_{u}Log~{\hat a}_0}{(\omega/T)}+\p_{u}({\hat a}_1/{\hat a}_0)(\omega/T)+\cdots\bigg].
\ee
This particular temperature dependence, $T^{d-3}$,  is in complete agreement with the result that follows from dimensional analysis, \cite{ks}.  Eq(\ref{eq_for_hat_a0_RN}) is not easy to solve, and in order to solve it, we make a series expansion in $ \mu/T$ for $\mu \ll T$, i.e., ${\hat a}_0={\hat a}_{0,0}+{\hat a}_{0,1}~q+{\hat a}_{0,2}~q^2+\cdots$. Finally, the conductivity is a double series expansion in $\omega/T$ and  $q\sim\mu/T$. So,  $Re\sigma\sim\f{(TL)^{d-3}}{g^2_{YM}}[{\hat a}_{0,0}(u)+{\hat a}_{0,1}(u)~q+\cdots+({\hat a}_{1,0}(u)+{\hat a}_{1,1}(u)~q+\cdots)~(\omega/T)^2+{\cal O}(\omega/T)^3] $. 

From eq(\ref{re_sigma_z_1_maxwell_d_dim}) and eq(\ref{im_sigma_z_1_maxwell_d_dim}) there follows that the phase angle of the conductivity at $\omega/T= 0$, matches with the numerical result presented in  fig(\ref{fig_4}), which is $\theta^{xx}=\pi/2$, but  is not clear why does that feature continues up to $\omega/T=2$.

\subsubsection{For any $z$}
Let us assume that the charged black hole solution with any dynamical exponent, $z$, has the form
$ds^2=L^2[-\f{{\tilde f}(r,q)}{r^{2z}}dt^2+\f{dx^2_i}{r^2}+\f{dr^2}{r^2f(r,q)}]$, where the function ${\tilde f}(r,q)$ has a simple zero at the horizon. 
Following the dimensional analysis presented
in Appendix B, the  temperature $T\sim r^{-z}_h$. Using the low frequency expansion of the dimensionless conductivity as  in eq(\ref{series_re_sigma_low_frequency_dim_less}) and  
redefining ${\tilde a}_i(u)={\tilde a}_0(r_h){\hat a}_i(u)$ with ${\tilde a}_0(r_h)=L^{3-d}\Sigma_A(r_h)$, suggests us to rewrite the low frequency expansion of the conductivity as
\be\label{conductivity_any_z}
Re\sigma=\f{L^{d-3}r^{3-d}_h}{g^2_{YM}}[{\hat a}_0(u,q)+{\hat a}_1(u,q) (\omega/T)^2+\cdots]\sim \f{L^{d-3}T^{\f{d-3}{z}}}{g^2_{YM}}[{\hat a}_0(u,q)+{\hat a}_1(u,q) (\omega/T)^2+\cdots]
\ee

This form of the conductivity for any dynamical exponent, $z$, matches with that predicted in  \cite{ks}.

\subsection{Mid-frequency range: $\omega >T$}

This frequency range, which  is strictly defined as $\omega >T$, can also be defined in an approximate way as the system that is at zero temperature with very small but finite frequency, i.e., $T=0$ and $\omega\neq 0$. In this frequency range, there occurs something special,  that is there occurs a geometric sum to the solution resulting in a very compact structure like Hankel function \cite{gd},  \cite{hr}, and \cite{hpst} etc.. This follows by looking at  several examples because, generically, it is  very difficult to solve the equation of motion to gauge field and  find any explicit structures of the conductivity. So we need to study on a case by case basis, i.e., through several examples. Moreover, we do not  know how to  solve the flow  eq(\ref{sol_im_sigma}) and eq(\ref{sol_re_sigma}). So, here we shall make a guess to the form of the gauge potential $A_x$ that will give us the desired structure of the conductivity.

It is not necessary that the conductivity equation like that in  eq(\ref{series_re_sigma_low_frequency}) should also work
in this regime of frequency, and to reproduce the real part of the conductivity that goes like $ \sigma \sim (i\omega)^{\beta}$ is very difficult. The symmetry of the flow equation for conductivity, $\sigma^{\star}(-\omega)=\sigma(\omega)$, suggests that the gauge field should obey the symmetry $A^{\star}_x(-\omega)=A_x(\omega)~ e^{c_0}$, where $c_0$ is a constant and for simplicity, we have set it to zero, i.e, $c_0=0$. Based on this, let us assume the  following series "solution" to gauge field  eq(\ref{diff_eom_gauge_field}), in the low frequency limit
\be\label{ansatz_gauge_field}
A_x(r,\omega)= A_0-i^{\alpha}\omega^{\alpha} A_1(r)+\cdots,
\ee
where $A_0$ is a real constant that provides the normalization of the gauge field, and ellipses stands for higher powers in frequency. We have assumed  that $\alpha$ is a positive quantity, $\alpha \geq 0$. The function $A_1$ is purely real. Upon substituting this solution into eq(\ref{diff_eom_gauge_field}), we find
\be
\omega^{\alpha}[A''_1+f_1 A'_1+\cdots]-i^{-\alpha}\f{g_{rr}}{g_{tt}}\bigg(\omega^2-\f{A'^2_0}{{\tilde g}^2_{YM}g_{rr}}\bigg)A_0+\omega^{\alpha}\f{g_{rr}}{g_{tt}}\bigg(\omega^2-\f{A'^2_0}{{\tilde g}^2_{YM}g_{rr}}\bigg)A_1+\cdots=0.
\ee

If we try setting each power of frequency to zero, i.e., solving the equation iteratively as we did in the last section, then   there exists a term with $\omega^0$, which  cannot be set  
 to zero. Moreover, if we set the constant $A_0$ to zero then it will not be a consistent choice because $A_1(r)$ satisfy the same equation of motion as $A_x(r,\omega)$, but the functions that appear in the equations  are different.  

Let us, now, calculate the conductivity, in the low frequency limit, using the definition eq(\ref{def_conductivity}) along with the "solution" eq(\ref{ansatz_gauge_field})
\be
\sigma^{xx}\simeq G\f{A'_1}{A_0}(i\omega)^{\alpha-1}
\ee

\subsubsection{An example}
In this subsection, we shall find the frequency dependence of the conductivity in $3+1$ dimensional bulk spacetime dimensions by solving the gauge field equation of motion at zero temperature as mentioned above, i.e., setting $h\simeq 1$ in the region $r_h\ll r \leq  r_0$, as in \cite{hpst}. So,  the equation of the gauge field in RN AdS black hole 
\be
A_x''(r)+\f{2}{r}A_x'(r)+\bigg(\f{L^4\omega^2}{r^4}-\f{\mu'^2}{r^6}\bigg)A_x(r)=0,
\ee 
where $\mu'^2\equiv \f{\mu^2L^2r^{2}_h}{{\tilde g}^2_{YM}}$. 
In the regime that we are interested in suggests  us to set the temperature to zero. It means the second term in the parentheses can be dropped. In which case,  the solution is in terms of the trigonometric functions
\be
A_x=c_1e^{i L^2\omega/r} + c_2e^{-i L^2\omega/r}.
\ee
Demanding the choice of the solution at the boundary to be outgoing, i.e., we set the constant $c_2$ to zero. So, upon setting $c_2=0$ and using eq(\ref{def_conductivity}) to calculate the conductivity,  in the low frequency limit,  results
\be
\sigma^{xx}(r_0)=G(r_0) L^2/r^2_0,
\ee 
 There follows the phase angle of the conductivity 
\be
tan~\theta^{xx}(r_0)=0 \quad \Longrightarrow\quad  \theta^{xx}(r_0)=n\pi,\quad n=0,1,2,\cdots.
\ee

\begin{figure}[htb]
\centering
\subfigure[$\theta^{xx}(u\simeq 1)$]{\includegraphics[width=3.0 in]
{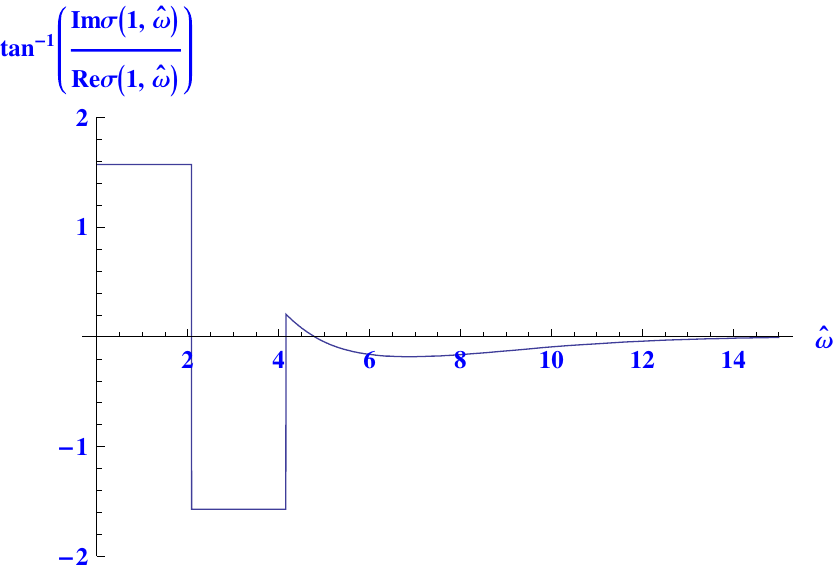}}
\subfigure[$\theta^{xx}(u\simeq 0.3)$]{\includegraphics[width=3.0 in]
{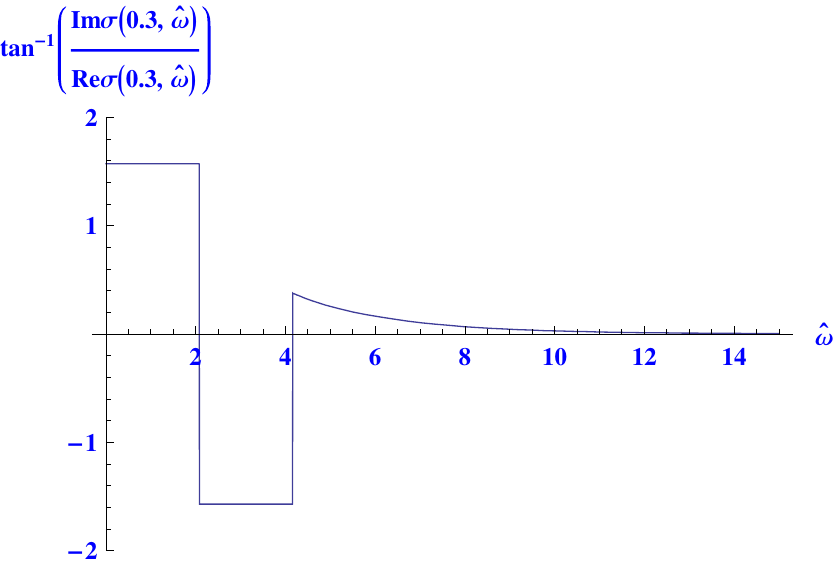}}
\caption{The behavior of the phase angle of the conductivity, $ \theta^{xx}$,  along the vertical direction, is plotted versus ${\hat\omega}$  for $q=0.7,~ ~{L/r_h}=1, 
~{\hat g}_{YM}=1$ and in $3+1$ dimensional RN black hole.}
\label{fig_4}
\end{figure}

It is easy to notice from fig(\ref{fig_4})  the result that the phase angle of the conductivity
computed numerically both very close to the boundary and away from it,   in the large frequency limit, i.e,. $\omega> T$, approaches, $0$,  asymptotically. From the fig(\ref{fig_4}), it also follows that there occurs a sharp jump to the phase angle of the conductivity at frequencies close to $\omega\simeq 2 T$ and $\omega\simeq 4T$, which  is not clear why does that happen ? So, the complete  understanding 
 deserves further study, i.e., the phase angle of the conductivity in the frequency region, $0 < \omega/T \leq 4$.

\section{Flow equation with DBI action  in a fixed geometry  }

In this section, we shall study the flow equation for the 
electrical conductivity in a fixed background geometry, where  the action is  described by the DBI type. The background is   a  $d+1$ dimensional bulk spacetime, with a trivial dilaton profile, $\phi=0$. 
\be\label{geometry_fixed}
ds^2_{d+1}=g_{MN}dx^Mdx^N=-g_{tt}(r)dt^2+g_{xx}(r)dx^2+g_{yy}(r)dy^2+\sum^{d-3}_{i=1}g_{ij}(r)
dz^idz^j+g_{rr}(r)dr^2,
\ee
The action is 
\be
S=-T_b\int \sqrt{-det([g]+\lambda F)_{ab}},
\ee 
where $[g]_{ab}$ is the induced metric on the brane world volume, $T_b$ is the tension of the brane and $\lambda$ is a dimension full object but we shall set it to unity. The dilaton is considered to be trivial as we want the scaling symmetry of the (closed string)  background fields and has been absorbed in the tension of the brane. In this paper, we shall be working in the simpler massless limit, i.e., the embedding fields are trivial. Using the static gauge choice, we can rewrite the induced metric 
in this special  massless case   $[g]_{ab}=g_{ab}$. Let us do the infinitesimal fluctuation to the gauge fields and the total field strength is
\bea\label{gauge_field_fixed}
F&=&F^{(0)}_{rt}dr\w dt+F^{(0)}_{xy}dx\w dy+
F^{(1)}_{rt}dr\w dt+F^{(1)}_{rx}dr\w dx+\nn&&
F^{(1)}_{ry}dr\w dy+F^{(1)}_{tx}dt\w dx+F^{(1)}_{yx}dy\w dx+F^{(1)}_{ty}dt\w dy,
\eea 
where we have put the  number $(1)$ in the superscripts to denote the fluctuations and the fluctuating fields depends on $t,~x,$ and $r$. In  the following we have set the gauge choice $A_r=0$, which means $F^{(0)}_{rt}=A'^{(0)}_t$ and shall denote $F^{(0)}_{xy}=B$, which is a constant. Let us  denote $ (g+F)_{ab}\equiv M_{ab}=(g^{(0)}+F^{(0)})_{ab}+(g^{(1)}+F^{(1)})_{ab}\equiv (M^{(0)}_++M^{(1)}_+)_{ab}$, where $M^{(n)}_{\pm ab}=(g^{(n)}\pm F^{(n)})_{ab},$ for $n=0$ and $1$. Note that $M^{(n)}_{+{ab}}=M^{(n)}_{-{ba}}$. In this section, since we do not consider  any fluctuation to the background geometry implies $M^{(1)}_{+ab}=F^{(1)}_{ab}.$

Expanding out the DBI action to quadratic order in the fluctuation
\bea\label{action_quadratic_dbi_fixed}
S^{(2)}&=&-T_b\int \sqrt{-det M^{(0)}_+}\bigg[1-\f{1}{2}{\bigg(M^{(0)}_+\bigg)^{-1}}^{ab}F^{(1)}_{ab}-
\f{1}{4}{\bigg(M^{(0)}_+\bigg)^{-1}}^{ab}F^{(1)}_{bc}{\bigg(M^{(0)}_+\bigg)^{-1}}^{cd}F^{(1)}_{da}+\nn&&
\f{1}{8}{\bigg(M^{(0)}_+\bigg)^{-1}}^{ab}F^{(1)}_{ab}{\bigg(M^{(0)}_+\bigg)^{-1}}^{cd}F^{(1)}_{cd}\bigg].
\eea

Now using the equation of motion to unperturbed gauge field, $A^{(0)}_t$
\be\label{eom_a00}
\p_a\Bigg[\sqrt{-det M^{(0)}_+} \Bigg({\bigg(M^{(0)}_+\bigg)^{-1}}^{ab}-
{\bigg(M^{(0)}_+\bigg)^{-1}}^{ba} \Bigg)\Bigg]=0,
\ee 
 the equation of motion to the fluctuating gauge field that results 
\bea
&&\p_a\Bigg[\sqrt{-det M^{(0)}_+}\Bigg( 
{\bigg(M^{(0)}_+\bigg)^{-1}}^{ac}F^{(1)}_{cd}{\bigg(M^{(0)}_+\bigg)^{-1}}^{db}+{\bigg(M^{(0)}_-\bigg)^{-1}}^{ac}F^{(1)}_{cd}{\bigg(M^{(0)}_-\bigg)^{-1}}^{db}\Bigg) \Bigg]-\nn
&&\sqrt{-det M^{(0)}_+} \Bigg(\bigg(M^{(0)}_+\bigg)^{-1}F^{(0)}
\bigg(M^{(0)}_-\bigg)^{-1}\Bigg)^{ab}\p_a\Bigg( {\bigg(M^{(0)}_+\bigg)^{-1}}^{cd}F^{(1)}_{cd}\Bigg)=0.
\eea

The current that follows from eq(\ref{action_quadratic_dbi_fixed}),  at a constant-r slice is  
\bea\label{generic_current_fixed}
J^{(1)\mu}&=&-T_b\f{\sqrt{-det M^{(0)}_+}}{2}\Bigg[\Bigg(\bigg(M^{(0)}_+\bigg)^{-1}F^{(1)}
\bigg(M^{(0)}_+\bigg)^{-1} +\bigg(M^{(0)}_-\bigg)^{-1}F^{(1)}
\bigg(M^{(0)}_-\bigg)^{-1}\Bigg)^{r\mu}+\nn
&&\f{1}{2}\Bigg(\bigg(M^{(0)}_+\bigg)^{-1}-
\bigg(M^{(0)}_-\bigg)^{-1}\Bigg)^{r\mu}
{\bigg(M^{(0)}_+\bigg)^{-1}}^{ab}F^{(1)}_{ab}\Bigg]
\eea

Using the explicit background geometry eq(\ref{geometry_fixed}) and 
gauge field eq(\ref{gauge_field_fixed}) in eq(\ref{generic_current_fixed}), results in the current as
\bea\label{current_dbi_b}
J^{(1)x}&=&-T_b\sqrt{\prod^{d-3}_{i=1} g_{z^iz^j}}\Bigg(\f{g_{tt}g_{yy}\p_r A^{(1)}_x-i\omega B \p_rA^{(0)}_t A^{(1)}_y}{\sqrt{g_{tt}g_{rr}-(\p_r A^{(0)}_t)^2}\sqrt{g_{xx}g_{yy}+B^2}}\Bigg)\nn
J^{(1)y}&=&-T_b\sqrt{\prod^{d-3}_{i=1} g_{z^iz^j}}\Bigg(\f{g_{tt}g_{xx}\p_r A^{(1)}_y+iB\p_rA^{(0)}_t E^{(1)}}{\sqrt{g_{tt}g_{rr}-(\p_r A^{(0)}_t)^2}\sqrt{g_{xx}g_{yy}+B^2}}\Bigg),
\eea
where we have done the Fourier transformation of the (fluctuating)  field  with the convention of $e^{i(kx-wt)}$ and $E^{(1)}\equiv\omega A^{(1)}_x+kA^{(1)}_t $. From now onwards, we shall set the tension of the brane to unity, $T_b=1$, for simplicity. The explicit form of the equations of motion to gauge fields 
\bea
&&\p_r\Bigg[\sqrt{\prod_i g_{z^iz^j}}\Bigg(\f{g_{tt}g_{yy}\p_r A^{(1)}_x-i\omega B \p_rA^{(0)}_t A^{(1)}_y}{\sqrt{g_{tt}g_{rr}-(\p_r A^{(0)}_t)^2}\sqrt{g_{xx}g_{yy}+B^2}}\Bigg) \Bigg]+\nn
&&\omega ~
\sqrt{\prod_i g_{z^iz^j}}\Bigg(\f{g_{rr}g_{yy} E^{(1)}+i B \p_rA^{(0)}_t \p_rA^{(1)}_y}{\sqrt{g_{tt}g_{rr}-(\p_r A^{(0)}_t)^2}\sqrt{g_{xx}g_{yy}+B^2}}\Bigg)=0,
\eea

\bea
&&\p_r\Bigg[\sqrt{\prod_i g_{z^iz^j}}\Bigg(\f{(g_{xx}g_{yy}+B^2)g_{tt}g_{rr}\p_r A^{(1)}_t+ik B \p_rA^{(0)}_t A^{(1)}_y(g_{tt}g_{rr}-(\p_r A^{(0)}_t)^2)}{(g_{tt}g_{rr}-(\p_r A^{(0)}_t)^2)^{3/2}\sqrt{g_{xx}g_{yy}+B^2}}\Bigg) \Bigg]-\nn
&&k~
\sqrt{\prod_i g_{z^iz^j}}\Bigg(\f{g_{rr}g_{yy} E^{(1)}+i B \p_rA^{(0)}_t \p_rA^{(1)}_y}{\sqrt{g_{tt}g_{rr}-(\p_r A^{(0)}_t)^2}\sqrt{g_{xx}g_{yy}+B^2}}\Bigg)=0,
\eea

\bea
&&\p_r\Bigg[\sqrt{\prod_i g_{z^iz^j}}\Bigg(\f{g_{tt}g_{xx}\p_r A^{(1)}_y}{\sqrt{g_{tt}g_{rr}-(\p_r A^{(0)}_t)^2}\sqrt{g_{xx}g_{yy}+B^2}}\Bigg) \Bigg]-\f{iBE^{(1)}\p_r A^{(0)}_t\sqrt{\prod_i g_{z^iz^j}}\p_r(g_{xx}g_{yy})}{\sqrt{g_{tt}g_{rr}-(\p_r A^{(0)}_t)^2}(g_{xx}g_{yy}+B^2)^{3/2}}\nn
&&+\f{\sqrt{\prod_i g_{z^iz^j}}}{\sqrt{g_{xx}g_{yy}+B^2}}\Bigg[\f{\omega^2g_{xx}g_{rr}}{\sqrt{g_{tt}g_{rr}-(\p_r A^{(0)}_t)^2}}- 
\f{g_{xx}g_{yy} k^2\sqrt{g_{tt}g_{rr}-(\p_r A^{(0)}_t)^2}}{g_{xx}g_{yy}+B^2}\Bigg]A^{(1)}_y=0,
\eea

and the constraint equation
\be
\omega g_{rr}(g_{xx}g_{yy}+B^2)\p_rA^{(1)}_t+kg_{yy}[g_{tt}g_{rr}-(\p_r A^{(0)}_t)^2]\p_r A^{(1)}_x=0.
\ee

\subsection{Zero magnetic field}

Here we shall take a pause and find the flow equation for 
the conductivity in the zero magnetic field case. 
In which case, the equation of motion to $E^{(1)}$ becomes 
\be
\p^2_r E^{(1)}+\f{\p_r E^{(1)} \p_r {\tilde X}}{{\tilde X}}+\Bigg[\f{\omega^2 g_{xx}g_{rr}-k^2(g_{tt}g_{rr}-(\p_r A^{(0)}_t)^2)}{g_{tt}g_{xx}}\Bigg]E^{(1)}=0,
\ee
where
\be
{\tilde X}=\f{\sqrt{(\prod_i g_{z^iz^j})g_{xx}g_{yy}}g_{tt}g_{rr}}{\sqrt{g_{tt}g_{rr}-(\p_r A^{(0)}_t)^2}[\omega^2 g_{xx}g_{rr}-k^2(g_{tt}g_{rr}-(\p_r A^{(0)}_t)^2)]}
\ee

Writing down the current using Ohm's law, we get
\be
J^{(1)x}=\sigma^{xx}F^{(1)}_{xt}=i\sigma^{xx} E^{(1)},
\ee
where we have  done the Fourier transformation to the gauge field with the convention of $e^{i(kx-wt)}$. From this there follows the flow equation for the conductivity
\bea
\p_r \sigma^{xx}&=&\f{i\omega}{\sqrt{(\prod_i g_{z^iz^j})g_{xx}g_{yy}}\sqrt{g_{tt}g_{rr}-(\p_r A^{(0)}_t)^2}g_{tt}g_{rr}g_{xx}} \Bigg[(g_{tt}g_{rr}-(\p_r A^{(0)}_t)^2)g_{xx}(\sigma^{xx})^2\nn &&\Bigg(g_{xx}g_{rr}-\f{k^2}{\omega^2}(g_{tt}g_{rr}-(\p_r A^{(0)}_t)^2)\Bigg) -(\prod_i g_{z^iz^j})g_{xx}g_{yy} g_{tt}g^2_{rr}\Bigg]
\eea

Solving the  equation of motion to unperturbed gauge field, $A^{(0)}_t$, gives
\be
g_{tt}g_{rr}-(\p_r A^{(0)}_t)^2=\f{g_{tt}g_{rr}(\prod_i g_{z^iz^j})g_{xx}g_{yy}}{\rho^2+(\prod_i g_{z^iz^j})g_{xx}g_{yy}},
\ee 
where $\rho$ is the constant of motion and can be interpreted as the charge density. Substituting this into the flow equation for the  conductivity results in
\bea
\p_r \sigma^{xx}&=&\f{i\omega}{g_{xx}\sqrt{\rho^2+(\prod_i g_{z^iz^j})g_{xx}g_{yy}}}\sqrt{\f{g_{rr}}{g_{tt}}}\Bigg[(\sigma^{xx})^2 g_{xx}\nn&&\Bigg(g_{xx}-\bigg(\f{k^2}{\omega^2}\bigg)\bigg(\f{g_{tt}(\prod_i g_{z^iz^j})g_{xx}g_{yy}}{\rho^2+(\prod_i g_{z^iz^j})g_{xx}g_{yy}}\bigg)\Bigg)-\Bigg(\rho^2+(\prod_i g_{z^iz^j})g_{xx}g_{yy}\Bigg)\Bigg]
\eea

Regularity at the horizon, where $g_{tt}(r_h)=0$,  requires that we need to impose the following boundary condition
\be\label{regularity_condition_B_0}
(\sigma^{xx})^2_{r_h}=\Bigg[\f{\rho^2+(\prod_i g_{z^iz^j})g_{xx}g_{yy}}{g^2_{xx}} \Bigg]_{r_h}.
\ee 

There follows an important solution that in $2+1$ dimensional field theory at zero momentum with non-zero charge density, the ac conductivity  depends non-trivially on the frequency, which is in contrast to that  found for the Maxwell system in  \cite{hkss}, i.e., eq(\ref{constant_cond}). In the high density limit, the dc  conductivity at the horizon  approximately becomes,  $\sigma^{xx}(r_h)\sim \f{\rho}{g_{xx}(r_h)}$. Let us assume that the black hole spacetime is described by 
\be\label{metric_example}
ds^2=-r^{2z}{\tilde f}(r)dt^2+r^{2w}dx^2+r^2 dy^2+r^2 dz_idz^i+\f{dr^2}{r^2{\tilde f}(r)},
\ee
where the function ${\tilde f}(r)$ has a zero at $r=r_h$, which means the Hawking temperature $T\sim r^z_h$. Then  the dc conductivity at high density becomes $\sigma^{xx}_{r_h}\sim  T^{\f{-2w}{z}}$ \cite{ssp}, which  for $w=1$  reproduces the result, $\sigma^{xx}_{r_h}\sim  T^{\f{-2}{z}}$, of \cite{hpst}, when we restrict to $3+1$ dimensional bulk spacetime. Whereas in the low density limit it approaches unity in the leading order in the density.

\subsubsection{Charge diffusion}

It is argued in \cite{il} that there can be a non-zero current at the boundary even in the absence of the external electric field, which leads to the appearance of a pole in the Green function. This is due to the fluctuation in the charge density.  This behavior can be studied from the gravity by going over to the hydrodynamic regime, where $\omega\sim k^2,~ \omega\ll T$ and $k\ll T$.  We can rewrite the flow equation of the conductivity as
\bea
\f{\p_r \sigma^{xx}}{T}&=&\f{i{\hat\omega}}{g_{xx}\sqrt{\rho^2+(\prod_i g_{z^iz^j})g_{xx}g_{yy}}}\sqrt{\f{g_{rr}}{g_{tt}}}\Bigg[(\sigma^{xx})^2 g_{xx}\nn&&\Bigg(g_{xx}-\bigg(\f{{\hat k}^2}{{\hat\omega}^2}\bigg)\bigg(\f{g_{tt}(\prod_i g_{z^iz^j})g_{xx}g_{yy}}{\rho^2+(\prod_i g_{z^iz^j})g_{xx}g_{yy}}\bigg)\Bigg)-\Bigg(\rho^2+(\prod_i g_{z^iz^j})g_{xx}g_{yy}\Bigg)\Bigg],
\eea
where ${\hat\omega}=\omega/T$ and  ${\hat k}=k/T$.  Considering $\sigma^{xx}\sim {\cal O}(1) $, and for small density we can rewrite it as
\be
\f{\p_r \sigma^{xx}}{T}\simeq -\f{i\sqrt{g_{rr}g_{tt}} (\sigma^{xx})^2}{\sqrt{\rho^2+(\prod_i g_{z^iz^j})g_{xx}g_{yy}}}\bigg(\f{{\hat k}^2}{{\hat\omega}}\bigg)\bigg(\f{(\prod_i g_{z^iz^j})g_{xx}g_{yy}}{\rho^2+(\prod_i g_{z^iz^j})g_{xx}g_{yy}}\bigg)
\ee

Now, integrating it from the horizon, $r_h$, to some generic point,  $r$
\be
\f{1}{\sigma^{xx}(r)}=\f{1}{\sigma^{xx}(r_h)}+i\f{k^2}{\omega}\int^r_{r_h} dr' \f{\sqrt{g_{rr}g_{tt}}(\prod_i g_{z^iz^j})g_{xx}g_{yy}}{[\rho^2+(\prod_i g_{z^iz^j})g_{xx}g_{yy}]^{3/2}}.
\ee

The retarded  Green function at the boundary, $G^{xx}_R(\omega,~k)=-i\omega \sigma^{xx}(r\rightarrow \infty)$ is
\be
G^{xx}_R(\omega,~k)=\f{\omega^2 \sigma^{xx}(r_h)}{i\omega-Dk^2},
\ee
where the diffusion constant at the boundary  is
\be
D= \sigma^{xx}(r_h) \int^{\infty}_{r_h}dr' \f{\sqrt{g_{rr}g_{tt}}(\prod_i g_{z^iz^j})g_{xx}g_{yy}}{[\rho^2+(\prod_i g_{z^iz^j})g_{xx}g_{yy}]^{3/2}}.
\ee

Now using the Einstein relation, $ \Xi=\sigma^{xx}(r_h)/D$
\be
\Xi=\Bigg(  \int^{\infty}_{r_h}dr' \f{\sqrt{g_{rr}g_{tt}}(\prod_i g_{z^iz^j})g_{xx}g_{yy}}{[\rho^2+(\prod_i g_{z^iz^j})g_{xx}g_{yy}]^{3/2}}\Bigg)^{-1}.
\ee

So, the Green function,  $G^{xx}_R(\omega,~k)$, in the hydrodynamic limit with small charge density has a pole 
\be
\omega=-iD k^2.
\ee

In the zero density limit, as expected, we do reproduce the diffusion constant for the Maxwell system. It would be interesting to have another independent derivation of the diffusion constant at the boundary. We leave the consistency of this way of finding the gapless spectrum at zero momentum to that found in Appendix D, for future studies.

As an example, the computation of the charge diffusion constant and the charge susceptibility for the Lifshitz $d+1$ dimensional spacetime with dynamical exponent, $z=1$, has the following temperature dependence 
\bea\label{charge_diffusion_dbi}
D&=&\f{L^{2-z}}{(d-1-z)} \bigg(\f{TL}{\alpha}\bigg)^{\f{z-2}{z}}\sqrt{1+\rho^2\bigg(\f{\alpha}{TL}\bigg)^{\f{2d-2}{z}}}{}_2F_1\Bigg[\f{3}{2},\f{1-d+z}{2-2d},\f{3-3d+z}{2-2d},-\rho^2\bigg(\f{TL}{\alpha}\bigg)^{\f{2-2d}{z}}\Bigg],\nn
\Xi&=&\f{(d-1-z)}{L^{2-z}}\bigg(\f{TL}{\alpha}\bigg)^{\f{d-1-z}{z}}\Bigg({}_2F_1\bigg[\f{3}{2},\f{1-d+z}{2-2d},\f{3-3d+z}{2-2d},-\rho^2\bigg(\f{TL}{\alpha}\bigg)^{\f{2-2d}{z}}\bigg]\Bigg)^{-1}
\eea
where the constant, $\alpha=(d+z-1)/(4\pi)$ and ${}_2F_1[a,b,c,x]$ is the hypergeometric function. 
In the zero density limit, it reproduces, precisely, the result of \cite{kr}, \cite{as}, \cite{bkls}. As expected, the  expansion of the diffusion constant in eq(\ref{charge_diffusion_dbi}), in the small density limit
\be
D=\f{L^{2-z}}{(d-1-z)} \bigg(\f{TL}{\alpha}\bigg)^{\f{z-2}{z}}\Bigg[1+\Bigg(\f{1}{2}\bigg(\f{\alpha}{TL}\bigg)^{\f{2d-2}{z}}+\f{3(d-z-1)}{2(3-3d+z)}\bigg(\f{TL}{\alpha}\bigg)^{\f{2-2d}{z}}\Bigg)\rho^2+\cdots\Bigg],
\ee
differs from the result of   the charge diffusion constant  for the RN AdS black hole in \cite{bkls}. 
Interestingly, for $d=4$ and $z=1$, eq(\ref{charge_diffusion_dbi}) reproduces the result\footnote{There looks to be some  error in the result of \cite{kz}, possibly  typographical.} of \cite{kz} and \cite{mst}.

\subsection{Non-zero magnetic field}

Here we shall turn on the magnetic field but consider the case of having rotational symmetry in the $x-y$ plane of the background geometry, i.e., by setting $g_{xx}=g_{yy}$. For simplicity of the calculation, we shall, also, set the momentum to zero $k=0$. In which case $E^{(1)}=\omega A^{(1)}_x$. Define, $E^{(1)}_{\pm}\equiv iE^{(1)}\mp\omega A^{(1)}_y=i\omega(A^{(1)}_x\pm iA^{(1)}_y)\equiv i\omega A^{(1)}_{\pm}$. Similarly, the currents and conductivity can be defined as
$J^{(1)}_{\pm}\equiv {J^{(1)}}^x\pm i{J^{(1)}}^y$ and $\sigma_{\pm}\equiv \sigma^{xy}\pm i\sigma^{xx}$. Also, $\sigma^{xx}=\sigma^{yy}$ and $\sigma^{xy}=-\sigma^{yx}$. In this case the Ohm's law becomes, $J^{(1)}_{\pm}=\mp i\sigma_{\pm} E^{(1)}_{\pm}$ as in \cite{hh}. The equation of motion to $A^{(1)}_{\pm}$
\be\label{eom_A_plus_minus}
\p_r\Bigg[\f{\sqrt{\prod_i g_{z^iz^j}}(g_{tt}g_{xx}\p_r A^{(1)}_{\pm}\mp \omega B A'_t A^{(1)}_{\pm})}{\sqrt{g_{tt}g_{rr}-(\p_r A^{(0)}_t)^2}\sqrt{g^2_{xx}+B^2}} \Bigg]+\f{\omega\sqrt{\prod_i g_{z^iz^j}}[\omega g_{rr}g_{xx}A^{(1)}_{\pm}\pm B A'_t \p_r A^{(1)}_{\pm}]}{\sqrt{g_{tt}g_{rr}-(\p_r A^{(0)}_t)^2}\sqrt{g^2_{xx}+B^2}}=0
\ee

and the expression to the flow equation for conductivity 
\bea\label{flow_dbi_b_plus_minus_cond}
\p_r \sigma_{\pm}&=&\mp \f{2\omega B \p_r A^{(0)}_t \sigma_{\pm}}{g_{tt}g_{xx}}\pm \f{\omega\sigma^2_{\pm}}{g_{tt}g_{xx}}\f{\sqrt{g_{tt}g_{rr}-(\p_r A^{(0)}_t)^2}\sqrt{g^2_{xx}+B^2}}{\sqrt{\prod_i g_{z^iz^j}}}\nn
&\pm&\f{\omega \sqrt{\prod_i g_{z^iz^j}}}{\sqrt{g_{tt}g_{rr}-(\p_r A^{(0)}_t)^2}\sqrt{g^2_{xx}+B^2}}
\Bigg[g_{rr}g_{xx}+\f{B^2(\p_r A^{(0)}_t)^2}{g_{tt}g_{xx}}\Bigg],
\eea

where $\sigma_{\pm}$ is 
\be\label{def_sigma_plus_minus_dbi_b}
\sigma_{\pm}=\pm \f{ \sqrt{(\prod_i g_{z^iz^j})}}{\sqrt{g_{tt}g_{rr}-(\p_r A^{(0)}_t)^2}\sqrt{g^2_{xx}+B^2}}\bigg[-\f{g_{tt}g_{xx}\p_r A^{(1)}_{\pm}}{\omega A^{(1)}_{\pm}}\pm B (\p_r A^{(0)}_t) \bigg]
\ee

It is easy to notice that the flow equation, eq(\ref{flow_dbi_b_plus_minus_cond}), shows an interesting property 
\be
\sigma_+(-B)=-\sigma_-(B),
\ee
where the explicit dependence on $r$ and $\omega$ is suppressed. In fact this property of the  conductivity follows very naturally from eq(\ref{def_sigma_plus_minus_dbi_b}).
From eq(\ref{flow_dbi_b_plus_minus_cond}), there follows the flow equation for the diagonal and Hall conductivity
\bea\label{flow_non_zero_B_cond}
\p_r\sigma^{xy}&=&-2i\bigg(\f{\omega\sigma^{xx}}{g_{tt}g_{xx}}\bigg)\Bigg[ B \p_r A^{(0)}_t -\sigma^{xy}\f{\sqrt{g_{tt}g_{rr}-(\p_r A^{(0)}_t)^2}\sqrt{g^2_{xx}+B^2}}{\sqrt{\prod_i g_{z^iz^j}}}\Bigg],\nn
\p_r\sigma^{xx}&=&i\f{\omega}{g_{tt}g_{xx}}[B h_1(\rho,r)\sigma^{xy}+h_2(B,\rho,r)((\sigma^{xy})^2-(\sigma^{xx})^2)+h_3(B,\rho,r)],
\eea
where the functions are
\bea
h_1(\rho,r)&=&2\p_r A^{(0)}_t,\quad h_2(B,\rho,r)=-\f{\sqrt{g_{tt}g_{rr}-(\p_r A^{(0)}_t)^2}\sqrt{g^2_{xx}+B^2}}{\sqrt{\prod_i g_{z^iz^j}}},\nn
h_3(B,\rho,r)&=&-\f{ \sqrt{\prod_i g_{z^iz^j}}}{\sqrt{g_{tt}g_{rr}-(\p_r A^{(0)}_t)^2}\sqrt{g^2_{xx}+B^2}}
\Bigg[g_{rr}g_{tt}g^2_{xx}+B^2(\p_r A^{(0)}_t)^2\Bigg].
\eea

From eq(\ref{flow_non_zero_B_cond}), it follows that for zero magnetic field, the Hall conductivity vanishes identically. 
Solving the  equation of motion to unperturbed gauge field, $A^{(0)}_t$, for non-zero magnetic field, gives
\be\label{sol_at_unperturbed_dbi_b}
(\p_r A^{(0)}_t)^2=\f{g_{tt}g_{rr}\rho^2}{\rho^2+(\prod_i g_{z^iz^j})(g^2_{xx}+B^2)} \Longrightarrow g_{tt}g_{rr}-(\p_r A^{(0)}_t)^2=\f{g_{tt}g_{rr}(\prod_i g_{z^iz^j})(g^2_{xx}+B^2)}{\rho^2+(\prod_i g_{z^iz^j})(g^2_{xx}+B^2)},
\ee 
where $\rho$ is  a constant of integration and identified as the charge density. Using this solution in the flow equation results in
\bea\label{flow_long_hall_cond_dbi_b}
\p_r\sigma^{xy}&=&-2i\omega\sigma^{xx}
\bigg(\f{\sqrt{g_{rr}}}{\sqrt{g_{tt}}g_{xx}}\bigg)\Bigg(\f{B\rho-\sigma^{xy}(g^2_{xx}+B^2)}{\sqrt{\rho^2+(\prod_i g_{z^iz^j})(g^2_{xx}+B^2)}}\Bigg),\nn
\p_r\sigma^{xx}&=&-i\omega\bigg(\f{\sqrt{g_{rr}}}{\sqrt{g_{tt}}g_{xx}}\bigg)
\Bigg(\f{-2B\rho\sigma^{xy}+
(g^2_{xx}+B^2)((\sigma^{xy})^2-(\sigma^{xx})^2)+\rho^2+g^2_{xx}(\prod_i g_{z^iz^j})}{\sqrt{\rho^2+(\prod_i g_{z^iz^j})(g^2_{xx}+B^2)}}\Bigg),\nn
\eea

Demanding the regularity condition at the horizon, $r=r_h,$ yields
\bea\label{long_hall_conductivity_at_horizon}
Re\sigma^{xy}(r_h)&=&\f{B\rho}{g^2_{xx}(r_h)+B^2},\quad 
Re\sigma^{xx}(r_h)=\pm\Bigg(\f{g_{xx}\sqrt{\rho^2+(\prod_i g_{z^iz^j})(g^2_{xx}+B^2)}}{g^2_{xx}+B^2}\Bigg)_{r_h},\nn
Im\sigma^{xy}(r_h)&=&0,\quad Im\sigma^{xx}(r_h)=0.
\eea

In $3+1$ dimensional bulk spacetime, in the high density and high magnetic field limit with the geometry of the form eq(\ref{metric_example}) and $w=1$, gives the conductivities as
\be\label{conductivity_dbi_high_density_high_b}
Re\sigma^{xy}\sim \f{\rho}{B},\quad Re\sigma^{xx}\sim T^{2/z}\f{\sqrt{\rho^2+B^2}}{B}.
\ee

It matches precisely with the dc Hall conductivity of \cite{hk} and \cite{hh}, but not  the dc conductivity, $\sigma^{xx}$. {\em A priori}, there is not any reason why the dc conductivity should match, because  \cite{hk} and \cite{hh} considers the Maxwellian type of action. However, if we take the high density and low magnetic field limit, then the conductivities are
\be\label{conductivity_dbi_high_density_low_b}
Re\sigma^{xy}\sim  \rho B T^{-4/z},\quad Re\sigma^{xx}\sim \rho T^{-2/z},
\ee 

which matches with the result of \cite{hpst}. It is interesting to note that the flow equation written down in eq(\ref{flow_long_hall_cond_dbi_b}), obeys the Onsager relation: $\sigma^{xy}(-B)=-\sigma^{xy}(B)=\sigma^{yx}(B)$. 
It follows from eq(\ref{conductivity_dbi_high_density_high_b}), so also from  eq(\ref{conductivity_dbi_high_density_low_b}),  that the ratio of the dc longitudinal conductivity to the dc  Hall conductivity has the following temperature dependence, $Re\sigma^{xx}/Re\sigma^{xy}\sim T^{ 2/z}$. 

The dc longitudinal  and Hall conductivities  at the horizon are given in eq(\ref{long_hall_conductivity_at_horizon}). Upon differentiating the Hall conductivity with respect to the  magnetic field 
\be
\f{\p Re\sigma^{xy}(r_h)}{\p B}=\rho\f{(g^2_{xx}(r_h)-B^2)}{(g^2_{xx}(r_h)+B^2)^2},
\ee
which in the low magnetic field limit becomes  $\f{\p Re\sigma^{xy}}{\p B}\sim \f{\rho}{g^2_{xx}(r_h)}$. If we identify it with $\f{\p Re\sigma^{xy}}{\p B}=T^{-1/\nu z}$, then we can determine the parameter, $\nu$. Using the choice,  $g_{xx}(r_h)=r^2_h\sim T^{2/z}$, where $T$ is the Hawking temperature, gives 
\be
\nu=1/4=0.25.
\ee

Generically, in $d+1$ bulk Lifshitz  spacetime dimensions, the real part of the conductivity at the horizon goes as
\be\label{generic_conductivity_dbi_b_lifshitz_horizon}
Re\sigma^{xy}(r_h)=\f{B\rho T^{-4/z}}{1+T^{-4/z}B^2},\quad Re\sigma^{xx}(r_h)=T^{\f{(d-3)}{z}}\f{\sqrt{1+\rho^2 T^{-2(d-1)/z}+B^2 T^{-4/z}}}{1+T^{-4/z}B^2}.
\ee

There follows in the low density and low magnetic field limit
\be
Re\sigma^{xy}(r_h)\sim B\rho T^{-4/z},\quad Re\sigma^{xx}(r_h)=T^{\f{(d-3)}{z}}.
\ee
So, the dc longitudinal conductivity of theories described by the  DBI action in the low charge density and zero magnetic field limit essentially mimics the features of the dc longitudinal conductivity of  theories described by the Maxwell type of action.

In $3+1$ spacetime dimensions the longitudinal and Hall conductivity  given in  eq(\ref{long_hall_conductivity_at_horizon}) at the horizon precisely matches with  eq(56) of \cite{Alanen:2009cn} for unit tension of the probe brane. It  matches even after  including the tension of brane following the recipe given in eq(\ref{recipe_tension_dbi}).

Let us define the following functions:
\bea
X_1&=&\f{2}{g_{xx}}\sqrt{\f{g_{rr}}{g_{tt}}},\quad X_2=\f{B\rho}{\sqrt{\rho^2+(\prod_i g_{z^iz^j})(g^2_{xx}+B^2)}},\nn  
X_3&=&\f{g^2_{xx}+B^2}{\sqrt{\rho^2+(\prod_i g_{z^iz^j})(g^2_{xx}+B^2)}},
\quad X_7=\f{\rho^2+g^2_{xx}(\prod_i g_{z^iz^j})}{\sqrt{\rho^2+(\prod_i g_{z^iz^j})(g^2_{xx}+B^2)}}.
\eea

Equating the real and imaginary parts of the conductivities
in the flow  eq(\ref{flow_long_hall_cond_dbi_b}), results
\bea\label{flow_re_im_long_hall_dbi_b}
\p_r Re\sigma^{xy}&=&\omega X_1 Im\sigma^{xx}(X_2-X_3 Re\sigma^{xy})-\omega X_1 X_3 Re\sigma^{xx}Im\sigma^{xy},\nn
\p_r Im\sigma^{xy}&=&-\omega X_1 X_3 Im\sigma^{xx} Im\sigma^{xy}-\omega X_1 Re\sigma^{xx}(X_2-X_3 Re\sigma^{xy}),\nn
\p_r Re\sigma^{xx}&=&-\omega X_1[X_2 Im\sigma^{xy}-X_3(Re\sigma^{xy}Im\sigma^{xy}-Re\sigma^{xx}Im\sigma^{xx}) ] ,\nn
\p_r Im\sigma^{xx}&=&\f{\omega}{2} X_1\bigg[2X_2 Re\sigma^{xy}-\nn && X_3\bigg((Re\sigma^{xy})^2-(Im\sigma^{xy})^2-
(Re\sigma^{xx})^2+(Im\sigma^{xx})^2\bigg)-X_7\bigg].
\eea

 Upon looking at the flow eq(\ref{flow_re_im_long_hall_dbi_b}), the regularity condition at the horizon gives  another possibility, other than eq(\ref{long_hall_conductivity_at_horizon}). That is
\bea\label{regularity_condition_sigma_xy}
Re\sigma^{xy}(r_h)&=&\f{B\rho}{g^2_{xx}(r_h)+B^2},\quad Im\sigma^{xy}(r_h)=\pm\Bigg(\f{g_{xx}\sqrt{\rho^2+(\prod_i g_{z^iz^j})(g^2_{xx}+B^2)}}{g^2_{xx}+B^2}\Bigg)_{r_h},\nn
Re\sigma^{xx}(r_h)&=&0,\quad Im\sigma^{xx}(r_h)=0.
\eea

Upon substituting  eq(\ref{sol_at_unperturbed_dbi_b}) into eq(\ref{def_sigma_plus_minus_dbi_b})   gives the expression of the conductivity as
\bea\label{def_sigma_plus_minus_dbi_b_metric_rho}
\sigma_{\pm}&=&\pm \f{\sqrt{\rho^2+(\prod_i g_{z^iz^j})(g^2_{xx}+B^2)}}{\sqrt{g_{tt}g_{rr}}(g^2_{xx}+B^2)}\bigg[-\f{g_{tt}g_{xx}\p_r A^{(1)}_{\pm}}{\omega A^{(1)}_{\pm}} \pm \f{B\rho\sqrt{g_{tt}g_{rr}}}{\sqrt{\rho^2+(\prod_i g_{z^iz^j})(g^2_{xx}+B^2)}}\bigg],\nn
&=&\mp \f{\sqrt{\rho^2+(\prod_i g_{z^iz^j})(g^2_{xx}+B^2)}}{\sqrt{g_{rr}}(g^2_{xx}+B^2)}\f{\sqrt{g_{tt}}g_{xx}\p_r A^{(1)}_{\pm}}{\omega A^{(1)}_{\pm}}+\f{B\rho}{(g^2_{xx}+B^2)}.
\eea

There follows that at the horizon, where $g_{tt}(r_h)=0$, gives back the dc conductivity  at the horizon as written in eq(\ref{long_hall_conductivity_at_horizon}).  So, it's the beauty of the flow equation, eq(\ref{flow_re_im_long_hall_dbi_b}), that gives the appropriate dc conductivity at the horizon without even the need to solve any differential equation.

The  symmetries of the flow equations, eq(\ref{flow_re_im_long_hall_dbi_b}), are of the following type
\bea\label{symmetries_long_hall_dbi_b}
&(1)& Re\sigma^{xx}(-\omega)=Re\sigma^{xx}(\omega),\quad Im\sigma^{xx}(-\omega)=-Im\sigma^{xx}(\omega),\quad Re\sigma^{xy}(-\omega)=Re\sigma^{xy}(\omega),\nn && Im\sigma^{xy}(-\omega)=-Im\sigma^{xy}(\omega);\nn
&(2)& B\rightarrow -B,\quad \rho \rightarrow -\rho;\nn
&(3)&B\rightarrow -B  ,\quad Re\sigma^{xx}\rightarrow Re\sigma^{xx},\quad Im\sigma^{xx}\rightarrow Im\sigma^{xx}, \quad Re\sigma^{xy}\rightarrow -Re\sigma^{xy},\nn && Im\sigma^{xy}\rightarrow -Im\sigma^{xy};\nn 
&(4)&\rho\rightarrow -\rho  ,\quad Re\sigma^{xx}\rightarrow Re\sigma^{xx},\quad Im\sigma^{xx}\rightarrow Im\sigma^{xx},\quad Re\sigma^{xy}\rightarrow -Re\sigma^{xy},\nn && Im\sigma^{xy}\rightarrow -Im\sigma^{xy};\nn 
&(5)&B\rightarrow -B,\quad Re\sigma^{xx}\rightarrow -Re\sigma^{xx},\quad Im\sigma^{xx}\rightarrow -Im\sigma^{xx},\quad Re\sigma^{xy}\rightarrow -Re\sigma^{xy},\nn && Im\sigma^{xy}\rightarrow -Im\sigma^{xy}, \quad \omega \rightarrow -\omega;\nn 
&(6)&\omega \rightarrow -\omega, \quad Re\sigma^{xx}\rightarrow -Re\sigma^{xx},\quad Im\sigma^{xx}\rightarrow -Im\sigma^{xx},\quad Re\sigma^{xy}\rightarrow Re\sigma^{xy},\nn && Im\sigma^{xy}\rightarrow Im\sigma^{xy}.
\eea

The symmetry of the type $(1)$ is nothing but the time reversal symmetry: $\sigma^{\star}(\omega)=\sigma({-\omega})$ as suggested in \cite{dvdm}. It is interesting to see that the boundary conditions, at the horizon, eq(\ref{long_hall_conductivity_at_horizon}) and eq(\ref{regularity_condition_sigma_xy}) does not respect the symmetries of the type $(5)$ and $(6)$. Moreover, the boundary condition, eq(\ref{regularity_condition_sigma_xy}), does not respect the symmetry of the type $(3)$, too. In what follows, we shall be considering only the symmetries of the type $(1),~(2),~(3)$, and $(4)$. Hence, the only consistent boundary condition   with this is eq(\ref{long_hall_conductivity_at_horizon}). It means, 
the regularity condition at the horizon, namely, eq(\ref{regularity_condition_sigma_xy}), is ruled out because of the violation of the symmetry of type $(3)$ and $(4)$. 

It is interesting to note that the low frequency expansion of the conductivity as written down in \cite{hh}, respects only the symmetries of the type $(1),~(2),~(3)$, and $(4)$, of eq(\ref{symmetries_long_hall_dbi_b}), even though    the authors of \cite{hh}, were only dealing with  the Einstein-Maxwellian type of action. 

{\bf Electro-Magnetic duality}:

 For the Einstein-Maxwell type of action,  in $3+1$ dimensional bulk spacetime, as considered in \cite{hh}, there exists a duality  under which  the charge density and magnetic fields are exchanged and the conductivity  goes to its inverse up to a sign
\be\label{s_duality}
\rho\rightarrow B,\quad B \rightarrow -\rho,\quad \sigma_{\pm}\rightarrow -\f{1}{\sigma_{\pm}}.
\ee

The flow equation of the conductivity for  the DBI type of action in a fixed background geometry does  show up the electro-magnetic duality symmetry, eq(\ref{s_duality}), for a $3+1$ dimensional bulk system. In order to see it explicitly, let us rewrite eq(\ref{flow_dbi_b_plus_minus_cond}), in terms of the metric components, the charge density and magnetic fields as
\bea
\p_r\sigma_{\pm}&=&\mp \f{2\omega B\rho \sqrt{g_{rr}}~ \sigma_{\pm}}{\sqrt{g_{tt}}g_{xx}{\sqrt{\rho^2+g^2_{xx}+B^2}}}\nn &\pm& \f{\omega\sqrt{g_{rr}}}{g_{xx}\sqrt{g_{tt}}}\f{1}{\sqrt{\rho^2+g^2_{xx}+B^2}}
\Bigg[\sigma^2_{\pm}(g^2_{xx}+B^2)+(g^2_{xx}+\rho^2)\Bigg]
\eea 

The left hand side of the flow equation goes,
 under the above mentioned  transformation as
\bea 
\p_r\sigma_{\pm}\rightarrow \f{\p_r\sigma_{\pm}}{\sigma^2_{\pm}}&=&\mp \f{2\omega B\rho \sqrt{g_{rr}}~ }{\sqrt{g_{tt}}g_{xx}{\sqrt{\rho^2+g^2_{xx}+B^2}}}\f{1}{\sigma_{\pm}}\nn&&\pm \f{\omega\sqrt{g_{rr}}}{g_{xx}\sqrt{g_{tt}}}\f{1}{\sqrt{\rho^2+g^2_{xx}+B^2}}
\Bigg[\f{(g^2_{xx}+\rho^2)}{\sigma^2_{\pm}}+(g^2_{xx}+B^2)\Bigg],
\eea
which is precisely  the right hand side of the flow equation upon using eq(\ref{s_duality}). It is interesting to note that
the interchange of the charge density and magnetic field is not a symmetry of the DBI type of action in contrast to the Maxwell type of action. The simplest way to see it as follows: The quantity $\sqrt{-det M^{(0)}_+}=\sqrt{\prod_i g_{z^iz^j}}{\sqrt{g_{tt}g_{rr}-(\p_r A^{(0)}_t)^2}}\sqrt{g^2_{xx}+B^2}$,  appears in the action eq(\ref{action_quadratic_dbi_fixed}) and hence in the equation of motion, eq(\ref{eom_A_plus_minus}), and in the current, eq(\ref{current_dbi_b}). Using the solution to $A^{(0)}_t$ from eq(\ref{sol_at_unperturbed_dbi_b}), there follows that  it is this term that breaks the $\rho\leftrightarrow B$ symmetry explicitly, i.e., $\sqrt{-det M^{(0)}_+}=\f{(g^2_{xx}+B^2)\sqrt{g_{tt}g_{rr}}}{\rho^2+ g^2_{xx}+B^2}$.  

\subsubsection{Regularity condition=In-falling boundary condition}

In this subsection we shall show that the imposition of the regularity condition on the flow equation of the conductivity is exactly same as putting the in-falling boundary condition on the   gauge field at the horizon. Evaluating the conductivity at the horizon using eq(\ref{def_sigma_plus_minus_dbi_b_metric_rho}) and eq(\ref{long_hall_conductivity_at_horizon}), we find
\be\label{near_horizon_dbi_b_ratio_gauge_field}
\Bigg(\f{\p_r A^{(1)}_+}{A^{(1)}_+}\Bigg)_{r_h}=\Bigg(\f{\p_r A^{(1)}_-}{A^{(1)}_-}\Bigg)_{r_h},\quad \Bigg(\f{\p_r A^{(1)}_+}{A^{(1)}_+}+\f{\p_r A^{(1)}_-}{A^{(1)}_-}\Bigg)_{r_h}=\mp 2i\omega \Bigg(\sqrt{\f{g_{rr}}{g_{tt}}}\Bigg)_{r_h}
\ee

There follows  the behavior of the gauge  field close to the horizon 
\be
A^{(1)}_+(r,\omega)\simeq A^{(1)}_+(r_h)e^{\mp i\omega \int^r dr \sqrt{\f{g_{rr}}{g_{tt}}}}=A^{(1)}_+(r_h) (r-r_h)^{\mp i\omega \sqrt{c_r/c_t}}.
\ee
In obtaining the second equality, we have used the following behavior to metric components close to the horizon, $g_{tt}=c_t(r-r_h)$ and $g_{rr}=c_r/(r-r_h)$. See Appendix C, for further details.

\subsubsection{Low frequency limit: $\omega\rightarrow 0$ }

Using the symmetry of type $(1)$ and the boundary condition eq(\ref{long_hall_conductivity_at_horizon}), we can expand the real and imaginary part of the dimensionless conductivities eq(\ref{flow_re_im_long_hall_dbi_b_dim_less}), in the low frequency limit,   as 
\bea\label{series_low_frequency_long_hall_dbi_b}
Re{\tilde\sigma}^{xx}&=&{\tilde a}_0(r,~\rho,~B,~T)+{\tilde a}_1(r,~\rho,~B,~T){\hat\omega}^2+
{\tilde a}_2(r,~\rho,~B,~T){\hat\omega}^4+\cdots,\nn Im{\tilde\sigma}^{xx}&=&\f{{\tilde b}_0(r,~\rho,~B,~T)}{{\hat\omega}}+{\tilde b}_1(r,~\rho,~B,~T){\hat\omega}+
{\tilde b}_2(r,~\rho,~B,~T){\hat\omega}^3+\cdots,\nn
Re{\tilde\sigma}^{xy}&=&{\tilde c}_0(r,~\rho,~B,~T)+{\tilde c}_1(r,~\rho,~B,~T){\hat\omega}^2+
{\tilde c}_2(r,~\rho,~B,~T){\hat\omega}^4+\cdots,\nn Im{\tilde\sigma}^{xy}&=&\f{{\tilde d}_0(r,~\rho,~B,~T)}{{\hat\omega}}+{\tilde d}_1(r,~\rho,~B,~T){\hat\omega}+
{\tilde d}_2(r,~\rho,~B,~T){\hat\omega}^3+\cdots.
\eea

From, now onwards, we shall drop the arguments of ${\tilde a}_i,{\tilde b}_i,{\tilde c}_i$ and ${\tilde d}_i$'s, for easy.
Inserting this low frequency expansion into eq(\ref{flow_re_im_long_hall_dbi_b}) and equating each power of the frequency, results
\bea
\p_u {\tilde c}_0&=&-{\tilde{\tilde X}}_1X_2 {\tilde b}_0+{\tilde{\tilde X}}_1 {\tilde X}_3({\tilde b}_0{\tilde c}_0+{\tilde a}_0{\tilde d}_0),\nn \p_u {\tilde c}_1&=&-{\tilde b}_1 {\tilde{\tilde X}}_1 X_2+ {\tilde{\tilde X}}_1 {\tilde X}_3({\tilde b}_1{\tilde c}_0+{\tilde c}_1{\tilde b}_0+{\tilde a}_0{\tilde d}_1+{\tilde a}_1{\tilde d}_0),\nn \p_u {\tilde d}_0&=&{\tilde{\tilde X}}_1{\tilde X}_3 {\tilde b}_0{\tilde d}_0,\quad\p_u {\tilde d}_1={\tilde{\tilde X}}_1{\tilde X}_3({\tilde b}_0{\tilde d}_1+{\tilde b}_1{\tilde d}_0-{\tilde a}_0{\tilde c}_0)+{\tilde{\tilde X}}_1X_2 {\tilde a}_0,\nn
\p_u {\tilde c}_2&=&-{\tilde{\tilde X}}_1X_2 {\tilde b}_2+{\tilde{\tilde X}}_1{\tilde X}_3({\tilde c}_0{\tilde b}_2+{\tilde b}_1{\tilde c}_1+{\tilde c}_2{\tilde b}_0+{\tilde a}_0{\tilde d}_2+{\tilde a}_1{\tilde d}_1+{\tilde a}_2{\tilde d}_0),\nn
\quad \p_u {\tilde a}_0&=&{\tilde{\tilde X}}_1X_2{\tilde d}_0+{\tilde{\tilde X}}_1{\tilde X}_3[{\tilde a}_0{\tilde b}_0- {\tilde c}_0{\tilde d}_0],\quad 
\p_u {\tilde b}_0= \f{{\tilde{\tilde X}}_1{\tilde X}_3}{2} [{\tilde b}^2_0-{\tilde d}^2_0],\nn \p_u {\tilde a}_1&=&{\tilde{\tilde X}}_1X_2 {\tilde d}_1+{\tilde{\tilde X}}_1{\tilde X}_3({\tilde a}_0{\tilde b}_1+{\tilde a}_1{\tilde b}_0-{\tilde c}_1{\tilde d}_0-{\tilde c}_0{\tilde d}_1),\nn
\p_u {\tilde d}_2&=& {\tilde{\tilde X}}_1{\tilde X}_3({\tilde b}_2{\tilde d}_0+{\tilde b}_1{\tilde d}_1+{\tilde b}_0{\tilde d}_2-{\tilde a}_1{\tilde c}_0-{\tilde a}_0{\tilde c}_1)+{\tilde{\tilde X}}_1X_2 {\tilde a}_2,\nn
\p_u {\tilde a}_2&=&{\tilde{\tilde X}}_1X_2 {\tilde d}_2+{\tilde{\tilde X}}_1{\tilde X}_3({\tilde a}_0{\tilde b}_2+{\tilde a}_1{\tilde b}_1+{\tilde a}_2{\tilde b}_0-{\tilde c}_2{\tilde d}_0-{\tilde c}_0{\tilde d}_2-{\tilde c}_1{\tilde d}_1),\nn \p_u{\tilde b}_1&=&\f{{\tilde{\tilde X}}_1{\tilde X}_7}{2}+\f{{\tilde{\tilde X}}_1{\tilde X}_3}{2}[{\tilde c}^2_0-{\tilde a}^2_0-2{\tilde d}_0{\tilde d}_1+2{\tilde b}_0{\tilde b}_1]+{\tilde{\tilde X}}_1X_2 {\tilde c}_0,\nn
\p_u {\tilde b}_2&=&\f{{\tilde{\tilde X}}_1{\tilde X}_3}{2}[2{\tilde c}_0{\tilde c}_1+{\tilde d}^2_1+2{\tilde d}_0{\tilde d}_2-2{\tilde a}_0{\tilde a}_1+{\tilde b}^2_1+2{\tilde b}_0{\tilde b}_2]-{\tilde{\tilde X}}_1X_2{\tilde c}_2,\quad \cdots\cdots\cdots
\eea

\begin{figure}[htb]
\centering
\subfigure[$Re{\tilde \sigma}^{xx}$]{\includegraphics[width=3.0 in]
{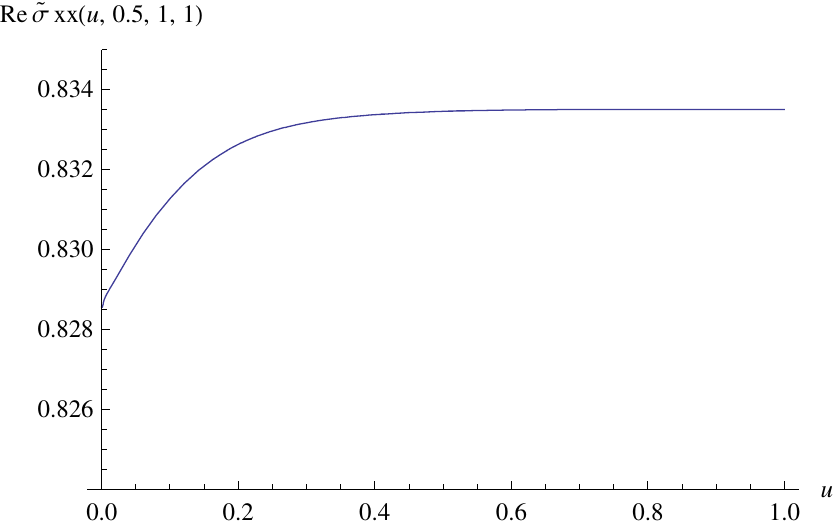}}
\subfigure[$Im{\tilde \sigma}^{xx}$]{\includegraphics[width=3.0 in]
{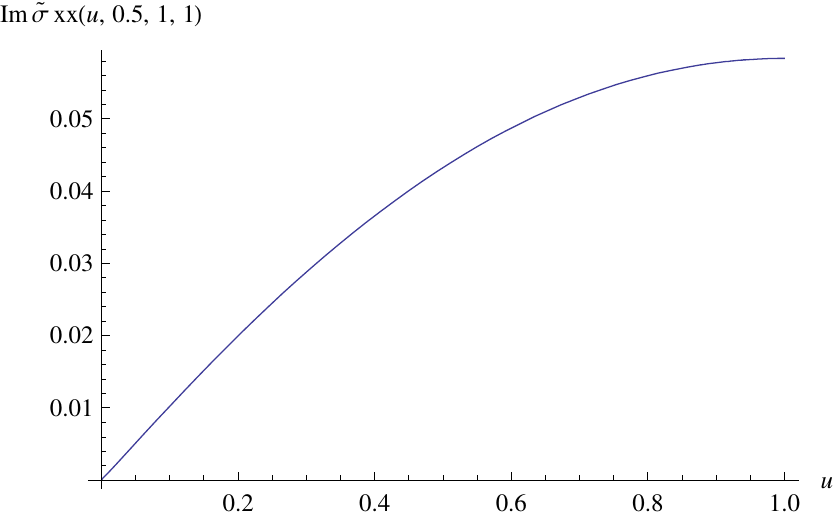}}
\caption{The behavior of the $ Re{\tilde \sigma}^{xx}$ and 
$ Im{\tilde \sigma}^{xx}$, along the vertical direction, is plotted versus $u$  for ${\hat\omega}=0.5,~ B=1=\rho,~{L/r_h}=1, 
~{\hat g}_{YM}=1$ for $3+1$ dimensional Schwarzschild black hole.}
\label{fig_5}
\end{figure}

\begin{figure}[htb]
\centering
\subfigure[$Re{\tilde \sigma}^{xy}$]{\includegraphics[width=3.0 in]
{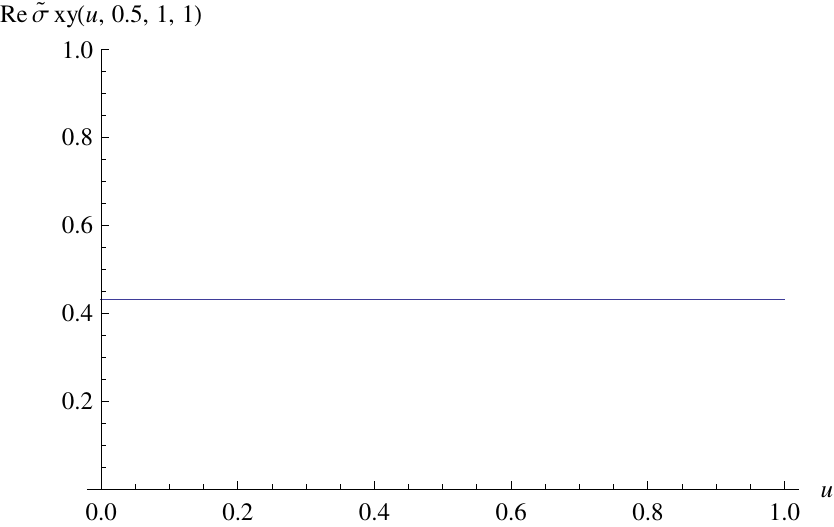}}
\subfigure[$Im{\tilde \sigma}^{xy}$]{\includegraphics[width=3.0 in]
{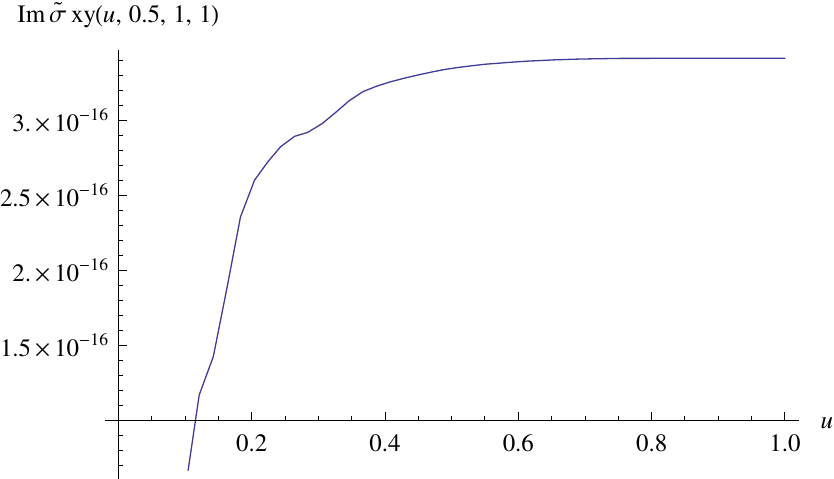}}
\caption{The behavior of the $ Re{\tilde \sigma}^{xy}$ and 
$ Im{\tilde \sigma}^{xy}$, along the vertical direction, is plotted versus $u$  for ${\hat\omega}=0.5,~ B=1=\rho,~{L/r_h}=1, 
~{\hat g}_{YM}=1$ for $3+1$ dimensional Schwarzschild black hole.}
\label{fig_6}
\end{figure}

It is not clear whether the real part of the conductivity shows up the Drude like behavior. In order to check that we need the exact solutions, which is very difficult to find. 
The numerical solutions are given in  fig(\ref{fig_5}) and fig(\ref{fig_6}) and it suggests that ${\tilde c}_0(u)={\tilde c}_0(u_h),~{\tilde c}_i(u)=0$, for $i=1,2,3,\cdots,~ {\tilde a}_0(u)={\tilde a}_0(u_h),~{\tilde b}_0(u)=0$, and ${\tilde d}_0(u)=0$. It means $Re{\tilde\sigma}^{xy}(u)={\rm constant}\equiv Re{\tilde\sigma}^{xy}(u_h).$ So, there follows that  $Re{\tilde\sigma}^{xy}$ does not show up the Drude like form.

\subsubsection{High frequency limit: $\omega\rightarrow \infty$ }

Once again following the same strategy as earlier in the very high frequency limit, i.e, demanding that the derivative of the conductivity evaluated at a critical frequency, $\omega_c$,  is independent of the frequency. Consistency then suggests that the critical frequency should better be very large, $\omega_c\rightarrow\infty$.  

The result of this  is consistent with the boundary condition 
\bea\label{high_frequency_long_hall_dbi_b}
&&Lim_{\omega\rightarrow\omega_c=\infty} Re\sigma^{xx}=\pm\f{\sqrt{X_3X_7-X^2_2}}{X_3},\quad Lim_{\omega\rightarrow\omega_c=\infty}Im\sigma^{xx}=0,\nn && Lim_{\omega\rightarrow\omega_c=\infty}Re\sigma^{xy}=\f{X_2}{X_3},\quad Lim_{\omega\rightarrow\omega_c=\infty}Im\sigma^{xy}=0.
\eea

\subsubsection{Phase angle of conductivity at high frequency}

Here, we shall try to find the phase angle of the Hall conductivity as well as that of the longitudinal conductivity at very high frequency i.e., using the result of eq(\ref{high_frequency_long_hall_dbi_b}). The flow equation  for the phase angle are
\bea
\p_rtan~\theta^{xy}&=&\omega X_1X_3 Re\sigma^{xx}sec^2\theta^{xy}-\omega X_1X_2 \f{Re\sigma^{xx}}{Re\sigma^{xy}}[1+tan~\theta^{xx}tan~\theta^{xy}],\nn
\p_rtan~\theta^{xx}&=&\omega X_1X_2 \f{Re\sigma^{xy}}{Re\sigma^{xx}}[1+tan~\theta^{xx}tan~\theta^{xy}]-\omega X_1X_3 \bigg[\f{tan~\theta^{xx}Re\sigma^{xy}Im\sigma^{xy}}{Re\sigma^{xx}}-\nn && \f{tan~\theta^{xx}Im\sigma^{xx}}{2}+\f{(Re\sigma^{xy})^2}{2Re\sigma^{xx}}-\f{(Im\sigma^{xy})^2}{2Re\sigma^{xx}}-\f{Re\sigma^{xx}}{2}\bigg]-\f{\omega X_1X_7}{2Re\sigma^{xx}}
\eea

Let us apply the approach as used in the previous sections, i.e., we shall impose the condition that the derivative of the phase angle with respect to the frequency vanishes as we evaluate it at $\omega_c=\infty$. By doing so and using the behavior of the conductivity at infinity frequency, eq(\ref{high_frequency_long_hall_dbi_b}), we ended up with
\be
sec^2~\theta^{xy}=1+tan~\theta^{xx}tan~\theta^{xy},\quad tan~\theta^{xx}tan~\theta^{xy}=0.
\ee

The first equation follows from the phase angle of the Hall conductivity and the second one, from the phase angle of the
longitudinal conductivity, respectively. Now, solving these equations for the phases, we find
\be
\theta^{xy}=n\pi,\quad \theta^{xx}={\rm Arbitrary},
\ee
that the  phase angle of the Hall conductivity is an integral multiple of $\pi$, whereas the phase angle of the
longitudinal conductivity remains arbitrary. The quantity, $n$, is an integer including zero.

\subsection{Mid-frequency region: $\omega > T$}
This particular region in frequency,  as was argued previously, is same as considering at zero temperature with  small  yet finite frequency. Generically, in this region, it is very difficult to solve the flow equation for conductivity, eq(\ref{flow_dbi_b_plus_minus_cond}) because of it's non-linear form. However, we can solve the differential equation for $A^{(1)}_{\pm}$, i.e., eq(\ref{eom_A_plus_minus})
for few special cases. Then we can use eq(\ref{def_sigma_plus_minus_dbi_b}) to read out the conductivity.

For easy of doing calculation, let us restrict to $3+1$ dimensional bulk spacetime dimensions. The explicit structure to geometry that we shall consider is of the Lifshitz type
\be
ds^2=-\f{{\tilde f}(r)dt^2}{r^{2z}}+\f{dx^2+dy^2}{r^2}+\f{dr^2}{r^2 {\tilde f}(r)},
\ee
where the zero of the function, ${\tilde f}(r)$, determines the location of the horizon, ${\tilde f}(r_h)=0$. As argued in \cite{hpst} by considering a region $ r_0 \leq  r \ll r_h$, we can set ${\tilde f}(r)\simeq 1$. Which essentially goes over to the argument suggested in the previous paragraph\footnote{ Note that in this subsection only, we assume that the boundary is at $r=0$.}.  Let us consider a very high density and small magnetic field limit, in which case the equation of motion to gauge field reduces to 
\be\label{eom_gauge_field_3+1_dbi_b}
\p_r [r^{3-z}\p_r A^{(1)}_{\pm}]+\omega r^3[\omega r^{z-2}\mp 4B]A^{(1)}_{\pm}=0.
\ee

It is very difficult to solve the differential equation for any generic choice of the  dynamical exponent $z$. So, we shall solve for $z=1$ and $z=2$ cases, exactly.  For $z=1$, the solution reads
\be\label{sol_z_1_dbi_b_mid_frequency}
A^{(1)}_{\pm}=\f{c_1}{r}~ Ai\bigg[\f{\pm 4B\omega r-\omega^2 }{2^{4/3}(\pm B\omega)^{2/3}}\bigg]+\f{c_2}{r}~ Bi\bigg[\f{\pm 4B\omega r-\omega^2 }{2^{4/3}(\pm B\omega)^{2/3}}\bigg],
\ee
where $Ai(x)$ and $Bi(x)$ are Airy functions and $c_i$'s are constants of integration. For $z=2$, the solution reads
\be\label{sol_z_2_dbi_b_mid_frequency}
A^{(1)}_{\pm}=c_1~ I_0\bigg(\f{r^2\sqrt{\pm 4B\omega-\omega^2}}{2}\bigg)+c_2~ K_0\bigg(\f{r^2\sqrt{\pm 4B\omega-\omega^2}}{2}\bigg),
\ee
where $I_n(x)$ and $K_n(x)$ are modified Bessel function of the first and second kind respectively.

In the absence of the  magnetic field, $B=0$, the solution  for any $z$ except $z=2$ is \cite{hpst}
\be
A^{(1)}_{\pm}=c_1~r^{\f{z}{2}-1}J_{\f{1}{2}-\f{1}{z}}\bigg(\f{\omega r^z}{z}\bigg)+c_2~r^{\f{z}{2}-1}J_{-\f{1}{2}+\f{1}{z}}\bigg(\f{\omega r^z}{z}\bigg),
\ee

and for $z=2$ the second $J_{\nu}$ is replaced by $K_{\nu}$.
However, if we consider a regime of parameter space for which the magnetic field term dominates over the frequency term in the second square bracket of eq(\ref{eom_gauge_field_3+1_dbi_b}), then the solutions for $z\neq 2$ are
\bea
A^{(1)}_+&=& c_1~r^{\f{z}{2}-1}I_{\f{2-z}{2+z}}\bigg(\f{4\sqrt{B\omega}}{2+z}r^{1+\f{z}{2}}\bigg)+c_2~r^{\f{z}{2}-1}I_{\f{z-2}{z+2}}\bigg(\f{4\sqrt{B\omega}}{2+z}r^{1+\f{z}{2}}\bigg),\nn
A^{(1)}_-&=& c_1~r^{\f{z}{2}-1}J_{\f{2-z}{2+z}}\bigg(\f{4\sqrt{B\omega}}{2+z}r^{1+\f{z}{2}}\bigg)+c_2~r^{\f{z}{2}-1}J_{\f{z-2}{z+2}}\bigg(\f{4\sqrt{B\omega}}{2+z}r^{1+\f{z}{2}}\bigg)
\eea 

for $z=2$, the $I_{\nu}$ is replaced by $K_{\nu}$ and $J_{\nu}$ is replaced by $Y_{\nu}$, which is the Bessel function of the second kind. 

Considering a $3+1$ dimensional bulk spacetime dimensions in eq(\ref{def_sigma_plus_minus_dbi_b_metric_rho}),  follows    the expression of the conductivity 
\be
\sigma_{\pm}=\sigma_{xy}\pm i\sigma_{xx} =\mp \f{\sqrt{\rho^2+g^2_{xx}+B^2}}{\sqrt{g_{rr}}(g^2_{xx}+B^2)}\f{\sqrt{g_{tt}}g_{xx}\p_r A^{(1)}_{\pm}}{\omega A^{(1)}_{\pm}}+\f{B\rho}{(g^2_{xx}+B^2)}.
\ee

The different  components of the conductivities are
\bea\label{conductivity_dbi_b_w_gt_T}
\sigma_{xx}&=&\f{\sigma_+-\sigma_-}{2i}=-\f{1}{2}\f{\sqrt{\rho^2+g^2_{xx}+B^2}}{\sqrt{g_{rr}}(g^2_{xx}+B^2)}\f{\sqrt{g_{tt}}g_{xx}}{i\omega }\bigg(\f{\p_r A^{(1)}_+}{A^{(1)}_+}+\f{\p_r A^{(1)}_-}{A^{(1)}_-} \bigg),\nn 
\sigma_{xy}&=&\f{\sigma_++\sigma_-}{2}=-\f{1}{2}\f{\sqrt{\rho^2+g^2_{xx}+B^2}}{\sqrt{g_{rr}}(g^2_{xx}+B^2)}\f{\sqrt{g_{tt}}g_{xx}}{\omega }\bigg(\f{\p_r A^{(1)}_+}{A^{(1)}_+}-\f{\p_r A^{(1)}_-}{A^{(1)}_-} \bigg)+\f{B\rho}{(g^2_{xx}+B^2)}.
\eea

As a warm up, for zero magnetic field with unity dynamical exponent, the solution to eq(\ref{eom_gauge_field_3+1_dbi_b}), $A_+=A_-\equiv A$
\be
A=\f{c_1}{r}~e^{ i\omega r}+\f{c_2}{r}~e^{- i\omega r},
\ee
where $c_i$'s are constants of integration. 
For this solution, $\f{\p_r A^{(1)}_+}{A^{(1)}_+}+\f{\p_r A^{(1)}_-}{A^{(1)}_-}=2\f{\p_r A}{A}=-1/r+i\omega(c_1 e^{i\omega r}-c_2 e^{-i\omega r})/(c_1 e^{i\omega r}+c_2 e^{-i\omega r})$.  If we set $c_1=0$, then 
\be
\sigma_{xx}(r_0, \omega, \rho,)= \Bigg(\f{\sqrt{\rho^2+g^2_{xx}}}{\sqrt{g_{rr}}g^2_{xx}}\sqrt{g_{tt}}g_{xx}\Bigg)_{r_0}\bigg[\f{1}{i\omega r_0}+1\bigg]=\Bigg(\f{g_{xx}\sqrt{\rho^2+g^2_{xx}}}{g^2_{xx}}\Bigg)_{r_0}\bigg[\f{1}{i\omega r_0}+1\bigg],
\ee
however, if we set $c_2=0$, then
\be
\sigma_{xx}(r_0, \omega, \rho)= \Bigg(\f{\sqrt{\rho^2+g^2_{xx}}}{\sqrt{g_{rr}}g^2_{xx}}\sqrt{g_{tt}}g_{xx}\Bigg)_{r_0}\bigg[\f{1}{i\omega r_0}-1\bigg]=\Bigg(\f{g_{xx}\sqrt{\rho^2+g^2_{xx}}}{g^2_{xx}}\Bigg)_{r_0}\bigg[\f{1}{i\omega r_0}-1\bigg],
\ee
where we have used $g_{tt}/g_{rr}\simeq 1$, in the regime of our interest $r_0 \leq  r <<r_h$. 
It is interesting to note that the frequency dependent part of the  conductivity remains the same irrespective of  setting either of the constant of integration to zero.
But the outgoing boundary condition at the boundary makes us to set $c_2=0$.
As expected there is not any ac conductivity to $\sigma_{xy}$ and the result of the frequency dependence of the conductivity  matches with that of \cite{hpst}.

\subsubsection{$z=1$ with non-zero magnetic field}

The solution for a $3+1$ dimensional bulk spacetime dimension with unity dynamical exponent is given in eq(\ref{sol_z_1_dbi_b_mid_frequency}). {\em A priori}, it is not clear how to fix the boundary condition. So, we shall proceed by setting either of the constant of integrations to zero and analyze the frequency dependence to  conductivity.   
The low frequency expansion of the gradient of the logarithm of the gauge potential, when $c_1=0$
\bea
\f{\p_r A^{(1)}_+}{A^{(1)}_+}&=&-\f{1}{r}+\f{2^{2/3} (3B)^{1/3}\Gamma(2/3)}{\Gamma(1/3)}\omega^{1/3}-\f{2^{4/3} (3B)^{2/3}\Gamma^2(2/3)}{\Gamma^2(1/3)}r\omega^{2/3}+{\cal O}(\omega, B),\nn
\f{\p_r A^{(1)}_-}{A^{(1)}_-}&=&-\f{1}{r}+\f{2^{2/3} (-3B)^{1/3}\Gamma(2/3)}{\Gamma(1/3)}\omega^{1/3}-\f{2^{4/3} (-3B)^{2/3}\Gamma^2(2/3)}{\Gamma^2(1/3)}r\omega^{2/3}+{\cal O}(\omega, B),\nn
\eea
but when  $c_2=0$
\bea
\f{\p_r A^{(1)}_+}{A^{(1)}_+}&=&-\f{1}{r}-\f{2^{2/3} (3B)^{1/3}\Gamma(2/3)}{\Gamma(1/3)}\omega^{1/3}-\f{2^{4/3} (3B)^{2/3}\Gamma^2(2/3)}{\Gamma^2(1/3)}r\omega^{2/3}+{\cal O}(\omega, B),\nn
\f{\p_r A^{(1)}_-}{A^{(1)}_-}&=&-\f{1}{r}-\f{2^{2/3} (-3B)^{1/3}\Gamma(2/3)}{\Gamma(1/3)}\omega^{1/3}-\f{2^{4/3} (-3B)^{2/3}\Gamma^2(2/3)}{\Gamma^2(1/3)}r\omega^{2/3}+{\cal O}(\omega, B).
\eea

Substituting this low frequency solution into eq(\ref{conductivity_dbi_b_w_gt_T}) and evaluating it at $r=r_0$ gives
\bea
\sigma_{xx}(c_2=0,r_0)&\simeq& \Bigg(\f{1}{r}+\f{\sqrt{\rho^2+g^2_{xx}+B^2}}{\sqrt{g_{rr}}(g^2_{xx}+B^2)}\sqrt{g_{tt}}g_{xx}\Bigg)_{r_0}\bigg(\f{(3B)^{1/3}\Gamma(2/3)}{2^{-1/3}\Gamma(1/3)}\bigg)(3+i\sqrt{3})\omega^{-2/3},\nn
\sigma_{xy}(c_2=0,r_0)&\simeq& \Bigg(\f{1}{r}+\f{\sqrt{\rho^2+g^2_{xx}+B^2}}{\sqrt{g_{rr}}(g^2_{xx}+B^2)}\sqrt{g_{tt}}g_{xx}\Bigg)_{r_0}\bigg(\f{(3B)^{1/3}\Gamma(2/3)}{2^{-1/3}\Gamma(1/3)}\bigg)(1-i\sqrt{3})\omega^{-2/3},\nn
\eea
where we have dropped the frequency independent piece to $\sigma_{xy}$. 
The phase of the conductivity is  defined as $tan~\theta=Im\sigma/Re\sigma$
\be
tan~\theta^{xx}(r_0)=\f{1}{\sqrt{3}},\quad tan~\theta^{xy}(r_0)=-\sqrt{3}  \quad \Longrightarrow \theta^{xx}(r_0)=\pi/6,\quad \theta^{xy}(r_0)=-\pi/3.
\ee

Similarly, when $c_1=0$, the conductivities are
\bea
\sigma_{xx}(c_1=0,r_0)&\simeq& \Bigg(\f{1}{r}+\f{\sqrt{\rho^2+g^2_{xx}+B^2}}{\sqrt{g_{rr}}(g^2_{xx}+B^2)}\sqrt{g_{tt}}g_{xx}\Bigg)_{r_0}\bigg(\f{(3B)^{1/3}\Gamma(2/3)}{2^{-1/3}\Gamma(1/3)}\bigg)(3+i\sqrt{3})\omega^{-2/3},\nn
\sigma_{xy}(c_1=0,r_0)&\simeq& \Bigg(\f{1}{r}+\f{\sqrt{\rho^2+g^2_{xx}+B^2}}{\sqrt{g_{rr}}(g^2_{xx}+B^2)}\sqrt{g_{tt}}g_{xx}\Bigg)_{r_0}\bigg(\f{(3B)^{1/3}\Gamma(2/3)}{2^{-1/3}\Gamma(1/3)}\bigg)(1-i\sqrt{3})\omega^{-2/3},\nn
\eea
and the phase of the conductivity at $r=r_0$
\be
tan~\theta^{xx}(r_0)=\f{1}{\sqrt{3}},\quad tan~\theta^{xy}(r_0)=-\sqrt{3}  \quad \Longrightarrow \theta^{xx}(r_0)=\pi/6,\quad \theta^{xy}(r_0)=-\pi/3.
\ee

It is suggested in \cite{hkmsy}, in another context, to choose the Airy function, $Ai(x)$, over  $Bi(x)$, as the previous function obeys the in-falling boundary condition at the horizon. 

\subsubsection{$z=2$ with non-zero magnetic field}

The solution for this example  is given in eq(\ref{sol_z_2_dbi_b_mid_frequency}). The modified Bessel function of the second kind for small argument has an expansion, 
$K_0(x)=-\gamma-Log~x+Log~2+\cdots$, where $\gamma$ is the EulerGamma constant. Using this, we find the following expansion in the low frequency limit,  $K_0((r^2\sqrt{\pm4B\omega-\omega^2})/2)=2Log~(2/r)-\gamma-
\f{1}{2}Log~\omega-\f{1}{2}Log~(\pm4B)\pm\f{\omega}{8B}+\cdots$, which gives
\be
\f{\p_r K_0\bigg(\f{r^2\sqrt{\pm4B\omega-\omega^2}}{2}\bigg)}{K_0\bigg(\f{r^2\sqrt{\pm4B\omega-\omega^2}}{2}\bigg)}\simeq \f{4}{r(2\gamma+Log(\pm B/4)+4Log~r+Log~\omega)}+\cdots,
\ee 
where the ellipses stand for the higher powers in frequency. 
Similarly, computing the gradient of the logarithm of the Bessel function of the first kind in the low frequency limit gives
\be
\f{\p_r I_0\bigg(\f{r^2\sqrt{\pm4B\omega-\omega^2}}{2}\bigg)}{I_0\bigg(\f{r^2\sqrt{\pm4B\omega-\omega^2}}{2}\bigg)}\simeq \pm 4 B r^3 \omega-\f{r^3(2+B^2 r^4)}{8}\omega^2\pm\f{B}{48}r^7(3+B^2 r^4)\omega^3 +\cdots.
\ee
Again the ellipses stand for the higher powers in frequency.
The conductivities in the low frequency limit are
\bea
\sigma_{xx}(r_0,~c_1=0)&\sim&\f{i}{\omega Log~\omega},\quad \sigma_{xx}(r_0,~c_2=0)\sim i\omega,\nn
\sigma_{xy}(r_0,~c_1=0)&\sim&\f{i}{\omega Log~\omega},\quad \sigma_{xy}(r_0,~c_2=0)\sim i \omega^2,
\eea

where we have written only the  leading frequency dependent part.

\subsubsection{The phase angle for $z=3$ with zero magnetic field}

From the study of \cite{hpst}, we have seen the behavior of the conductivity with non-zero frequency in the $\omega > T$ regime.
Here we shall try to find the phase angle of the conductivity and see whether it matches with the experimental result \cite{dvdm} or not.  The solution with proper boundary condition is $A^{(1)}_{\pm}\equiv A=c_1 r^{\f{z}{2}-1}H^{(1)}_{\f{1}{2}-\f{1}{z}}\bigg(\f{\omega r^z}{z}\bigg)$, where $H^{(1)}_{\nu}(x)$ is the Hankel function of the first kind and $c_1$ is  the constant of integration. In which case, $\sigma_{xy}=0$ and $\sigma_{xx}(r_0)=-\bigg(\f{\sqrt{\rho^2+g^2_{xx}}}{g_{xx}}\f{\sqrt{g_{tt}}}{\sqrt{g_{rr}} }\bigg)_{r_0}\bigg(\f{\p_r A}{i\omega A} \bigg)_{r_0}$. 
By a suitable choice of, $c_1$, the solution can be re-expressed as $A(x)={\tilde c}_1 (x/2)^{\nu} H^{(1)}_{\nu}(x)$, where $\nu=1/2-1/z$ and $x=\omega/z r^z$.
 Using the expansion of $(x/2)^{\nu}H^{(1)}_{\nu}(x)=-i\f{\Gamma(\nu)}{\pi}+\bigg(\f{x}{2}\bigg)^{2\nu}\f{[1+icot(\pi\nu)]}{\Gamma(1+\nu)}$, for small $x$ and $0< \nu <1$, leads to 
\be
A(x)=1-\f{\pi[cot(\pi\nu)-i]}{\Gamma(\nu)\Gamma(1+\nu)}(x/2)^{2\nu}+\cdots. \Longrightarrow \f{1}{i}\f{dA}{dx}=\f{\pi\nu[1+icot(\pi\nu)]}{\Gamma(\nu)\Gamma(1+\nu) }(x/2)^{2\nu-1}
\ee
where we have chosen the constant of integration, ${\tilde c_1}=i\pi/\Gamma(\nu)$.  Finally, the conductivity for $z=3$ is
\be
\sigma_{xx}(r_0)\propto \omega^{-2/3}(1+i\sqrt{3}),
\ee 
where the proportionality constant is  real and an unimportant quantity. From this, the phase angle of the conductivity goes as
\be
tan~\theta_{xx}(r_0)=\Bigg(\f{Im\sigma_{xx}}{Re\sigma_{xx}}\Bigg)_{r_0}=\sqrt{3}  \quad \Longrightarrow\quad \theta_{xx}(r_0)=\pi/3.
\ee 

So, after repeating the study  of $z=3$, we conclude that it's not just the precise power of frequency, as found in \cite{hpst}, but the proper phase angle of \cite{dvdm} can also be reproduced.

\subsection{Flow equation with non-trivial dilaton}

The inclusion of the dilaton, $\phi$, certainly\footnote{The dc and ac behavior of the conductivity with non-trivial dilaton  has been studied in \cite{cgkkm}, \cite{kkp},  \cite{lpp} and for trivial dilaton in \cite{kst} using the prescription of \cite{kob} and \cite{ssp}.} changes the structure of the flow equation of the conductivity. Instead of going through  the complete derivation of it,  we shall adopt a couple of simple steps to get it. Step 1: Bring the tension of the brane, $T_b$, under the integral and replace it by, $T_b\rightarrow T_b~e^{-\phi}$ and then take   $T_b$ out of the integral. Step 2:  Replace the metric components by, $g_{ab}\rightarrow g_{ab} e^{\f{4\phi}{d-1}}$. Using this we get the desired action, $S=-T_b\int e^{-\phi}\sqrt{-det(e^{\f{4\phi}{d-1}}[g]+F)_{ab}}$, where $[g]_{ab}$ is the pull back of the metric in Einstein frame.

The flow equation of the conductivity for $g_{xx}=g_{yy}$,  becomes
\bea\label{flow_long_hall_cond_dbi_b_dilaton}
\p_r\sigma^{xy}&=&-2i\omega\sigma^{xx}
\bigg(\f{\sqrt{g_{rr}}~e^{\f{-4\phi}{d-1}}}{\sqrt{g_{tt}}g_{xx}}\bigg)\Bigg(\f{B\rho-\sigma^{xy}(e^{\f{8\phi}{d-1}}g^2_{xx}+B^2)}{\sqrt{\rho^2+e^{-2\phi}(\prod^{d-3}_{i=1}e^{\f{4\phi}{d-1}} g_{z^iz^j})(e^{\f{8\phi}{d-1}}g^2_{xx}+B^2)}}\Bigg),\nn
\p_r\sigma^{xx}&=&-i\omega\bigg(\f{\sqrt{g_{rr}}e^{\f{-4\phi}{d-1}}}{\sqrt{g_{tt}}g_{xx}}\bigg)\times\nn &&
\Bigg(\f{-2B\rho\sigma^{xy}+
(e^{\f{8\phi}{d-1}}g^2_{xx}+B^2)((\sigma^{xy})^2-(\sigma^{xx})^2)+\rho^2+e^{\f{8\phi}{d-1}-2\phi}g^2_{xx}(\prod^{d-3}_{i=1}e^{\f{4\phi}{d-1}} g_{z^iz^j})}{\sqrt{\rho^2+e^{-2\phi}(\prod^{d-3}_{i=1}e^{\f{4\phi}{d-1}} g_{z^iz^j})(e^{\f{8\phi}{d-1}}g^2_{xx}+B^2)}}\Bigg),\nn
\eea

Demanding regularity condition at the horizon, $r=r_h,$ yields
\bea\label{long_hall_conductivity_at_horizon_dilaton}
Re\sigma^{xy}(r_h)&=&\f{B\rho}{e^{\f{8\phi}{d-1}}g^2_{xx}(r_h)+B^2},\quad Im\sigma^{xy}(r_h)=0,\quad Im\sigma^{xx}(r_h)=0, \nn 
Re\sigma^{xx}(r_h)&=&\pm\Bigg(\f{e^{\f{4\phi}{d-1}}g_{xx}\sqrt{\rho^2+e^{-2\phi}(\prod^{d-3}_{i=1}e^{\f{4\phi}{d-1}} g_{z^iz^j})(e^{\f{8\phi}{d-1}}g^2_{xx}+B^2)}}{e^{\f{8\phi}{d-1}}g^2_{xx}+B^2}\Bigg)_{r_h}.
\eea

Upon restricting to  $3+1$ dimensional Lifshitz geometry with dynamical exponent, $z$, we get in the high density and low magnetic field limit

\be\label{long_hall_conductivity_at_horizon_dilaton_d_3}
Re\sigma^{xy}(r_h)\sim B\rho e^{-4\phi} T^{-4/z},~ Im\sigma^{xy}(r_h)=0,~ Im\sigma^{xx}(r_h)=0, \quad
Re\sigma^{xx}(r_h)\sim \pm \rho e^{-2\phi} T^{-2/z} .
\ee

This matches with the result reported in \cite{ssp}. Moreover, in this limit the ratio of the dc Hall  conductivity of the dc longitudinal conductivity obeys, the Drude type behavior of dc conductivity,   $Re\sigma^{xy}/Re\sigma^{xx}\sim B Re\sigma^{xx} $.

We can move away from the Drude type behavior to dc  conductivity. 
In the low density and low magnetic field limit for $3+1$ dimensional Lifshitz geometry with dynamical exponent, $z$, we can have a  non-trivial temperature dependence to $Re\sigma^{xy}$ and $Re\sigma^{xx}$
\be
Re\sigma^{xy}(r_h)\sim B\rho e^{-4\phi} T^{-4/z},~ Im\sigma^{xy}(r_h)=0,~ Im\sigma^{xx}(r_h)=0, \quad
Re\sigma^{xx}(r_h)\sim \pm  e^{-\phi}. 
\ee

For the specific choice of dilaton, $e^{-\phi}\sim T^{-1}$, we can have 
\be
Re\sigma^{xy}(r_h)\sim B\rho T^{-4-4/z},\quad Re\sigma^{xx}(r_h)\sim \pm T^{-1},
\ee
and for negative dynamical exponent, $z=-4,~ Re\sigma^{xy}(r_h)\sim B\rho T^{-3}$, which in turn give the precise Hall  
angle, $Re\sigma^{xx}(r_h)/Re\sigma^{xy}(r_h)\sim T^2$ of \cite{cwo},\cite{pwa},\cite{tm}. It matches with the result of \cite{ssp}.

\section{Flow equation in a fluctuating geometry without magnetic field}

The effective action involving gravity and gauge field is
\be\label{action_dbi_EH}
S=\f{1}{2\kappa^2}\int d^{d+1}x\bigg[\sqrt{-g}\bigg(R-2\Lambda\bigg)-{ T}\sqrt{-det(g+F)}\bigg],
\ee
where $T=T_{b} 2\kappa^2$ and $T_{b}$ is the tension of the brane and $\Lambda$ is the cosmological constant. The equation of motion that results 
\bea\label{eom_dbi}
&&R_{MN}-\f{2\Lambda}{d-1}g_{MN}-\f{{ T}g_{MN}}{4(d-1)}\f{\sqrt{-det(g+F)}}{\sqrt{-det(g)}}\bigg[(g+F)^{-1}+(g-F)^{-1}\bigg]^{KL}g_{KL}+\nn &&\f{{T}}{4}\f{\sqrt{-det(g+F)}}{\sqrt{-det(g)}}\bigg[(g+F)^{-1}+(g-F)^{-1}\bigg]^{KL}g_{MK}g_{NL}=0,\nn &&
\p_M\bigg[\sqrt{-det(g+F)}\bigg((g+F)^{-1}-(g-F)^{-1}\bigg)^{MN}\bigg]=0.
\eea

We assume that the solution to the equation of motion is of the diagonal form 
\be
ds^2_{d+1}=-g_{tt}(r)dt^2+g_{xx}(r)dx^2+\sum^{d-2}_{a=1}g_{ab}(r)
dy^ady^b+g_{rr}(r)dr^2,~~~A=A_t(r) dt,
\ee
and $A'_t$ obeys 

\be\label{A_t_back_reaction}
\p_r\Bigg[\f{\sqrt{g_{xx}(\prod g_{y^ay^a})}}{\sqrt{g_{tt}g_{rr}-A'^2_t}}A'_t\Bigg]=0.
\ee
Let us do the following metric and gauge field fluctuations, $g^{(1)}_{tx}(t,x,r)$ and $A^{(1)}_x(t,x,r)$, respectively. In which case the $x-r$ component to the equation of motion, from eq(\ref{eom_dbi}), gives
\be
\f{i\omega}{g_{tt}g_{xx}}(g'_{xx}g^{(1)}_{xt}-g_{xx}g^{'(1)}_{xt})-T \sqrt{\f{g_{rr}}{g_{tt}}}\f{A'_t}{\sqrt{g_{tt}g_{rr}-A'^2_t}}F^{(1)}_{tx}=0,
\ee
where prime denotes derivative with respect to the radial coordinate, $r$, and the Fourier transformation is done with respect to $e^{-i\omega t}$ at zero momentum. So, we get a constraint that relates the metric fluctuation and gauge field fluctuation as
\be\label{constraint_dbi}
\sqrt{g_{tt}g_{rr}-A'^2_t}\bigg[g'_{xx}g^{(1)}_{xt}-g_{xx}g^{'(1)}_{xt}\bigg]+T g_{xx}\sqrt{g_{tt}g_{rr}}A'_t A^{(1)}_x=0.
\ee

This equation will be used latter to decouple the gauge field fluctuation from the metric fluctuation so as to obtain the equation of motion to the gauge field, $A^{(1)}_x$. Now, we are interested to find the equation of motion to  the gauge field fluctuation, in order to do so we must expand the DBI part of the action, eq(\ref{action_dbi_EH}).
\bea
S_{DBI}&=&-T_b\int \sqrt{-detM^{(0)}_+}\bigg[1-\f{1}{4}{\bigg(M^{(0)}_+\bigg)^{-1}}^{ab}\bigg(M^{(1)}_+\bigg)_{bc}{\bigg(M^{(0)}_+\bigg)^{-1}}^{cd}\bigg(M^{(1)}_+\bigg)_{da}+\nn 
&&\f{1}{8}{\bigg(M^{(0)}_+\bigg)^{-1}}^{ab}\bigg(M^{(1)}_+\bigg)_{ba}{\bigg(M^{(0)}_+\bigg)^{-1}}^{cd}\bigg(M^{(1)}_+\bigg)_{dc}+
\cdots\bigg],\nn 
&=&-T_b\int \sqrt{-detM^{(0)}_+}\bigg[1-\f{1}{4}{\bigg(M^{(0)}_+\bigg)^{-1}}^{ab}g^{(1)}_{bc}{\bigg(M^{(0)}_+\bigg)^{-1}}^{cd}g^{(1)}_{da}-\nn&&
\f{2(g_{rr}F^{(1)}_{tx}F^{(1)}_{tx}-g_{tt}F^{(1)}_{rx}F^{(1)}_{rx})-4A'_t g^{(1)}_{tx}F^{(1)}_{rx}}{4g_{xx}(g_{tt}g_{rr}-A'^2_t)} +\cdots\bigg],\nn 
&\equiv&-T_b\int \sqrt{-detM^{(0)}_+}\bigg[1-\f{1}{4}{\bigg(M^{(0)}_+\bigg)^{-1}}^{ab}g^{(1)}_{bc}{\bigg(M^{(0)}_+\bigg)^{-1}}^{cd}g^{(1)}_{da}+S^{(2)}_A
+\cdots\bigg]
\eea
where we have used the definition $M_{ab}=(g^{(0)}+F^{(0)})_{ab}+(g^{(1)}+F^{(1)})_{ab}\equiv (M^{(0)}_++M^{(1)}_+)_{ab}$, and $M^{(n)}_{\pm ab}=(g^{(n)}\pm F^{(n)})_{ab},$ for $n=0$ and $1$. Note that  $M^{(n)}_{+{ab}}=M^{(n)}_{-{ba}}$. The quadratically fluctuated gauge field action  
\be
S^{(2)}_A=-\f{T_b}{2}\int \f{\sqrt{g_{xx}(\prod g_{y^ay^a})}}{g_{xx}\sqrt{g_{tt}g_{rr}-A'^2_t}}\Bigg[g_{tt}\p_r A^{(1)}_x \p_r A^{(1)}_x-\bigg(\omega^2 g_{rr}+\f{2T_bA'^2_t\sqrt{g_{tt}g_{rr}}}{\sqrt{g_{tt}g_{rr}-A'^2_t}}\bigg)A^{(1)}_x A^{(1)}_x\Bigg].
\ee

The equation of motion that results  is
\be
\p^2 A^{(1)}_x+\p_r log\Bigg(\f{g_{tt}\sqrt{g_{xx}(\prod g_{y^ay^a})}}{g_{xx}\sqrt{g_{tt}g_{rr}-A'^2_t}}\Bigg)\p_r A^{(1)}_x+\bigg(\omega^2 g_{rr}+\f{2T_bA'^2_t\sqrt{g_{tt}g_{rr}}}{\sqrt{g_{tt}g_{rr}-A'^2_t}}\bigg)A^{(1)}_x=0,
\ee
and the current at a constant-r slice is
\be
J^{(1)}_x=-T_b\f{g_{tt}\sqrt{g_{xx}(\prod g_{y^ay^a})}}{g_{xx}\sqrt{g_{tt}g_{rr}-A'^2_t}}\p_r A^{(1)}_x.
\ee

Using Ohm's law $J^{(1)}_x=i\omega \sigma^{xx}A^{(1)}_x $, where we have done the Fourier transformation to both the field strength, $F^{(1)}_{xt}$, and the current, $J^{(1)}_x$.  This gives the flow equation for the conductivity as
\be
\p_r \sigma^{xx}=i\omega \Bigg[(\sigma^{xx})^2\f{g_{xx}\sqrt{g_{tt}g_{rr}-A'^2_t}}{T_b g_{tt}\sqrt{g_{xx}(\prod g_{y^ay^a})}}-T_b \f{g_{tt}\sqrt{g_{xx}(\prod g_{y^ay^a})}}{g_{xx}\sqrt{g_{tt}g_{rr}-A'^2_t}}\Bigg(\f{g_{rr}}{g_{tt}}-\f{2T_b A'^2_t\sqrt{g_{rr}}}{\omega^2\sqrt{g_{tt}}\sqrt{g_{tt}g_{rr}-A'^2_t}}\Bigg)\Bigg]
\ee 

Using the solution to eq(\ref{A_t_back_reaction}) in the flow equation for the conductivity results in 
\be\label{flow_cond_back_reaction_dbi}
\p_r \sigma^{xx}=i\omega\sqrt{\f{g_{rr}}{g_{tt}}}\Bigg[\f{(\sigma^{xx})^2}{\Sigma^{DBI}_A}-\Sigma^{DBI}_A\Bigg(1-\f{2T_b}{\omega^2} \f{g_{tt}\rho^2}{\sqrt{g_{xx}(\prod g_{y^ay^a})}\sqrt{\rho^2+g_{xx}(\prod g_{y^ay^a})}}
\Bigg) \Bigg],
\ee
where $\rho$ is the constant of integration to eq(\ref{A_t_back_reaction}) and is interpreted as the charge density. The function $\Sigma^{DBI}_A$ is defined as
\be
\Sigma^{DBI}_A\equiv T_b\f{\sqrt{g_{tt}g_{rr}}\sqrt{g_{xx}\prod g_{y^ay^a}}}{g_{xx}\sqrt{g_{tt}g_{rr}-A'^2_t}}.
\ee

The first order differential equation, eq(\ref{flow_cond_back_reaction_dbi}), requires a boundary condition and we impose the regularity condition at the horizon, $r_h$, which means  
\be\label{bc_cond_back_reaction_dbi}
\sigma^{xx}(r_h)=\Sigma^{DBI}_A(r_h)=T_b\Bigg[\f{\sqrt{\rho^2+g_{xx}
(\prod g_{y^ay^a})}}{g_{xx}}\Bigg]_{r_h}.
\ee

It follows from eq(\ref{flow_cond_back_reaction_dbi}) that in $3+1$ dimensional bulk spacetime, the ac conductivity in the dual field theory cannot be a constant, in particular, the  independence of frequency, for non-zero charge density. However, in the small density limit,  the leading order behavior to dc conductivity becomes independent of temperature, only when there is a rotational symmetry in the $x-y$ plane.

\section{Flow for thermoelectric and thermal conductivity}

In this section, we shall try to find the flow equation for the longitudinal thermoelectric coefficient, $\alpha$, and the longitudinal thermal conductivity, $ \kappa$. It is known that there arises electric current, $J^x$, and heat current, $Q_x$, which is related to   electric field, $E_x$ and thermal gradient, $\p_x T$ as 

\bea\label{matrix_conds}
\left(
    \begin{array}{c}
      J^x \\
      Q_x
    \end{array}
  \right) =  \left(
    \begin{array}{cc}
      \sigma & \alpha  \\
      \alpha T &{\bar\kappa}
    \end{array}
  \right) \left(
    \begin{array}{c}
      E_x \\
      -\nabla_x T
    \end{array}
  \right)
\eea

The heat current is defined as $Q_x=T^{FT}_{tx}-\mu J^x$, where $\mu$ is the chemical potential and $T^{FT}_{tx}$ is the $t-x$ component of the energy momentum tensor. The quantity, ${\bar\kappa}$, is related to the thermal conductivity as, ${\bar\kappa}=\kappa+T\alpha^2/\sigma$ \cite{hkms}.
The approach that we shall adopt is to find the flow equation for electrical conductivity in the absence of thermal gradient and use this information to calculate the flow of the  thermoelectric and thermal conductivity. In fact, we have already done the calculation for the flow equation for the electrical conductivity in eq(\ref{flow_fluctuating_geometry_pll}) of  the previous sections. 

To get going, we need to know the exact expression of the $t-x$ component of the energy momentum tensor. It is suggested in \cite{bk} to consider the Brown-York stress tensor along with the desired counter terms as the   proper stress tensor that describes the boundary  energy momentum tensor.
\be\label{by_tensor}
T^{BY}_{\mu\nu}=\f{1}{8\pi G}[K_{\mu\nu}-K\gamma_{\mu\nu}],
\ee
The total  stress tensor is the sum of $T^{BK}_{\mu\nu}=T^{BY}_{\mu\nu}+T^{ct}_{\mu\nu}$, 
where\footnote{ BK stands for Balasubramanian-Kraus paper \cite{bk}.}  the counter terms to energy momentum tensor has the form
\be
T^{ct}_{\mu\nu}= \f{1}{8\pi G}[c_1\gamma_{\mu\nu}+c_2 G_{\mu\nu}+\cdots].
\ee
The coefficients $c_1$ and $c_2$ are constants, $G_{\mu\nu}$
is the Einstein tensor made out of the induced boundary metric $\gamma_{\mu\nu}$ \cite{dhss}. The ellipses stands for the higher derivatives to Ricci tensor, Ricci scalar, Riemann tensors and the combinations thereof.  However, as we shall see $T^{BK}_{\mu\nu}$ is not the full field theory energy momentum tensor as it is not non-zero finite as we approach the boundary. To match with the proper field theory energy momentum tensor,  as in \cite{dhss},  which is finite, we needed to multiply an over all radial dependent factor.
\be\label{en_mo_ft}
T^{FT}_{\mu\nu}=\Sigma(r) ~~~T^{BK}_{\mu\nu},
\ee
where $T^{FT}_{\mu\nu}$ is the proper field theory energy momentum tensor defined at a constant-r slice and $\Sigma(r)$ is the radial dependent factor. For asymptotically $AdS_3$ spacetime, $\Sigma(r)$ becomes unity, and is of the form $\Sigma(r)=-r^{d-2}$, for $d+1$ dimensional bulk theory. 

For the choice of the bulk metric as in eq(\ref{metric}), we choose 
the fluctuation along the $t-x$ component of the  metric, $g_{tx}(t,r)$. In the fluctuation,  only the time and radial dependence is kept as we are working in the zero momentum limit. In our case the boundary is at $r\rightarrow\infty$. The extrinsic curvature is defined as $K_{\mu\nu}=-\f{1}{2}[\nabla_{\mu}n_{\nu}+\nabla_{\nu}n_{\mu}]$, for unit spacelike vector $n_{\mu}$ with outward pointing normal  to the boundary. The $t-x$ component of the Brown-York stress  tensor along with the counter terms give
\be
T^{BK}_{tx}=\f{1}{16\pi G}\f{1}{\sqrt{g_{rr}}}\bigg[-g'_{tx}+g_{tx}\bigg(\f{g'_{tt}}{g_{tt}}+\f{g'_{xx}}{g_{xx}}+g^{ab}g'_{ab}-2 c_1 \sqrt{g_{rr}}\bigg)\bigg],
\ee
where $c_1=-(d-1)$. It means the energy momentum tensor in the boundary field theory is
\be 
T^{FT}_{tx}=-\f{r^{d-2}}{16\pi G}\f{1}{\sqrt{g_{rr}}}\bigg[-g'_{tx}+g_{tx}\bigg(\f{g'_{tt}}{g_{tt}}+\f{g'_{xx}}{g_{xx}}+g^{ab}g'_{ab}-2 (d-1) \sqrt{g_{rr}}\bigg)\bigg].
\ee

For an asymptotically AdS spacetime for which,  $g_{tt}=g_{ab}=g_{xx}=g_{rr}^{-1}=r^2$, means 
$T^{FT}_{tx}=-\f{r^{d-1}}{16\pi G}[-g'_{tx}+\f{2}{r}g_{tx}]$. Upon using the asymptotic expansion of the metric component as $g_{tx}=r^2 g^{(0)}_{tx}+\f{g^{(1)}_{tx}}{r^{d-2}}+\cdots$, results in a finite expression at the boundary ($r\rightarrow\infty$) i.e.,   $T^{FT}_{tx}(r\rightarrow\infty)=-\f{d}{16\pi G} g^{(1)}_{tx}$. This particular form of the energy momentum tensor exactly reproduces the result of 1-pt function i.e., eq(4.12) of \cite{hhh} in the unit for which $16\pi G=1$.

The heat current becomes 
\bea\label{heat_current}
Q_x&=&-\f{r^{d-2}}{16\pi G}\f{1}{\sqrt{g_{rr}}}\bigg[-g'_{tx}+g_{tx}\bigg(\f{g'_{tt}}{g_{tt}}+\f{g'_{xx}}{g_{xx}}+g^{ab}g'_{ab}-2 (d-1) \sqrt{g_{rr}}\bigg)\bigg]-\mu J^x\nn
&\equiv& X(r) g'_{tx}+Y(r) g_{tx}-\mu J^x=X(r) g'_{tx}+Y(r) g_{tx}-\mu \sigma E_x,
\eea
where we have considered the energy momentum tensor as written in  eq(\ref{en_mo_ft}), $X(r)$ and $Y(r)$ are functions that depend on the radial coordinate, $r$, through the metric components.  
The thermal conductivity is defined as the ratio of the heat  current to the thermal gradient in the absence of electric current \cite{hbc}
\be\label{thermal_conductivity_callen}
\kappa=-\f{Q_x}{\nabla_x T},\quad {\rm For\quad J^x=0}\Rightarrow \nabla_x T=\f{\sigma}{\alpha} E_x,
\ee
where we have used eq(\ref{matrix_conds}), and the expression of the thermal conductivity can be rewritten as 
\be\label{thermal_conductivity}
\kappa=-\f{T^{FT}_{tx}}{\nabla_x T}+\mu\alpha=-\f{\alpha}{\sigma}\bigg[\f{T^{FT}_{tx}}{E_x}-\mu\sigma\bigg].
\ee 

The thermoelectric coefficient, $\alpha$, can be defined as
the ratio of the heat current to electric field in the absence of thermal gradient \cite{hh}
\be
T\alpha=\f{Q_x}{E_x}, \quad {\rm  For \quad \nabla_x T=0},
\ee
which means 
\be\label{thermoelectric}
T\alpha=\f{T^{FT}_{tx}}{E_x}-\mu\sigma.
\ee

Since we are working at zero momentum means $E_x=i\omega A_x$. Using eq(\ref{heat_current}) and eq(\ref{coupled_eom}) in eq(\ref{thermoelectric}) and eq(\ref{thermal_conductivity}), we get 
\bea\label{thermal_conds}
T\alpha&=&(X\p_r log~g_{xx}+Y)\f{g_{tx}}{i\omega A_x}+i\f{XA'_0}{\omega {\tilde g}^2_{YM}}-\mu\sigma,\nn
\kappa&=&-T\alpha^2/\sigma.
\eea

Here in the first equation of eq(\ref{thermal_conds}), we have not used the condition of zero thermal gradient and  kept it for the calculation of the flow equation, latter. Let us recall a formula that relate the thermal gradient to the fluctuation of the metric component, eq(72) of  \cite{hh1}
\be
g_{tx}=-\f{\nabla_x T}{i\omega T}.
\ee    
Using this information and recall that  the thermoelectric coefficient is computed in the absence of thermal gradient, means 
\be\label{thermo_electric}
T\alpha=i\f{XA'_0}{\omega {\tilde g}^2_{YM}}-\mu\sigma,\quad X=-\f{r^{d-2}}{16\pi G}\f{1}{\sqrt{g_{rr}}}=-\f{r^{d-1}}{16\pi G}\f{\sqrt{h}}{L}, 
\ee
where we have used the geometry of RN black hole as written in eq(\ref{RN_black_hole}). The flow equation of the thermoelectric coefficient and the thermal conductivity is completely determined by the flow equation for the electrical conductivity. 
\bea
T\p_r\alpha&=&\f{A'_0}{i\omega {\tilde g}^2_{YM}}\bigg[ X \p_r Log\bigg(\f{\sqrt{-g}}{g_{tt}g_{rr}g_{xx}}\bigg)-\p_rX-Y\bigg]-\mu\p_r\sigma,\nn
\f{\p_r\kappa}{\bigg(i\mu+\f{XA'_0}{\sigma\omega {\tilde g}^2_{YM}}\bigg)}&=&\bigg[\f{2A'_0}{T\omega {\tilde g}^2_{YM}}\bigg(\p_r X+Y-X\p_rLog\bigg(\f{\sqrt{-g}}{g_{tt}g_{rr}g_{xx}}\bigg)\bigg)+\f{\p_r\sigma}{T}\bigg(i\mu-\f{XA'_0}{\sigma\omega {\tilde g}^2_{YM}}\bigg)\bigg]\nn
\eea

{\bf Horizon behavior:}\\

The behavior of the thermoelectric coefficient and the thermal conductivity at the horizon is
\be\label{thermoelectric_thermal_conductivity_horizon}
T\alpha(r_h)=-\mu\sigma(r_h),\quad \kappa(r_h)=-T\alpha^2(r_h)/\sigma(r_h)=-\mu^2 \sigma(r_h)/T,~{\rm as}~\quad X(r_h)=0.
\ee

Recall that for the configuration described by the  Maxwell type of action, the conductivity goes as $\sigma(r_h)\sim T^{(d-3)/z}$, eq(\ref{conductivity_any_z}). It means the thermoelectric coefficient and the thermal conductivity at the horizon is
\be
\alpha(r_h)\sim -\mu~ T^{(d-3-z)/z},\quad \kappa(r_h)\sim -\mu^2~ T^{(d-3-z)/z}.
\ee

If the configuration is described by the exotic DBI type of action, then depending on the parameters like magnetic field and charge density, we can have different powers to temperature. Assuming  the form of the energy-momentum tensor remains same. We find, for example, in $3+1$ dimensional bulk spacetime with  high density and high magnetic field, the real part of the conductivity goes as, $Re\sigma(r_h) \sim T^{2/z}\f{\sqrt{\rho^2+B^2}}{B}$, in which case 
\be
Re\alpha(r_h)\sim - \mu~T^{(2-z)/z}\f{\sqrt{\rho^2+B^2}}{B},\quad Re\kappa(r_h)\sim -\mu^2~ T^{(2-z)/z}\f{\sqrt{\rho^2+B^2}}{B}. 
\ee  

If we consider the high density and low magnetic field limit, the real part of the conductivity goes as, $Re\sigma(r_h) \sim \rho T^{-2/z}$, in which case 
\be
Re\alpha(r_h)\sim - \mu~\rho~T^{-(2+z)/z},\quad Re\kappa(r_h)\sim -\mu^2~\rho~ T^{-(2+z)/z}. 
\ee  

In  the case of the  low density and low magnetic field limit, the real part of the conductivity goes as, $Re\sigma(r_h) \sim $ constant, it means 
\be
Re\alpha(r_h)\sim - \mu/T,\quad Re\kappa(r_h)\sim -\mu^2/T. 
\ee  
\subsection{A Puzzle}

In this subsection, we shall consider the conductivities $\sigma,~\alpha$, and $\kappa$ as defined in eq(\ref{matrix_conds}) as matrices.  
Let us assume  the following structure to retarded correlators of heat current, $Q_i=T_{ti}-\mu J_i$, and electric current, $J_i$
\be
G^R_{J_iJ_j}\equiv<J_i,J_j>=-i\omega \sigma_{ij},\quad G^R_{J_iQ_j}\equiv<J_i,Q_j>=-i\omega T \alpha_{ij},\quad G^R_{Q_iQ_j}\equiv<Q_i,Q_j>=-i\omega T \kappa_{ij},
\ee
   
where $T_{ti}$ is the $t-i$'th component of the energy momentum tensor. From these correlation function, there follows $<J_i,T_{tj}>=-i\omega[\mu\sigma_{ij}+T\alpha_{ij}].$ and 
\be
\kappa_{ij}=\f{i}{\omega T}<T_{ti},T_{tj}>-\mu (\alpha_{ij}+\alpha_{ji})-\f{\mu^2}{T}\sigma_{ji}
\ee

From which there follows 
\be\label{longitudinal_hall_conductivity}
\kappa_{xx}=\f{i}{\omega T}<T_{tx},T_{tx}>-2\mu \alpha_{xx}-\f{\mu^2}{T}\sigma_{xx},~~~
\kappa_{xy}=\f{i}{\omega T}<T_{tx},T_{ty}>+\f{\mu^2}{T}\sigma_{xy},
\ee

where we have used the Onsager's reciprocity relation $\sigma_{xy}=-\sigma_{yx}$ and $\alpha_{xy}+\alpha_{yx}=0$.
Let us consider a situation for which the 2-pt function of $T_{tx}$ is real. Then  the real part of $\kappa_{xx}$, using eq(\ref{thermo_electric}), results in \cite{kachru}
\be\label{thermal_cond}
Re~\kappa_{xx}=\bigg(\f{\mu}{T}\bigg)^2 T Re~\sigma_{xx}.
\ee

Up to an overall sign, it matches with eq(\ref{thermoelectric_thermal_conductivity_horizon}). There will be a perfect matching, if we change the sign of the heat current correlator, i.e, $<Q_i,Q_j>=i\omega T\kappa_{ij}$, but then there will be a mismatch of sign with the hydrodynamic result of \cite{ss} in the zero entropy density limit.
We know from the study   of the thermal conductivity using  hydrodynamics, in such a system without magnetic field, eq(5.1) of \cite{ss}, that the  thermal conductivity obey
\be\label{hydro_thermal_cond}
Re~\kappa_{xx}=\bigg(\f{s}{\rho}+\f{\mu}{T}\bigg)^2 T Re\sigma_{xx},
\ee
where $s,~\rho$ are entropy density and charge density respectively. It is very difficult to reconcile eq(\ref{thermal_cond}) and eq(\ref{hydro_thermal_cond}) unless   the entropy density  vanishes or $s/\rho$ is very small incomparision  to $\mu/T$. However, for non-vanishing entropy density  with $s/\rho \geq \mu/T$, it is not clear how these two approaches give the same result.

Once again assuming that $<T_{tx},T_{ty}>$ correlator is real, eq(\ref{longitudinal_hall_conductivity}) gives
\be
Re~\kappa_{xy}=(\mu/T)^2 ~T Re~\sigma_{xy}.
\ee

{\bf Plausible resolution:}\\

There exists two plausible ways to avoid this puzzle. First, if we modify the definition of  the heat current  as 
\be
Q_i=T_{ti}-(sT/\rho+\mu)J^x.
\ee 
then we do reproduce eq(\ref{hydro_thermal_cond}). By doing this we go with the assumption that the $<T_{tx},T_{tx}>$ correlator is real. Second, if we relax the  assumption that the correlator is real and allows it to be a complex quantity then consistency with eq(\ref{hydro_thermal_cond}) suggests that there should be an imaginary part to this correlator 
\be
Im <T_{tx},T_{tx}>=-\omega \f{T^2s}{\rho}\bigg(\f{s}{\rho}+\f{\mu}{T}\bigg)  Re\sigma_{xx}.
\ee





\section{Summary and Outlook}

In this paper, we have reproduced the result of the dc electrical conductivity \cite{dvdm} for a specific choice of the the dynamical exponent, through the study of the flow of the conductivity along the lines of \cite{il}. This interesting result of the conductivity is obtained in \cite{hpst} using the method of \cite{kob}, and using another approach in \cite{ssp}. In the approach of \cite{kob}, there occurs several disadvantages: (1) The conductivity should be evaluated at a holographic scale for which  the action takes a  $\f{0}{0}$ form or the vanishing of the Legendre transformed action in
\cite{ssp}. (2) More importantly, there does not seem to exists any way to calculate it at the boundary, which is essential for  the complete understanding   of the conductivity in the dual field theory. (3) It does not say how to calculate the conductivity for theories that are described by the Maxwell type of action. (4) It does not offer any information about the ac conductivity. 

Apart from getting the result of the dc  conductivity in the RG flow approach, we  tried to study the whole conductivity versus frequency plane along the lines of \cite{dvdm}. 
It seems from experimental result,  nothing special happens  when the frequency is of the order of the temperature, $\omega\sim T$. In our theoretical study, 
even though the flow equation captures all the regions of the parameter space, we won't be able to answer the $\omega\sim T$ region, analytically.  The main reason  of our inability is to solve it,  because of the complexity of the flow equation. 

In the low frequency limit, because of the existence of a naturally small parameter, $\omega \ll T$, allows us to make a Taylor series expansion of the conductivity in $\omega/T$. Moreover, we did not find any  evidence to the existence of the Drude like form.  Kramers-Kronig relation suggests that there should a,  $\delta(\omega=0)$, to the real part of the conductivity if there is a simple pole to the imaginary part of the conductivity. However, the flow equation, somehow, did not show it explicitly and  is the case with the other studies, e.g., \cite{hhh}.

In the mid-frequency region,  in order for  the ac conductivity to show the interesting  behavior of \cite{dvdm}, namely, $\sigma\sim \omega^{-2/3}$, we needed to consider the dynamical exponent to be  $z=3$ in \cite{hpst}. Moreover, we calculated  the phase angle of the conductivity for $z=3$ and is in perfect agreement with \cite{dvdm}. Here on the other hand, we reproduced this particular   frequency dependence of the conductivity for unit dynamical exponent with a non-zero magnetic field along with the charge density. This is obtained by directly solving the equation of motion to the gauge field as in \cite{hpst}. In this case, unfortunately, we could not fix the boundary condition precisely. However, this result of the conductivity is independent of the choice of the boundary condition. The phase angle of the conductivity goes as, $tan^{-1}(Im\sigma^{xx}/Re\sigma^{xx})=\pi/6$ and $tan^{-1}(Im\sigma^{xy}/Re\sigma^{xy})=-\pi/3$.

Generically, for the system described by the Einstein-Maxwell 
action without the magnetic field and with dynamical exponent $z$, the conductivity has the following temperature dependence at the horizon, $\sigma(r_h)\sim T^{(d-3)/z}$, in $d+1$ dimensional bulk spacetime. For the DBI type of action, the result is very different, see eq(\ref{generic_conductivity_dbi_b_lifshitz_horizon}),  but as expected, for very small magnetic field and charge density, the conductivity at the horizon has the same behavior  as that of the Maxwell action.

In the extremely high frequency region, upon neglecting the potential energy term in comparision to the square of the frequency in  the Schr\"{o}dinger equation for the gauge field, gives us the frequency independent result of the conductivity. This result becomes precise in $3+1$ dimensional bulk spacetime.

There exists a universal structure, which is seen in the phase angle of the conductivity evaluated at the horizon.  This result follows by demanding the regularity condition on the flow equation of the conductivity.  It gives either  the phase angle to vanishes or an integral multiple of $\pi$.

There arises an issue to the result of the conductivity  both for the Maxwell and DBI type of action.  Using the  holographic dictionary, we find it  goes as, $N^2_c$, apart from the appropriate temperature dependence, in any spacetime dimensions.  The easiest way to see it, for the Maxwell case, is as follows. From eq(\ref{bc_flow_fluctuating_geometry_pll}), we find $\sigma^{xx}(r_h)=g_{xx}^{(d-3)/2}(r_h)/g^2_{YM}$. Now using 
the guage-gravity duality, $g^2_{YM}\sim N^{-2}_c$, which  holds for the asymptotically $AdS_5$ spacetime \cite{fmmr}, here on the other hand, we have made a naive generalization of it to arbitrary spacetime dimensions. So, there follows the $N^2_c$ dependence of the conductivity. In order to get the $N_c$ independent result of the conductivity,  
the approach of \cite{liu} could be useful to understand it i.e., by evaluating the conductivity at one-loop.

The Green function for the DBI action, in the fixed background case, and   in the hydrodynamic region without the magnetic field, displays a pole. The quasinormal frequency showing the gapless structure is same as found in the Maxwell system \cite{il}. From which we read out the generic form of the charge diffusion constant for any arbitrary background geometry.

The horizon behavior of the thermoelectric and thermal conductivity is computed in the absence of the magnetic field. The result at the horizon is determined by the horizon property of the electrical conductivity. In terms of equations, $\alpha=-\mu~\sigma/T$ and 
$\kappa=-\mu^2~\sigma/T$. 

There are few other things that we left for future  studies. The flow equation of the electrical conductivity,  thermoelectric and thermal conductivity for anisotropic background geometries \cite{ssp1}, i.e., metric components that explicitly break the rotational symmetry. 
Study of these transport quantities for the whole parameter space, and to have a better understanding of the Nernst coefficient. It would be interesting to find  a way to study the sum rules through the RG flow of the transport quantities and then investigate its property. The other important thing of the transport quantities that we did not study is the determination of the pole structure   in terms of the   parameters like 
magnetic field and charge density etc. 

The flow equations found in the fixed background  geometry case in \cite{il} and \cite{flr}  are same for  fluctuating scalar degrees of freedom (dof), i.e., the dof's obey  the scalar field equation of motion.  For the vector type dof, the flow equations are same  only when the fields satisfy the Neumann type conditions at the boundary. Probably the approach of \cite{flr}, \cite{ns} and \cite{hp} could be useful to re-derive  the flow equations presented here, which we leave for future studies.

\section{Acknowledgment}

We would like to thank APCTP, Pohang,   for providing  a warm hospitality at the finishing stage of this work.
This work was supported by the Korea Science and Engineering Foundation (KOSEF) grant funded by the Korea
government (MEST) through the Center for Quantum Spacetime (CQUeST) of Sogang University with Grant No, R11-2005-021.

\section{Appendix A}

In this appendix we shall first change the dimension full flow equation for the  conductivity,  into a dimensionless  flow equation, which  we shall use for the numerical integration of the eq(\ref{flow_diff_separate}). But before that let us do some dimensional analysis for   RN AdS black hole.

\subsection{Flow equation for conductivity  in  RN AdS black hole}

The action for a Einstein-Hilbert-Maxwell action is 
\be
S=\f{1}{2\kappa^2}\int d^{d+1}x \sqrt{-g}[R+\f{d(d-1)}{L^2}-\f{L^2}{g^2_F}F_{MN}F^{MN}],
\ee
whose  asymptotically AdS solution 
\be\label{RN_black_hole}
ds^2=\f{r^2}{L^2}[-h(r)dt^2+dx^2_i]+\f{L^2}{r^2h(r)}dr^2,\quad A_0=\mu\Bigg[1-\bigg(\f{r_h}{r}\bigg)^{d-2}\Bigg],
\ee
with $h(r)=1+\f{Q^2}{r^{2d-2}}-\f{r^d_h}{r^d}-\f{Q^2}{r^{d-2}_hr^d}$, $\mu=\f{g_F Q}{c_d L^2 r^{d-2}_h}$ and $L$ is the AdS radius. The constant quantity, $c_d=\sqrt{\f{2(d-2)}{d-1}}$, with the horizon and the boundary is located at  $r=r_h$ and $r=\infty$, respectively.  The temperature $T=\f{dr_h}{4\pi L^2}[1-\f{(d-2)Q^2}{dr^{2d-2}_h}]$. 
Upon comparing with eq(\ref{ehm}), we find that the coupling as defined in the section 2, is related to $g_F$ as $4{\tilde g}^2_{YM}=g^2_F/L^2$. 

From the action and the solution, it is very easy to notice the dimension of the following quantities. In what follows, we shall be writing down the length dimensions and denote it using the square bracket. 
\be
[r]=[r_h]=[L]=[t]=[x_i]=[L^1]; \quad [\kappa^2]=[Q]=[L^{d-1}];\quad [g_F]=[L^0];\quad [\mu]=[A_M]=[L^{-1}]
\ee

The dimension of the YM's coupling constant and the quantities that appear in the flow equation for the conductivity, eq(\ref{flow_diff_separate})
\be
[{\tilde g}^2_{YM}]=[L^{-2}];\quad [g^2_{YM}=2\kappa^2 {\tilde g}^2_{YM}]=[L^{d-3}];\quad [\Sigma_A]=[L^{3-d}].
\ee

The quantity, $G$,  defined in the  conductivity,  eq(\ref{def_conductivity}), and the frequency, $\omega$, has the dimension
\be
[G]=[L^{3-d}];\quad [\omega]=[L^{-1}].
\ee

It is easy to find out that the dimension of the conductivity 
\be
[\sigma^{xx}]=[L^{3-d}].
\ee

Now, to do numerical computation, we first, need to transform the dimension full flow equation as written in eq(\ref{flow_diff_separate}) into a   dimensionless  flow equation.  To do so, let us define a new dimensionless coordinate, $u$, which is  related to $r$ as $r=r_h/(1-u)$. In which case the horizon and the boundary is moved to $u=0$ and $u=1$, respectively.

Let us evaluate the following quantities,  using the RN AdS black hole solution
\be
\Sigma_A=(r/L)^{d-3}\f{1}{g^2_{YM}}=(r_h/L)^{d-3}\f{(1-u)^{3-d}}{g^2_{YM}};\quad \sqrt{g_{rr}/g_{tt}}=L^2/(r^2h)=L^2/(r^2_h ~h)(1-u)^2,
\ee 
where we have dropped the argument of the function, $h$, for simplicity and has a structure
\be
h=1-(1-u)^d-q^2(1-u)^d+q^2 (1-u)^{2d-2},\quad q\equiv Q/r^{d-1}_h.
\ee

Note that $q$ is dimensionless. So, the LHS and the RHS of the first flow eq(\ref{flow_diff_separate}) becomes
\bea
\p_r(Re\sigma^{xx})&=&\f{(1-u)^2}{r_h}\p_{u}(Re\sigma^{xx})\nn -\f{2\omega}{\Sigma_A} \sqrt{\f{g_{rr}}{g_{tt}}}\bigg(Re\sigma^{xx}\bigg)~\bigg(Im\sigma^{xx}\bigg)
&=&-2\omega \bigg(\f{L}{r_h}\bigg)^{d-1}\f{g^2_{YM}}{h} (1-u)^{d-1}\bigg(Re\sigma^{xx}\bigg)~
\bigg(Im\sigma^{xx}\bigg)
\eea

Combining everything together
\be
\p_{u}(Re\sigma^{xx})=-\f{2\omega L^{d-1}g^2_{YM}}{r^{d-2}_hh}(1-u)^{d-3}\bigg(Re\sigma^{xx}\bigg)~
\bigg(Im\sigma^{xx}\bigg).
\ee 

Let us define the dimensionless conductivity, the YM's coupling and the frequency as
\be
\sigma^{xx} L^{d-3}={\tilde\sigma}^{xx};\quad g^2_{YM}=r^{d-3}_h {\hat g}^2_{YM} ;\quad {\bar\omega}=\omega L
\ee

Using this set of dimensionless variables, we find the first of the flow eq(\ref{flow_diff_separate})

\be
\p_{u}(Re{\tilde\sigma}^{xx})=-\f{2{\bar\omega} L{\hat g}^2_{YM}}{r_hh}(1-u)^{d-3}\bigg(Re{\tilde\sigma}^{xx}\bigg)~
\bigg(Im{\tilde\sigma}^{xx}\bigg).
\ee

Instead of studying the flow as a function of the dimensionless frequency, ${\bar\omega}$, we can  study it as a function of ${\hat\omega}\equiv\omega/T=4\pi (L/r_h) {\bar\omega}/[d-(d-2)q^2]\equiv  (L/r_h) ({\bar\omega}/m_1)$. It means the flow equation becomes
\be
\p_{u}(Re{\tilde\sigma}^{xx})=-\f{2{\hat\omega}}{h} m_1{\hat g}^2_{YM}(1-u)^{d-3}\bigg(Re{\tilde\sigma}^{xx}\bigg)~
\bigg(Im{\tilde\sigma}^{xx}\bigg).
\ee

The on-shell value of the quantity $A'^2_0/({\tilde g}^2_{YM}g_{rr}\omega^2)$ is
\be
\f{A'^2_0}{{\tilde g}^2_{YM}g_{rr}\omega^2}=\f{\mu^2(d-2)^2h}{{\tilde g}^2_{YM}L^2\omega^2}(1-u)^{2(d-2)}=\f{\mu^2(d-2)^2h}{{\tilde g}^2_{YM}{\bar\omega}^2}(1-u)^{2(d-2)}.
\ee

Using this, we get the other flow equation as
\bea
\p_{u}(Im{\tilde\sigma}^{xx})&=&\f{{\hat\omega}m_1}{h}{\hat g}^2_{YM}(1-u)^{d-3}\Bigg[\bigg(Re{\tilde\sigma}^{xx}\bigg)^2-
\bigg(Im{\tilde\sigma}^{xx}\bigg)^2-\f{(1-u)^{2(3-d)}}{{\hat g}^4_{YM}}\Bigg]+\nn
&&\f{2(d-1)(d-2)q^2}{{\hat g}^2_{YM} m_1{\hat\omega}}(1-u)^{d-1}.
\eea

The boundary condition, eq(\ref{bc_flow_fluctuating_geometry_pll}),  takes the following form
\bea
Re{\tilde\sigma}^{xx}(r_h,{\hat\omega})&=&\bigg(\f{L}{r_h}\bigg)^{d-3}\f{g^{(d-3)/2}_{xx}(r_h)}{{\hat g}^2_{YM}}=\Bigg(\f{(1-u)^{3-d}}{{\hat g}^2_{YM}}\Bigg)_{u=0}=\f{1}{{\hat g}^2_{YM}(u=0)},\nn
Im{\tilde\sigma}^{xx}(r_h,{\hat\omega})&=&0.
\eea

Finally, the real part of $\sigma^{xx}$ is determined by the following differential equation
\bea\label{z_1_flow_sigma_rn}
&&\p_{u}\bigg[\f{h(1-u)^{3-d}}{{\hat g}^2_{YM}}\f{\p_{u}(Re{\tilde\sigma}^{xx})}{Re{\tilde\sigma}^{xx}} \bigg]=-\f{2{\hat\omega}^2m^2_1}{h}{\hat g}^2_{YM}(1-u)^{d-3}\bigg[(Re{\tilde\sigma}^{xx})^2-
\f{h^2(1-u)^{2(3-d)}}{4{\hat \omega}^2{\hat g}^4_{YM}m^2_1}\times\nn
&&\bigg(\f{\p_{u}(Re{\tilde\sigma}^{xx})}{Re{\tilde\sigma}^{xx}}\bigg)^2-\f{(1-u)^{2(3-d)}}{{\hat g}^4_{YM}}\bigg]-\f{4q^2(d-1)(d-2)}{{\hat g}^2_{YM}}(1-u)^{d-1}
\eea

\section{Appendix B: Dimensional analysis }

Let us do a little bit of  the dimensional analysis for various  physical quantities. If the $d$ dimensional field theory spacetime coordinates (i.e the   $d+1$ dimensional bulk spacetime has an invariant length, $ds^2=L^2[-r^{-2z}dt^2+r^{-2}dx_i^2+r^{-2}dr^2]$) behaves under scaling as
\be
t~\rightarrow~ \lambda^z ~t,~~~x_i~\rightarrow~ \lambda~x_i,
\ee

then  the physical quantities possesses the following length dimension 
\bea
&&[t]=z,~[x_i]=1,~[J^t]=1-d,~[J^i]=2-z-d,~[A_t]=-z,
~[A_i]=-1, ~
 [E_i]=-1-z,\nn &&~[B_i]=-2,~ [T]=-z=[\omega],~[F]=-z,~[\sigma^{ij}]=3-d,
\eea 
where $J^t,~J_i,~A_t,~A_i,~E,~B,~T,~\omega,~F,~\sigma$ are charge density, current density, time component of the gauge 
potential,  $x_i$-component of the gauge potential, electric field, magnetic field, temperature, frequency,  free energy and conductivity respectively. The two form field strength has the following form i.e, $F_2=-E_{x_i} dt\w dx_i+B_{x_k}  dx_i\w dx_j$.

Using these observables one can construct an expression to spatial current density using dimensional analysis as follows
\bea\label{generic_current}
J&=& E^{\f{d-2+z}{1+z}} Y_1[E/T^{1+\f{1}{z}},~B/T^{\f{2}{z}},~\omega/T,~\mu/T],\nn
&=& B^{\f{d-2+z}{2}} Y_2[E/T^{1+\f{1}{z}},~B/T^{\f{2}{z}},~\omega/T,~\mu/T],\nn
&=& T^{\f{d-2+z}{z}} Y_3[E/T^{1+\f{1}{z}},~B/T^{\f{2}{z}},~\omega/T,~\mu/T],\nn
&=& \bigg(J^t\bigg)^{\f{d-2+z}{d-1}} Y_4[E/T^{1+\f{1}{z}},~B/T^{\f{2}{z}},~\omega/T,~\mu/T],
\eea

where the  functions $Y_i$'s are dimensionless objects whose explicit forms  are not known. 

\subsection{Dimensionless flow of conductivity in DBI action}

In writing down the  flow equation for the conductivity, we have set a quantity $\lambda$ to unity as well as the tension of the brane. To reinstate it, we need to make the following replacement
\be\label{recipe_tension_dbi}
A'_0\rightarrow \lambda A'_0,\quad B\rightarrow \lambda B,\quad (\prod_ig_{z^iz^j})\rightarrow T^2_b (\prod_ig_{z^iz^j}).
\ee

Let us assume the background metric is of the form
\be
ds^2_{d+1}=L^2\bigg[-\f{h(r)dt^2}{r^{2z}}+\f{dx^2+dy^2+dz^2_i}{r^2}+\f{dr^2}{r^2h(r)} \bigg],\quad h(r)=1-\bigg(\f{r}{r_h}\bigg)^{d+z-1}, \quad T=\alpha r^{-z}_h,
\ee
where $r_h$ and $T$ are the size of the horizon and temperature, respectively and $\alpha=\f{d+z-1}{4\pi}$. Let us define a dimensionless coordinate, $u$, as $r=r_h(1-u)$. In this new coordinate system, the horizon is at $u=u_h=0$ and the boundary is at $u=1$. Define a dimensionless frequency as ${\hat \omega}=\omega/T=\f{4\pi}{d+z-1}{\bar\omega} (r_h/L)^z$, where ${\bar\omega}=\omega L^z$. 

The dimensionless conductivity is defined as $\sigma={\tilde\sigma} L^{3-d}$ and the dimensionless RG flow equation takes the following form
\bea\label{flow_re_im_long_hall_dbi_b_dim_less}
\p_u Re{\tilde\sigma}^{xy}&=&-{\hat\omega}{\tilde{\tilde X}}_1[  Im{\tilde\sigma}^{xx}(X_2-{\tilde X}_3 Re{\tilde\sigma}^{xy})-  {\tilde X}_3 Re{\tilde\sigma}^{xx}Im{\tilde\sigma}^{xy}],\nn
\p_u Im{\tilde\sigma}^{xy}&=&-{\hat\omega}{\tilde{\tilde X}}_1[-  {\tilde X}_3 Im{\tilde\sigma}^{xx} Im{\tilde\sigma}^{xy}-  Re{\tilde\sigma}^{xx}(X_2-{\tilde X}_3 Re{\tilde\sigma}^{xy})],\nn
\p_u Re{\tilde\sigma}^{xx}&=&{\hat\omega} {\tilde{\tilde X}}_1[X_2 Im{\tilde\sigma}^{xy}-{\tilde X}_3(Re{\tilde\sigma}^{xy}
Im{\tilde\sigma}^{xy}-Re{\tilde\sigma}^{xx}Im{\tilde\sigma}^{xx}) ] ,\nn
\p_u Im{\tilde\sigma}^{xx}&=&-\f{{\hat\omega}}{2} {\tilde{\tilde X}}_1\bigg[2X_2 Re{\tilde\sigma}^{xy}-\nn && {\tilde X}_3\bigg((Re{\tilde\sigma}^{xy})^2-(Im{\tilde\sigma}^{xy})^2-
(Re{\tilde\sigma}^{xx})^2+(Im{\tilde\sigma}^{xx})^2\bigg)-
{\tilde X}_7\bigg],
\eea
where ${\tilde{\tilde X}}_1=\f{2\alpha (1-u)^{z+1}}{h(u) (L/r_h)^{2}},~{\tilde X}_3=X_3 L^{3-d},~ {\tilde X}_7=X_7 L^{d-3}$ and $h(u)=1-(1-u)^{d+z-1}$. The dimensionless boundary condition reads
\bea
Re{\tilde\sigma}^{xy}(u_h=0)&=&-\f{\lambda B{\tilde\rho}}{(L/r_h)^4+\lambda^2 B^2},\nn
 Re{\tilde\sigma}^{xx}(u_h=0)&=&\pm \bigg(\f{L}{r_h}\bigg)^2\f{\sqrt{{\tilde\rho}^2+{\tilde T}^2_b (\f{L}{r_h})^{2(d-3)}((L/r_h)^4+\lambda^2 B^2)}}{(L/r_h)^4+\lambda^2 B^2},
\eea
where ${\tilde\rho}=\rho L^{d-3}$ and ${\tilde T}_b=T_b L^{d-3}$. The form of the ${\tilde X}_i$'s are
\bea
 X_2&=&\f{\lambda B {\tilde\rho}}{\sqrt{{\tilde\rho}^2+{\tilde T}^2_b (\f{L}{r_h})^{2(d-3)}(1-u)^{2(3-d)}[(L/r_h)^4(1-u)^{-4}+\lambda^2 B^2]}},\nn {\tilde X}_3&=&\f{(L/r_h)^4(1-u)^{-4}+\lambda^2 B^2}{\sqrt{{\tilde\rho}^2+{\tilde T}^2_b (\f{L}{r_h})^{2(d-3)}(1-u)^{2(3-d)}[(L/r_h)^4(1-u)^{-4}+\lambda^2 B^2]}},\nn
{\tilde X}_7&=&\f{{\tilde\rho}^2+{\tilde T}^2_b (\f{L}{r_h})^{2(d-1)}(1-u)^{2(d-1)}}{\sqrt{{\tilde\rho}^2+{\tilde T}^2_b (\f{L}{r_h})^{2(d-3)}(1-u)^{2(3-d)}[(L/r_h)^4(1-u)^{-4}+\lambda^2 B^2]}}.
\eea

\section{Appendix C: The in-falling boundary condition}

Here we shall show the solution to guage field equation of motion, eq(\ref{diff_eom_gauge_field}), close to the horizon satisfies eq(\ref{horizon_conductivity_maxwell_gauge_field}), i.e., the in-falling boundary condition. Let us assume that the metric components close to the horizon goes as 
\be\label{near_horizon_metric}
g_{tt}=c_t (r-r_h),\quad g_{rr}=c_r/(r-r_h),\quad g_{xx}=g_{yy}=g_{z^iz^i}=c_x.
\ee 
The equation of motion to gauge field,  eq(\ref{diff_eom_gauge_field}), can be rewritten as
\be
\p_r[GA'_x]+Gf_2A_x=0
\ee
The quantities, $G$ and $f_2 G$, close to the horizon behaves as
\be
G=\f{\sqrt{-g}}{g^2_{YM}g_{rr}g_{xx}}=\sqrt{\f{c_t}{c_r}} c^{(d-3)/2}_x (r-r_h)/g^2_{YM},\quad f_2 G=\omega^2\sqrt{\f{c_r}{c_t}}c^{(d-3)/2}_x [(r-r_h)g^2_{YM}]^{-1}
\ee

Hence the equation to gauge field close to the horizon obeys
\be\label{eom_field}
(r-r_h)A''_x+A'_x+\omega^2 (c_r/c_t)~ A_x/(r-r_h)=0
\ee

Let us, also,  look at the linearized equation of motion to gauge field written in eq(\ref{eom_A_plus_minus}), which upon using eq(\ref{sol_at_unperturbed_dbi_b}), can be rewritten as
\bea
&&\p_r\Bigg[\f{\sqrt{\rho^2+(\prod g_{z^iz^j})(g^2_{xx}+B^2)}}{\sqrt{g_{tt}g_{rr}}(g^2_{xx}+B^2)}\Bigg(g_{tt}g_{xx}\p_r A^{(1)}_{\pm} \mp \f{\omega B\rho\sqrt{g_{tt}g_{rr}}A^{(1)}_{\pm}}{\sqrt{\rho^2+(\prod g_{z^iz^j})(g^2_{xx}+B^2)}}\Bigg)\Bigg]+\nn &&
\f{\sqrt{\rho^2+(\prod g_{z^iz^j})(g^2_{xx}+B^2)}}{\sqrt{g_{tt}g_{rr}}(g^2_{xx}+B^2)}\Bigg(\omega^2 g_{rr}g_{xx} A^{(1)}_{\pm}\pm \f{\omega B\rho\sqrt{g_{tt}g_{rr}}\p_rA^{(1)}_{\pm}}{\sqrt{\rho^2+(\prod g_{z^iz^j})(g^2_{xx}+B^2)}}\Bigg)\Bigg]=0.
\eea

Upon using the near horizon behavior to metric components as written in eq(\ref{near_horizon_metric}) gives us eq(\ref{eom_field}). In fact the massless scalar field close to the horizon also obeys the same equation. The solution to eq(\ref{eom_field}) is $A_x=A_{\pm}\equiv A(r,\omega)=A^{(0)}(r_h)(r-r_h)^{\pm i\omega \sqrt{c_r/c_t}}$, where $A^{(0)}(r_h)$ is an overall normalization factor.  There follows,  $\p_r A/A=(\pm i\omega \sqrt{c_r/c_t}) 1/(r-r_h)$. It is easy to check that this ratio is consistent with eq(\ref{horizon_conductivity_maxwell_gauge_field}) and eq(\ref{near_horizon_dbi_b_ratio_gauge_field}). The in-falling boundary condition corresponds to choosing the negative sign in the solution. Thus, the in-falling boundary condition is same as the regularity condition at the horizon. 

\section{Appendix D: Green function}

In this section, we shall present a huristic derivation of the scattering time and self energy by comparing the fermionic Green function with that of the bosonic. 
Even though this is puzzling, however, if we use the idea of the formation of  cooper pairs by some hidden mechanism then it looks fine and,  interestingly,  gives the correct scattering time.
   
Let us assume that a bunch of free or weakly coupled fermions are interacting  strongly with another set of fermions and the  form of the Green's function of these  fermions remain the  same as  in the weakly coupled regime, because of the factorization in the large N limit as argued in \cite{fp}.  We take the  Green's function as \cite{dhs} 
\be
G^R(k,\omega)=\f{Z_k}{\omega-{\varepsilon}_k+i\Gamma_k},
\ee
where $Z_k=(1-\f{\p Re\Sigma}{\p\omega})^{-1}$, ${\varepsilon}_k=Z_k(\epsilon_k+Re\Sigma)$, and $\Gamma_k=Z_k |Im\Sigma|$. The self energy is defined as $\Sigma=Re\Sigma+i Im \Sigma$. The quantity $\epsilon_k$ and ${\varepsilon}_k$
are the bare and renormalized fermi energy, respectively.
Now using $G^R(k,\omega)=<J_i,J_i>=-i\omega\sigma(k,\omega)$ and equating the Green's function, results in the imaginary part of the self-energy as
\be\label{im_self_energy}
|Im\Sigma|=-\f{Im G^R}{(ReG^R)^2+(ImG^R)^2}=\f{Re\sigma}{\omega[(Re\sigma)^2+(Im\sigma)^2]},
\ee
where we have used $Re G^R=\omega Im \sigma$ and $Im G^R=-\omega Re \sigma$. The real part of the self-energy goes as
\be\label{re_self_energy}
\f{\p Re \Sigma}{\p\omega}+\f{\epsilon_k+Re\Sigma}{\omega}-1=-\f{Re G^R}{\omega[(ReG^R)^2+(ImG^R)^2]}=-\f{Im\sigma}{\omega^2[(Re\sigma)^2+(Im\sigma)^2]}
\ee

In order to determine the $Re\sigma$ and the $Im \sigma$ in terms of the self energy, $Re\Sigma$ and $Im \Sigma$, we have to solve eq(\ref{im_self_energy}) and eq(\ref{re_self_energy}). The solution reads for
the $Im \sigma$ as
\be\label{cubic_eq_im_sigma}
Im\sigma=\f{\bigg(1-\f{\p Re \Sigma}{\p\omega}-\f{\epsilon_k+Re\Sigma}{\omega}\bigg)}{|Im\Sigma|^2+\omega^2\bigg(1-\f{\p Re \Sigma}{\p\omega}-\f{\epsilon_k+Re\Sigma}{\omega}\bigg)^2},
\ee

and for the $Re\sigma$ is 
\be\label{re_sigma}
 Re\sigma=\f{|Im\Sigma|}{\omega\bigg[\omega^2\bigg(1-\f{\p Re \Sigma}{\p\omega}-\f{\epsilon_k+Re\Sigma}{\omega}\bigg)^2+|Im\Sigma|^2\bigg] }.
\ee

The scattering rate, whose inverse determines the  life time of the quasi particles, is determined by
\bea\label{decay_rate}
\Gamma_k&=&-(\omega-\varepsilon_k)Im G^R/Re G^R=-\f{\omega|Im \Sigma|Im G^R}{|Im \Sigma| Re G^R-(\epsilon_k+Re\Sigma)Im G^R},\nn
&=&\f{\omega|Im \Sigma|Re \sigma}{|Im \Sigma| Im \sigma+(\epsilon_k+Re\Sigma)Re \sigma}=\f{\omega Re \sigma}{Im \sigma+\omega[(Re\sigma)^2+(Im\sigma)^2](\epsilon_k+Re\Sigma)}.
\eea

It matches with the definition of $1/\tau^{\star}\simeq\omega Re\sigma/Im \sigma$ in \cite{dvm}, when $(\epsilon_k+Re\Sigma)Re \sigma \ll |Im \Sigma| Im \sigma$ in eq(\ref{decay_rate}) and the approximate equality becomes exact equality for $\epsilon_k=-Re\Sigma$. However, when $(\epsilon_k+Re\Sigma)Re \sigma \gg |Im \Sigma| Im \sigma$, then $1/\tau\simeq \f{\omega|Im \Sigma|}{(\epsilon_k+Re\Sigma)}=\f{Re\sigma}{(\epsilon_k+Re\Sigma)[(Re\sigma)^2+(Im\sigma)^2]}$.
It is interesting to note that for $\epsilon_k+Re\Sigma\simeq 1$, and $\omega[(Re\sigma)^2+(Im\sigma)^2]> Im\sigma$, the frequency dependent scattering rate becomes 
\be\label{time_dvdm}
\tau^{-1}_{DvdM}\simeq \f{Re\sigma}{(Re\sigma)^2+(Im\sigma)^2},
\ee 
which matches with that used in \cite{dvdm}.  The quantity $Z_k$ is
\bea
Z_k&=&(\omega-\varepsilon_k)\f{[(Re G^R)^2+(Im G^R)^2]}{Re G^R}=\omega(\omega-\varepsilon_k)\f{[(Re \sigma)^2+(Im \sigma)^2]}{Im \sigma},\nn
&=&\f{\omega^2[(Re\sigma)^2+(Im\sigma)^2]}{Im \sigma+\omega[(Re\sigma)^2+(Im\sigma)^2](\epsilon_k+Re\Sigma)}
\eea

If we assume that the conductivity at zero momentum and at low frequency has an algebraic  behavior like
\be\label{power_law_type_conductivity_greens_function}
Re\sigma\sim \omega^{2\nu_1-1},~~~Im\sigma\sim \omega^{2\nu_2-1}.
\ee
then from eq(\ref{im_self_energy}), there follows that 
\be
|Im \Sigma|=\f{1}{\omega^{2\nu_1}+\omega^{2(2\nu_2-\nu_1)}}
,
\ee
and from eq(\ref{re_self_energy})
\be
\f{\p Re \Sigma}{\p\omega}+\f{\epsilon_k+Re\Sigma}{\omega}-1=-\f{1}{\omega^{4\nu_1-2\nu_2+1}
+\omega^{2\nu_2+1}},
\ee

where we have set the phase of conductivity to unity for simplicity.Let us try to find the form of the self energy by assuming the form of the conductivity to be of the form eq(\ref{power_law_type_conductivity_greens_function}).\\

{\bf choice 1}: $\nu_1\equiv\nu >0$,  $\nu_2=0$ and with $\epsilon_k=0$\\

In this case the  self energy is
\be
|Im \Sigma|\simeq\f{\omega^{2\nu}}{1+\omega^{4\nu}},\quad Re \Sigma=\f{{\rm constant}}{\omega}+\f{\omega}{2}-{}_2F_1\bigg(\f{1}{4\nu},1,1+\f{1}{4\nu},-\omega^{4\nu}\bigg),
\ee
where ${}_2F_1[a,b,c,z]$ is the hypergeometric function. 
\\

{\bf choice 2}: $\nu_1=\nu_2\equiv\nu >0$,   and with $\epsilon_k=0$\\

Upon using the experimental result  \cite{dvdm}, especially from the mid frequency regime eq(\ref{mid_frequency}),   it gives
\be
|Im\Sigma|\sim \omega^{-2\nu},\quad Re \Sigma=\f{{\rm constant}}{\omega}+\f{\omega^{-2\nu}}{2(2\nu-1)}+\f{\omega}{2}\\
\ee

{\bf choice 3:} $\nu_1\equiv\nu >0$,  $\nu_2=0$ and with $\epsilon_k=-Re\Sigma$\\

For  $\varepsilon_k=0$, implies $\Gamma_k=\f{\omega Re\sigma}{Im\sigma}$ and $Z_k=\f{\omega^2}{Im \sigma}[(Re \sigma)^2+(Im \sigma)^2]$. The self energy is
\be
|Im \Sigma|\simeq\omega^{2\nu},\quad Re\Sigma={\rm constant}+\omega-\f{Log~\omega}{4\nu}+\f{(4\nu-1)}{4\nu}Log\omega+\f{Log(1+\omega^{4\nu})}{4\nu}
\ee 

{\bf choice 4}: $\nu_1=\nu_2\equiv\nu >0$,   and with $\epsilon_k=-Re\Sigma$

The self energy goes as
\be
|Im\Sigma|\sim \omega^{-2\nu},\quad Re \Sigma={\rm constant}+\omega+\f{\omega^{-2\nu}}{4\nu}.\\
\ee

{\bf choice 5}: $\nu_1=\nu_2\equiv\nu >0$,   and with $\epsilon_k+Re\Sigma=1$

The self energy goes as
\be
|Im\Sigma|\sim \omega^{-2\nu},\quad Re \Sigma={\rm constant}-Log\omega+\omega+\f{\omega^{-2\nu}}{4\nu}.\\
\ee

\end{document}